\documentclass{aa}
\usepackage{graphicx}
\DeclareGraphicsExtensions{.eps,.ps,.pdf}
\usepackage{psfrag}
\usepackage{txfonts}
\usepackage{subfigure}
\usepackage{float}
\usepackage{natbib}
\bibpunct{(}{)}{;}{a}{}{,}
\usepackage{balance}
\begin{document}

\newcommand{\XMM}{XMM-\textit{Newton} }
\newcommand{\IRAS}{\textit{IRAS} }
\newcommand{\ISO}{\textit{ISO} }
\newcommand{\Spitzer}{\textit{Spitzer} }

   \title{Spectral Energy Distribution of Hyper-Luminous Infrared Galaxies}


   \author{A. Ruiz
          \inst{1,2}
	  \and
	  G. Miniutti
          \inst{3}
	  \and
	  F. Panessa
          \inst{4}
          \and
	  F.J. Carrera
          \inst{1}
         }

   \institute{Instituto de F\'isica de Cantabria (IFCA), CSIC-UC,
              Avda. de los Castros, 39005 Santander, Spain
	  \and
	     Istituto Nazionale di Astrofisica (INAF), Osservatorio Astronomico di Brera,
	     via Brera 21, 20121 Milano, Italy
	  \and
	      LAEX, Centro de Astrobiolog\'ia (CSIC-INTA);
              LAEFF, P.O: Box 78, E-28691 Villanueva de la Ca\~nada, Madrid, Spain
	  \and
	     Istituto Nazionale di Astrofisica (INAF), IASF-Roma,
             Via Fosso del Cavaliere 100, I-00133 Rome, Italy \\
          \email{angel.ruiz@brera.inaf.it}
             }

  \date{Received March 31, 2009; accepted March 2, 2010}

  \abstract
   {} 
    {The relationship between star formation and super-massive black
    hole growth is central to our understanding of galaxy formation
    and evolution. Hyper-Luminous Infrared Galaxies (HLIRGs) are
    unique laboratories to investigate the connection between starburst (SB)
    and Active Galactic Nuclei (AGN), since they exhibit extreme star
    formation rates, and most of them show evidence of harbouring powerful AGN.}
    {Our previous X-ray study of a sample of HLIRGs shows that the X-ray 
    emission of most of these sources is dominated by AGN activity. To 
    improve our estimate of the relative contribution of the AGN and SB 
    emission to its total bolometric output, we have built multi-wavelength
    (from radio to X-rays) spectral energy distributions (SEDs) for these HLIRGs,
    and we have fitted standard empirical AGN and SB templates to these SEDs.}
   {In broad terms, most sources are well fitted using this method, and we found AGN and
    SB contributions similar to those obtained by previous studies of HLIRGs. We
    have classified the HLIRGs SEDs in two groups, named class A and class B.
    Class A HLIRGs show a flat SED from the optical to the infrared energy range. Three out
    of seven class A sources can be modelled with a pure luminosity-dependent QSO
    template, while the rest of them require a type 1 AGN template and a SB template.
    The SB component is dominant in three out of four class A objects. Class B HLIRGs show 
    SEDs with a prominent and broad IR bump.
    These sources can not trivially be modelled with a combination of pure AGN and
    pure SB, they require templates of composite objects, suggesting that $\gtrsim50\%$
    of their emission comes from stellar formation processes.}
   {We propose that our sample is actually composed by three different populations:
    very luminous QSO (class A objects with negligible SB contribution), young 
    galaxies going through their maximal star formation period (class A objects with 
    significant SB emission) and the high luminosity tail of ULIRG population
    distribution (class B sources).}


   \maketitle

\section{Introduction}
\label{sec:intro}
During the last decade, the hypothesis that Active Galactic Nuclei (AGN) and 
galaxy formation and evolution are closely related has been supported by a 
growing body of observational evidences. On one hand, most galaxies have been 
shown to harbour a central super-massive black hole \citep{Kormendy01} whose mass 
is correlated with that of the host galaxy spheroid \citep{Magorrian98,McLure02} 
and, on the other hand, the evolution of cosmic star formation and of luminous 
AGN activity appear rather similar \citep{Fran99,Silverman05}. These hints
clearly suggest a connection between the growth of the central black
hole through accretion and the growth of the spheroid through star
formation.

The observational study of these two phenomena needs penetrating radiation 
like X-rays, mid-infrared (MIR), far-infrared (FIR) or sub-mm. On one hand,
star formation takes place in heavily obscured environments. Primary
radiation is then reprocessed by dust and re-emitted in the MIR-FIR
band. X-ray emission from starburst (SB) activity is enhanced by
energetic phenomena related to the final stages of stellar evolution,
e.g. supernova remnants or X-ray binaries \citep{Persic02}.
On the other hand, X-ray emission is the signature of AGN activity,
produced by black hole (BH) growth through accretion. However, synthesis models of
the X-ray background require that most AGNs in the Universe are
obscured \citep{Ueda03,Gilli07}, i.e. most of the accretion power in the
Universe is absorbed and then re-emitted in the infrared (IR) bands \citep{Fabian99}.

IR and X-ray observations are therefore essential to understand the phenomena of
star formation and AGN, as well as their interplay and connection. Fortunately, 
nowadays we have powerful tools to observe
the Universe in both energy ranges, like \textit{Chandra}, XMM-\textit{Newton},
\textit{Spitzer}, AKARI or \textit{Suzaku}. Different strategies can be
employed to investigate the IR/X-ray synergy and its effect on the AGN-galaxy
co-evolution, e.g. by multi-wavelength surveys like GOODS, AEGIS or COSMOS
\citep{Dickinson03,Davis07,Scoville07}, by targeted MIR observations of
peculiar X-ray sources like X-ray absorbed broad line QSO
\citep{Stevens05,Page07}, and by targeted X-ray observations of MIR/FIR-emitting
objects like Ultraluminous Infrared Galaxies \citep[ULIRGS,][]{Fran03,Teng05} or
Hyperluminous Infrared Galaxies \citep[HLIRGS,][]{Wilman98,Ruiz07}.

ULIRGs are a family of galaxies with IR luminosity $L_{\mathrm{IR}} \geq
10^{12}~L_{\sun}$, whose bolometric output is dominated by the emission in the IR
waveband (see \citealt{Lonsdale06} for a complete review). X-ray and IR
data clearly suggest that these sources are powered by SB and, in some
cases ($\sim50\%$), by AGNs \citep{Farrah03,Fran03,Teng05,Nardini08}. The
fraction of ULIRGs hosting an AGN increases with increasing IR
luminosity \citep{Veilleux95,Veilleux99}. Most of these objects are in
interacting systems, i.e. ULIRGs are most likely triggered by mergers of galaxies
\citep{Farrah01,Veilleux02}.

HLIRGs present an IR luminosity $L_\mathrm{IR} \geq 10^{13}~L_{\sun}$. 
These are among the most luminous objects in the Universe. Assuming that 
the FIR emission above $50~\mu$m is dominated by SB, their estimated 
star formation rates (SFR) are $\gtrsim1000~M_{\sun}~\mathrm{yr}^{-1}$ \citep{Rowan00}. 
IR and optical observations support that most harbour an AGN \citep{Verma02,Farrah02submm},
although the main power source is still controversial. As HLIRGs could
represent the most vigorous stage of galaxy formation, they are unique
laboratories to investigate extremely high stellar formation, and its
connection to super-massive black hole growth.

Only about a third of HLIRGs are located in interacting systems
\citep{Farrah02hst}, so a considerable number of these objects can not be classified just as
objects in the brightest end of the ULIRG population. They could be very young
galaxies experiencing their major episode of star formation \citep{Rowan00}, or
may be a completely new class of objects, e.g. a transient IR-luminous phase in
quasar evolution \citep{Farrah02submm,Stevens05}.

X-rays are a very convenient tool to disentangle the relative contribution 
of SB and AGN to the total output of HLIRGs. Only a few of these objects
had been studied in X-rays \citep{Wilman98,Wilman03,Iwasawa05} before
\citet{Ruiz07} presented the first systematic study of these sources 
in the X-ray band.

A sample of 14 HLIRGs was observed by \XMM and 10 were detected \citep{Ruiz07}. 
All of them show an AGN-dominated X-ray spectrum. We find X-ray thermal emission
associated with SB for just one source, while all ULIRGs show a SB 
component in their X-ray spectra \citep{Fran03}. The much brighter AGN emission probably
hides the X-rays originated in the SB (if this component actually exists). The IR 
luminosity of most HLIRGs of the sample is consistent with an AGN origin, but it 
is systematically over that expected for a local QSO \citep{Elvis94,Risaliti04} of 
the same X-ray luminosity. This IR excess could be due to X-ray obscuration, SB emission or 
may be due to an intrinsic difference between the spectral energy distribution (SED)
of AGN in HLIRGs and the SED of local QSO. 

To clarify these questions a proper study of the SED of these objects is
needed. Several studies of HLIRGs SED have been published
\citep{Rowan00,Verma02,Farrah02submm}, but they were always limited to the IR
energy range. These studies apply a two component model (AGN+SB) to reproduce the IR
emission, using radiative transfer models (RTM) for the AGN dust torus
\citep{Efstathiou95,Rowan95} and the SB \citep{Efstathiou00} components.
\citet{Rowan00} studied a sample of 45 HLIRGs, finding a continuum distribution in the
relative contribution of the AGN and SB components, from pure starburst to pure
AGN, with most objects being composite. On the other hand, \citet{Farrah02submm}
selected a complete sample of HLIRGs in a manner independent of obscuration,
inclination or AGN content and included sub-mm data (sub-mm data introduce a tight 
constraint on the SB luminosities), finding that all HLIRGs in the sample were 
composite objects.

In this paper we present a study of HLIRGs SED with two majors improvements
and one limitation compared with the earlier studies commented above: (a) we have greatly enlarged
the wavelength coverage, from radio to X-rays, and (b) we have significantly
increased the photometric data coverage. However, as a self consistent analytical model
able to reproduce the whole SED at so broad frequency range would be very
complex to compute (and beyond the scope of this paper), we have compared our
constructed SEDs with empirical AGN and SB templates, instead of using analytical RTM as 
in previous studies.

The paper is organized as follows. Section~\ref{sec:sample} briefly describes
the HLIRG sample. Section~\ref{sec:data} explains how we built the SED and the
data used to this end, and Sect.~\ref{sec:fitting} the methods we have employed
to model the SEDs. Results are presented in Sect.~\ref{sec:results}, compared
with previous studies of HLIRGs in Sect.~\ref{sec:comp} and discussed in
Sect.~\ref{sec:discuss}. Section \ref{sec:conc} summarizes our conclusions.

The \textit{Wilkinson Microwave Anisotropy Probe} (\textsl{WMAP})
concordance cosmology has been adopted throughout this paper: $H_0=70$ km
s$^{-1}$ Mpc$^{-1}$, $\Omega_m=0.27, \Omega_{\Lambda}=0.73$ \citep{Spergel03}.

\section{The HLIRG sample}
\label{sec:sample}
The sample studied here is the one investigated in \citet{Ruiz07}.
From the \citet{Rowan00} sample of HLIRGs we selected those sources with 
public data available in the \XMM Science Archive as of December 2004, 
and we added our own \XMM AO-5 observations.

We limited this sample to sources with redshift less than $\sim2$ to avoid
strong biasing towards high redshift quasars. Nevertheless, selecting the sample
by using the availability of X-ray data probably introduces a selection effect
in favour of the presence of an AGN. We also rejected one source from the
original sample, IRAS\,13279+3401. Using recent optical and MIR spectra, we have
determined its redshift to be $z\sim0.02$ (see Appendix A), much lower than the
one presented in the literature ($z= 0.36$, \citealt{Rowan00}). Therefore, our
estimate of its IR luminosity is $3\times10^{10}~L_{\sun}$, even below that
necessary to classify it as a LIRG. Hence, we have thirteen objects in our final
sample (see Table~\ref{tab:fits}).

According to their optical spectra (derived from the literature),
two sources are classified as starburst galaxies and twelve sources
present AGN features. Among the latter eight are classified as `type I',
and four of them as `type II'. All type II and one NL-SB galaxy are
Compton-Thick (CT) candidates. See \citet{Ruiz07} for a further discussion
on this sample.

\section{Data compilation}
\label{sec:data}
Our goal is to construct a well sampled SED for each object in a broad frequency range,
from radio to X-rays. To this end, we have carefully searched in the 
literature and in several astronomical databases. See Appendix~\ref{sec:tables} 
for a complete description on the origin of the photometry data for each HLIRG.

All data included in the SEDs (presented in Tables B.1-B.13, see Appendix~\ref{sec:tables})
have been converted to monochromatic flux density units, corrected for 
the Galactic reddening and blue-shifted to rest-frame.

\subsection{Radio}
Most of the HLIRGs in the sample have at least one observation in the radio range. These 
data come from different observations by VLA, ATCA, IRAM and other radio-telescopes.

\subsection{Infrared}
Our sources are well observed in the IR band. There are photometric data from \IRAS (Point 
Source Catalogue, \citealp{irasPSC}; Faint Source Catalogue, \citealp{irasFSC}) or \ISO for
all the objects. Most of them also have been observed with SCUBA in the sub-mm band \citep{Farrah02submm}.

In addition, there are public \Spitzer MIR data for several sources: IRAC photometric 
data and IRS spectra. We have reduced the IRAC data and we made our own photometric 
measurements. We have re-binned the IRS spectra of our HLIRG in broad bands, 
avoiding known emission and absorption features (a further analysis of these MIR spectra will be
presented in Ruiz et al., in preparation). Most of these sources also have NIR data from the
2MASS survey\footnote{http://www.ipac.caltech.edu/2mass/} \citep{2MASS}.

\subsection{Optical and UV}
Most of the optical data were obtained from the 
Sloan Digital Sky Survey-Data Release 5\footnote{http://www.sdss.org/dr5} \citep[SDSS-DR5,][]{sloan5} 
and SuperCOSMOS Sky Survey\footnote{http://www-wfau.roe.ac.uk/sss/} (SSS). A few data 
in V and B bands were taken from the \XMM Optical Monitor (OM).

We have only a few data in the UV range, mostly from the OM. Other data come from 
IUE and FUSE observations.

\subsection{X-ray}
The XMM-Newton spectra previously studied in \citet{Ruiz07} are
available. We have corrected each X-ray spectrum for the line of sight
Galactic absorption \citep{nh} and we have re-binned the data in just a few
energy bands\footnote{Through our X-ray data reduction we did not detect the
source IRAS\,14026+4341. Even so, this source has a counterpart in the 2XMMi
catalogue \citep{2xmmi}. We have considered the five energy band fluxes as in
the 2XMMi catalogue.}. In addition, the X-ray and the OM data come from
simultaneous observations, allowing us to check any variability effects.

\subsection{Overall description of the SEDs}
Figure~\ref{fig:seds} shows the SEDs we have built for our sources. We
have divided the sample in two classes accordingly to their optical spectral
classification. On one hand we grouped objects classified as type I AGN
(named class A sources) and on the other hand objects classified
as type II AGN and SB (named class B sources).

From a purely phenomenological point of view, class A and B sources seem
to show a different SED shape. Class A objects have a SED approximately flat
from the FIR to the optical range (the typical shape of quasars' SED), while
class B objects show a prominent broad IR bump dominating the emission over the
rest of the spectrum.

\begin{figure*}
\begin{center}
    \mbox{
     \subfigure[$F_\mathrm{X} / F_\mathrm{IR}$]{
      \includegraphics[width=.45\linewidth]{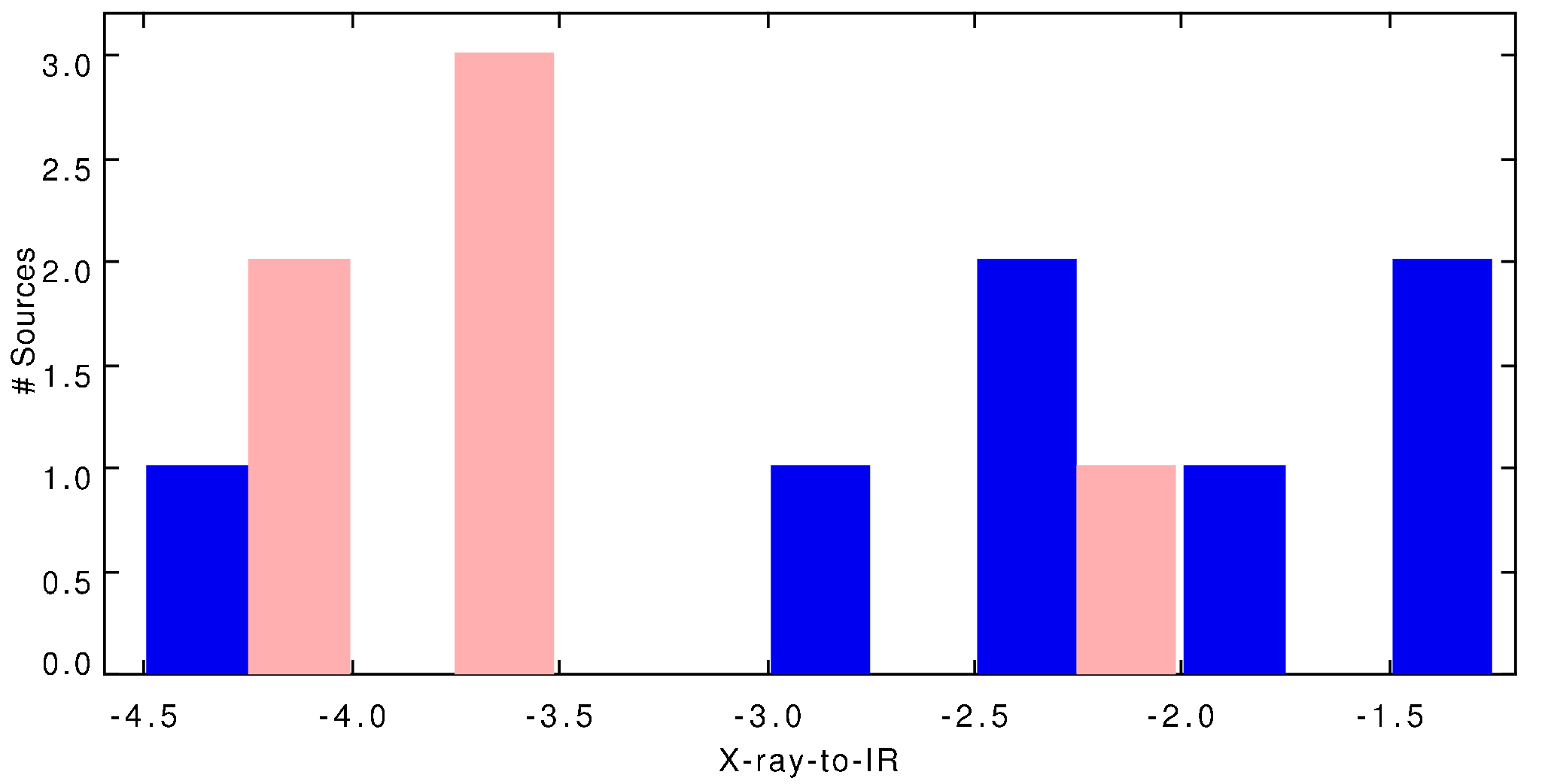}
      \label{fig:histxir}
     }
    \quad
     \subfigure[$F_\mathrm{opt} / F_\mathrm{IR}$]{
      \includegraphics[width=.45\linewidth]{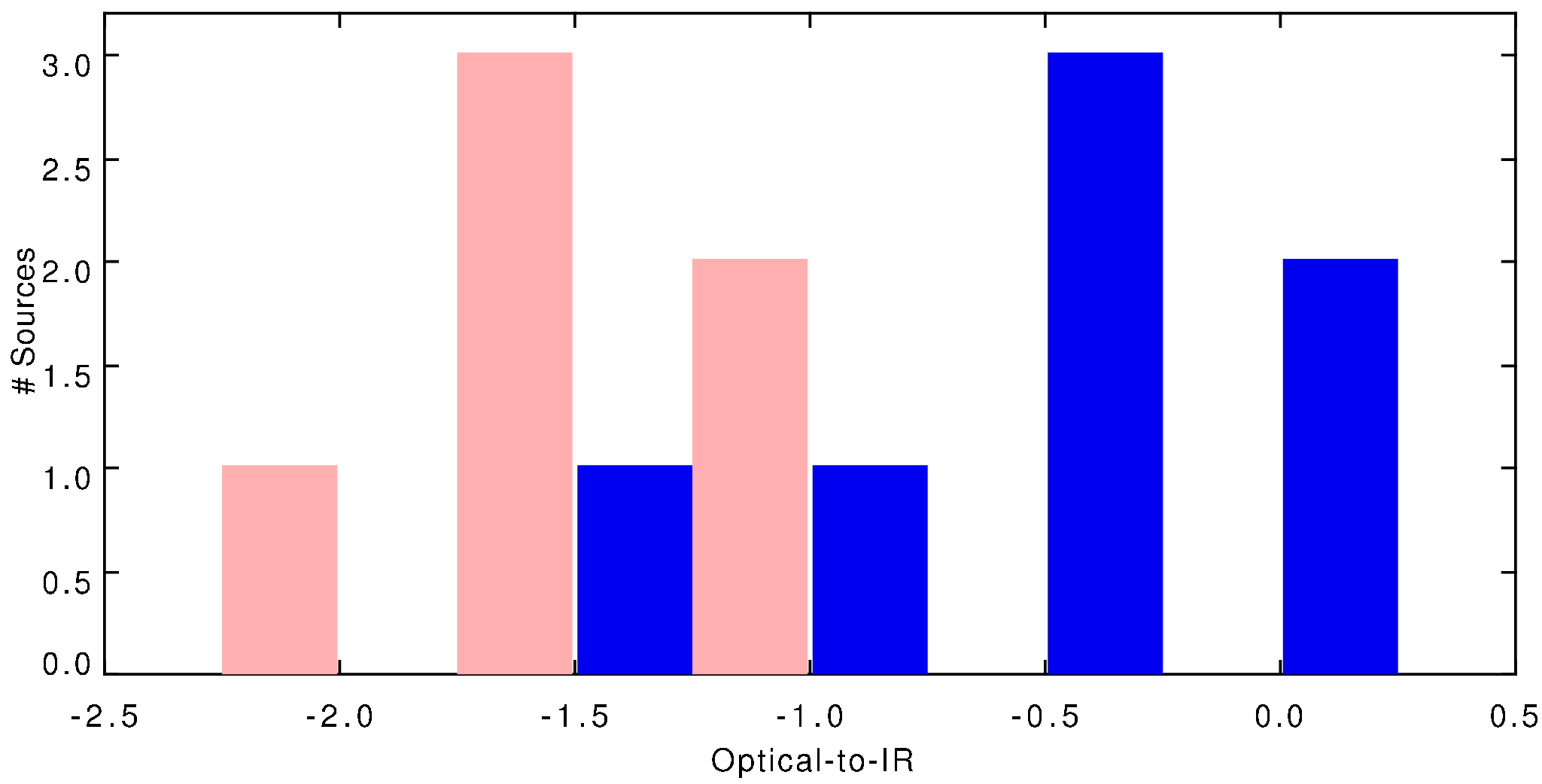}
      \label{fig:histoptir}
     }
    }

\caption{Distribution of (a) X-ray-to-IR and (b) Optical-to-IR flux ratios for class A (dark 
grey, blue in the colour version) and class B (light grey, pink in the colour version) HLIRGs.}
\label{fig:histograms}
\end{center}
\end{figure*}

To check if the above distinction holds quantitatively, we compared the
distribution of X-ray-to-IR and optical-to-IR flux ratios for class A and class
B sources. We estimated the monochromatic fluxes at three different rest-frame
wavelengths, in the IR ($30~\mu m$), optical (4400~\AA) and X-rays (2~keV)
through a linear interpolation of the SED (these points lie in well-sampled
regions of the SEDs, so these are reasonable estimates of the continua
at those energies). Fig~\ref{fig:histograms} shows the distribution of the
X-ray-to-IR ($F_\mathrm{X}/F_\mathrm{IR}$) and optical-to-IR
($F_\mathrm{opt}/F_\mathrm{IR}$) flux ratios for the class A (blue histogram)
and class B (pink histogram) sources. The distributions seem to be different for
both classes of HLIRGs.  By using a Kolmogorov-Smirnov test, the
probability that class A and class B samples come from different parent
populations is $92.6\%$ for the $F_\mathrm{X}/F_\mathrm{IR}$ distribution and
$\sim99.7\%$ for the $F_\mathrm{opt}/F_\mathrm{IR}$ distribution.

This rough analysis of the SED properties is clearly limited, but the results
seem to support our classification of HLIRGs in two classes. We suggest that,
since the SED classification is directly related to the optical spectra
classification, the distinct SED shape of HLIRGs could be explained by different
levels of obscuration in the line of sight and/or the relative contribution of
the SB emission to the total output.

\section{SED fitting}
\label{sec:fitting}
Once all the SEDs were built, our aim was to check for the presence of AGN
and/or SB emission in these sources and estimate the contribution of these
components to the total output. We fitted all SEDs by using the $\chi^2$
minimization technique with a simple model based on the use of convenient templates (see 
Sect.~\ref{sec:temp} for details). The fitting procedure and the SED templates
were implemented using the modelling and fitting tool \texttt{Sherpa} \citep{sherpa}, included in 
the software package CIAO 3.4\footnote{http://cxc.harvard.edu/ciao3.4/}.

Our model comprises two additive components, one associated to the AGN emission and the 
other associated to the SB emission. We can express this model as follows:
\begin{equation}
F_{\nu} = F_\mathrm{BOL} \left( \alpha~u_{\nu}^\mathrm{AGN} + (1-\alpha)~u_{\nu}^\mathrm{SB} \right), 
\end{equation}
where $F_\mathrm{BOL}$ is the total bolometric flux, $\alpha$ is the relative
contribution of the AGN to $F_\mathrm{BOL}$, $F_{\nu}$ is the total flux at the frequency
$\nu$, while $u_{\nu}^\mathrm{AGN}$ and $u_{\nu}^\mathrm{SB}$ are the normalized AGN and SB
templates (i.e., the value of the integral over the whole range of frequencies
is unity for each SED template). This model contains only two free parameters,
$F_\mathrm{BOL}$ (the normalization) and $\alpha$. The bolometric luminosity can be
estimated as $L_\mathrm{BOL}=4\pi D_{L}^2 F_\mathrm{BOL}$, where $D_{L}$ is the luminosity
distance.

The model we are adopting to fit the SED is somehow rough and does not provide a
precise description of the SED features, so we expect a poor fit in terms of
$\chi^2$ value. However, the entire SED shape, from the radio to soft gamma
rays, depends on a large number of physical parameters which produce different
SED shapes even among the same class of sources (AGN, SB, etc.). Moreover, the
impact of the different individual physical quantities on the overall SED and,
perhaps most importantly, the effect of their interplay and interaction on the
overall SED shape is far from being robustly settled from a theoretical point of
view. The development of an analytical or semi-analytical model would be of
great importance, but given that such models are difficult to build and likely
not unique, they clearly are beyond the scope of this work. We propose instead
the simpler template-fitting approach to discriminate, as a zeroth-order
approximation, the relative component contribution (AGN and/or SB) to the
overall bolometric luminosity of each source.

We have chosen the fit with the lowest reduced $\chi^2$ as our ``best fit''
model. As we said above, the value of $\chi^2/\mathrm{d.o.f.} >> 1$ even for
these best fits. Nevertheless, this quantity varies significantly for most sources
between the different combinations of templates we tested during the
$\chi^2$ minimization process. In those objects where different types of
templates obtained similar $\chi^2$ values, we have chosen the template most
consistent with previous results in the literature.

Our templates were chosen to minimize the contribution of the host galaxy's
non-SB stellar emission (see Sect.~\ref{sec:temp}), but there could still be a
remnant of this emission in the templates. Therefore, by adding two
different templates we could have summed twice this stellar emission. We checked
this effect adding a stellar template to the model\footnote{The SED of the
elliptical galaxy M~87 was employed to model the stellar emission.}. The
normalization of this component was free and negative, in order to subtract the
``second'' stellar contribution. The addition of the new component
did not change the final results of the SED fitting, so we can reject any
important stellar contamination in our templates.

\subsection{Templates}
\label{sec:temp}
The templates we have employed to model the SEDs of our sources are 
empirical SEDs of well observed SB and Seyfert (Sy) galaxies in the local 
universe (see Table~\ref{tab:temp}).

\begin{figure*}
\begin{center}
    \mbox{
     \subfigure[AGN I templates.]{
      \includegraphics[angle=-90,width=.45\linewidth]{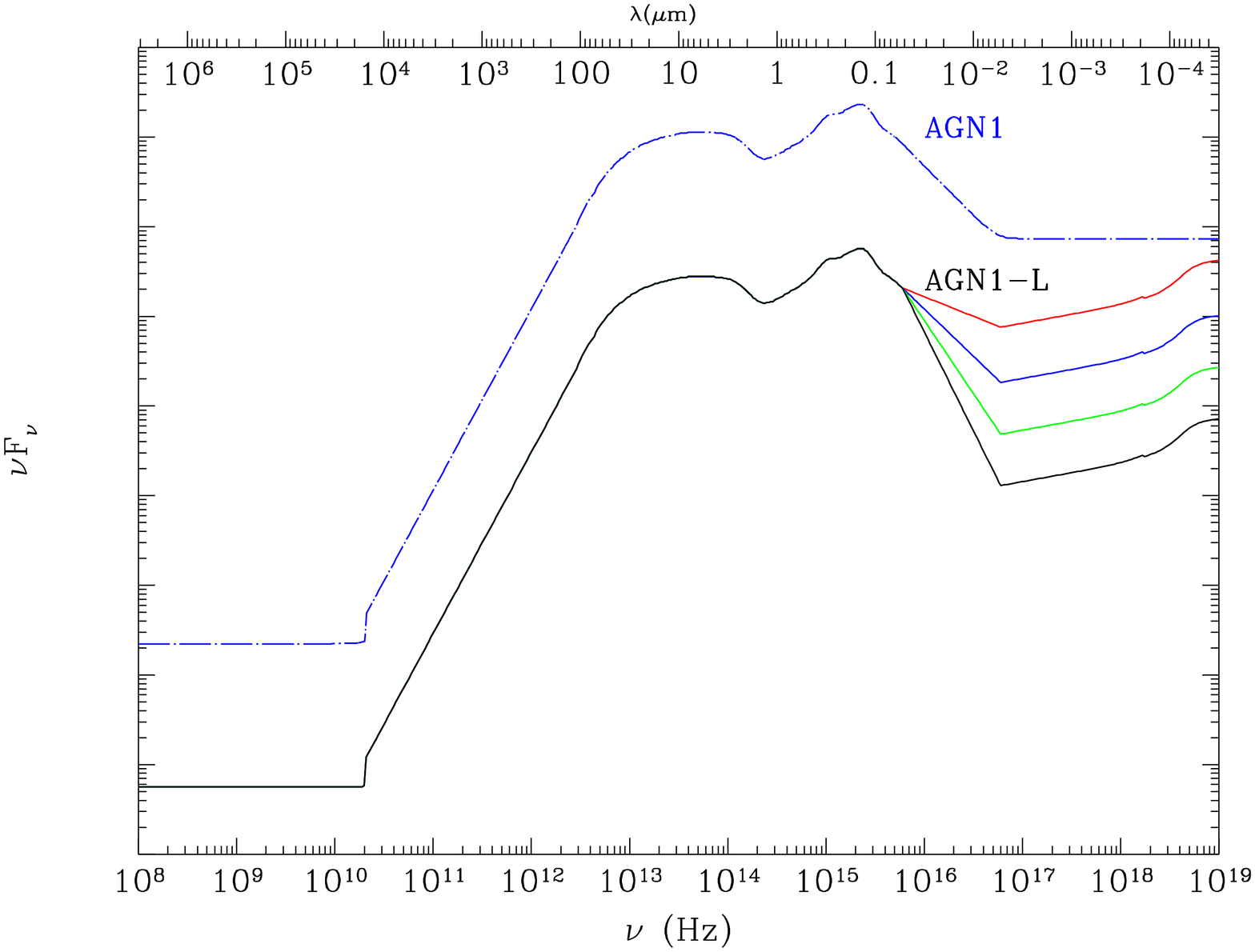}
      \label{fig:agnItemp}
     }
    \quad
     \subfigure[AGN II templates.]{
      \includegraphics[angle=-90,width=.45\linewidth]{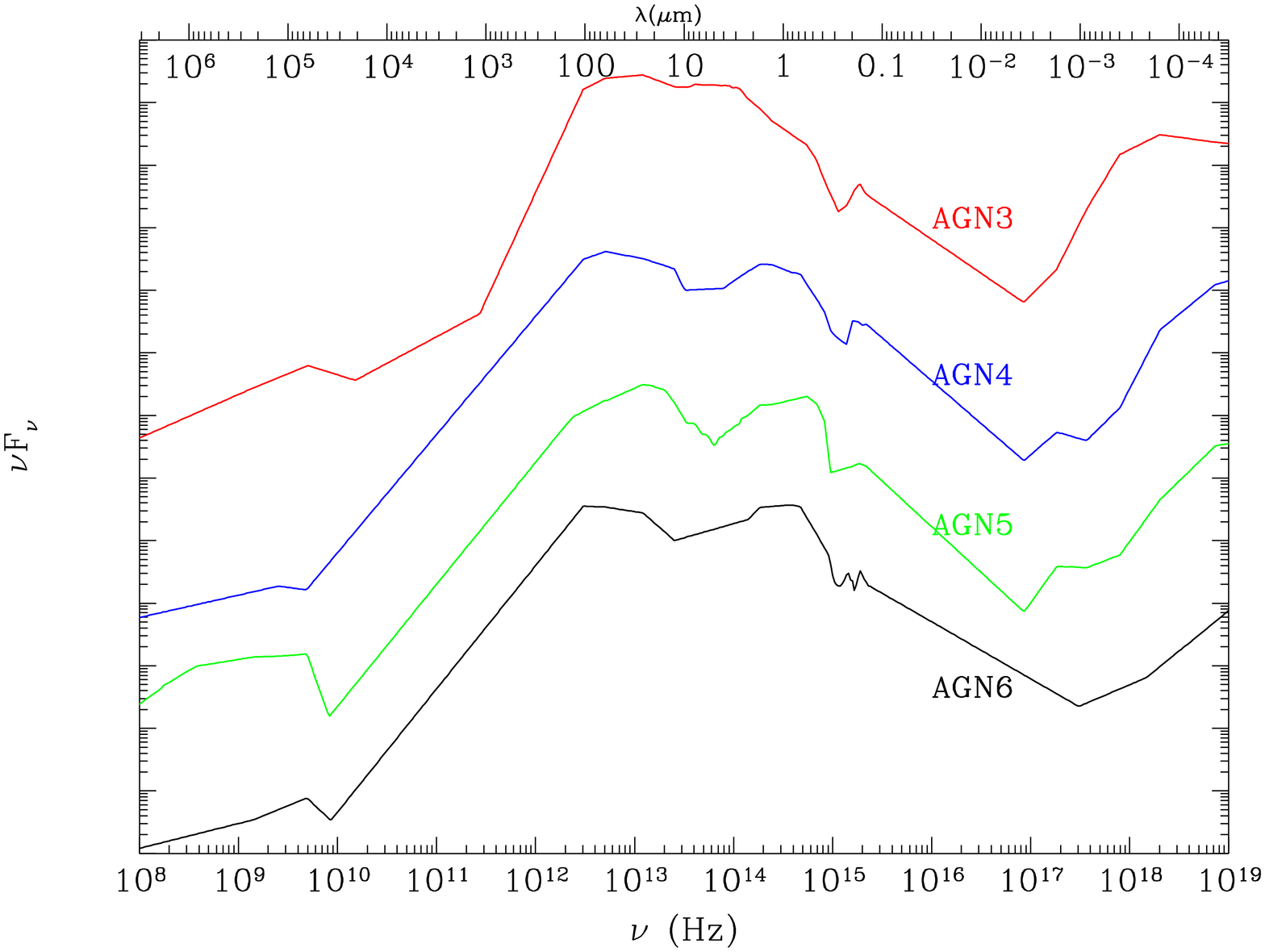}
      \label{fig:agnIItemp}
     }
    }

\caption{AGN templates.
(a) The top line (blue in the colour version) is the standard SED for radio quiet quasar
\citep[AGN1,][]{Richards06}. The group below is the luminosity-dependent SED for
quasar \citep[AGN1-L,][]{Hopkins07}, plotted for several bolometric luminosities
(the top line - red in the colour version - is for $10^{10}~L_{\sun}$ and the bottom 
black line is for $10^{16}~L_{\sun}$).
(b) Listed downwards: 
NGC\,5506 (AGN3, $N_\mathrm{H}=3\times10^{22}\mathrm{cm^{-2}}$), 
NGC\,4507 (AGN4, $N_\mathrm{H}=4\times10^{23}\mathrm{cm^{-2}}$), 
Mrk\,3 (AGN5, $N_\mathrm{H}=1.4\times10^{24}\mathrm{cm^{-2}}$), 
NGC\,3393 (AGN6, $N_\mathrm{H}>1\times10^{25}\mathrm{cm^{-2}}$).
The SED fluxes are shifted for clarity. See Sect.~\ref{sec:temp} for details.}
\label{fig:agntemps}
\end{center}
\end{figure*}

To reproduce the AGN contribution we used six AGN templates: 
\begin{enumerate}
 \item Two mean SED of radio quiet local QSO (Fig.~\ref{fig:agnItemp}): a luminosity independent SED
 \citep{Elvis94,Richards06} and a luminosity-dependent one \citep{Hopkins07}. The
 latter template is similar to the standard SED of QSO from \citet{Elvis94}, but
 the value of $\alpha_\mathrm{OX}$ depends on the bolometric luminosity
 \citep{Steffen06}, and the X-ray emission above 0.5~keV is modelled by a power
 law ($\Gamma=1.8$) with a cut-off at 500~keV and a reflection component
 generated with the \texttt{PEXRAV} model \citep{pexrav}. Therefore, this
 template has two parameters: normalization (the bolometric flux of the AGN) and
 redshift. For a given flux and redshift, the bolometric luminosity is calculated
 and, hence, the value of $\alpha_\mathrm{OX}$. The first parameter was left free to
 vary during the fitting, while the second was fixed accordingly 
 to the redshifts obtained in the literature.

 \item Four Sy2 galaxies (Fig.~\ref{fig:agnIItemp}): these objects have hydrogen
 column densities ($N_\mathrm{H}$) varying from $10^{22}~\mathrm{cm}^{-2}$ (Compton thin objects)
 to greater than $10^{25}~\mathrm{cm}^{-2}$ (Compton thick objects). They were
 selected from a sample of Sy2 galaxies with minimal starburst contribution
 \citep{Bianchi06}. The AGN templates show two bumps, in the FIR and in the
 NIR-optical, except for the AGN3 template, which only present a broad IR bump. The
 differences between them are the relative height of these bumps, the position of
 their peaks and the ratio between the optical and X-ray fluxes.
\end{enumerate}

\begin{figure*}
    \mbox{
     \subfigure[SB templates.]{
      \includegraphics[angle=-90,width=.45\linewidth]{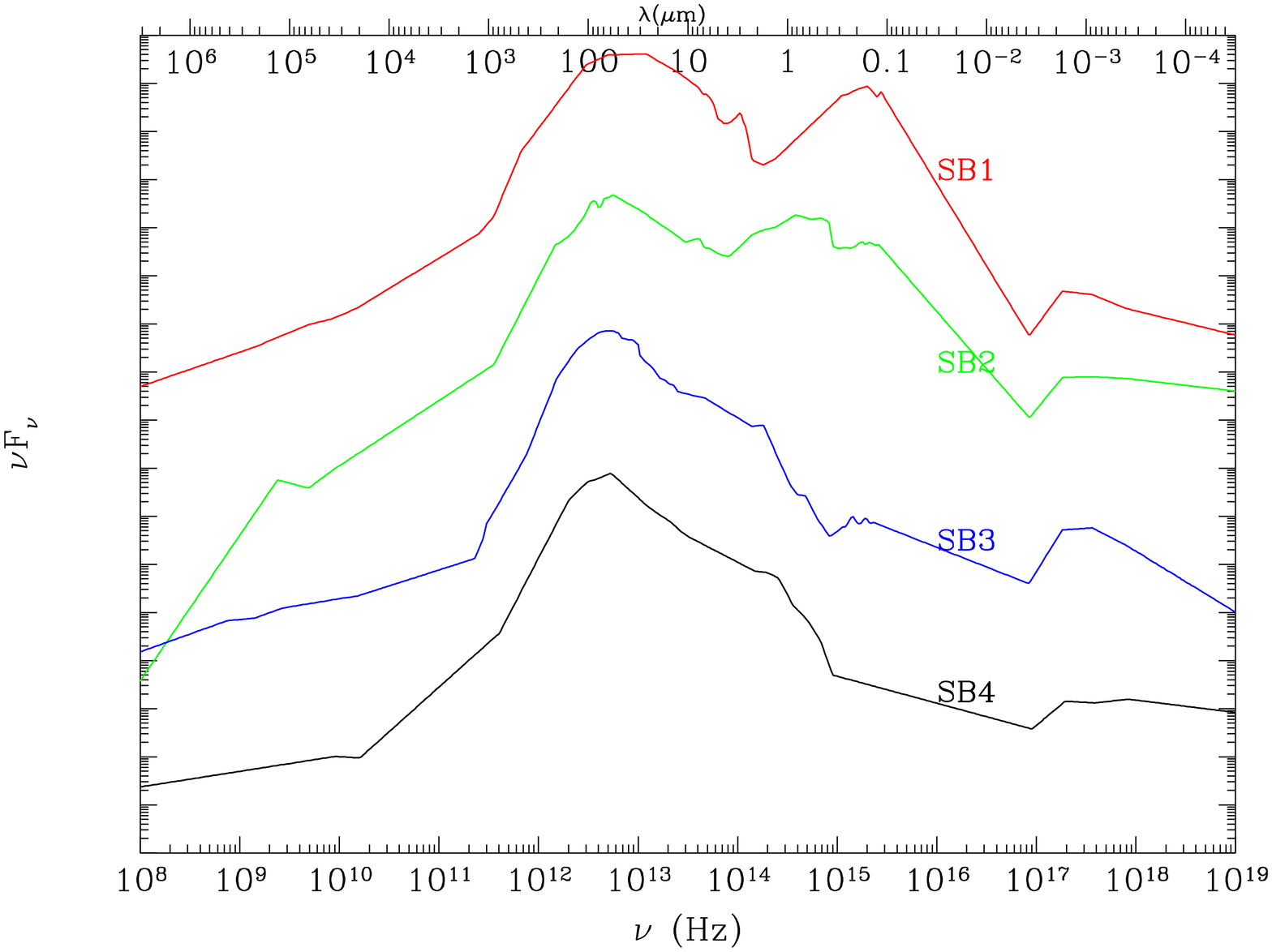}
      \label{fig:sbtemp}
     }
    \quad
     \subfigure[Composite templates.]{
      \includegraphics[angle=-90,width=.45\linewidth]{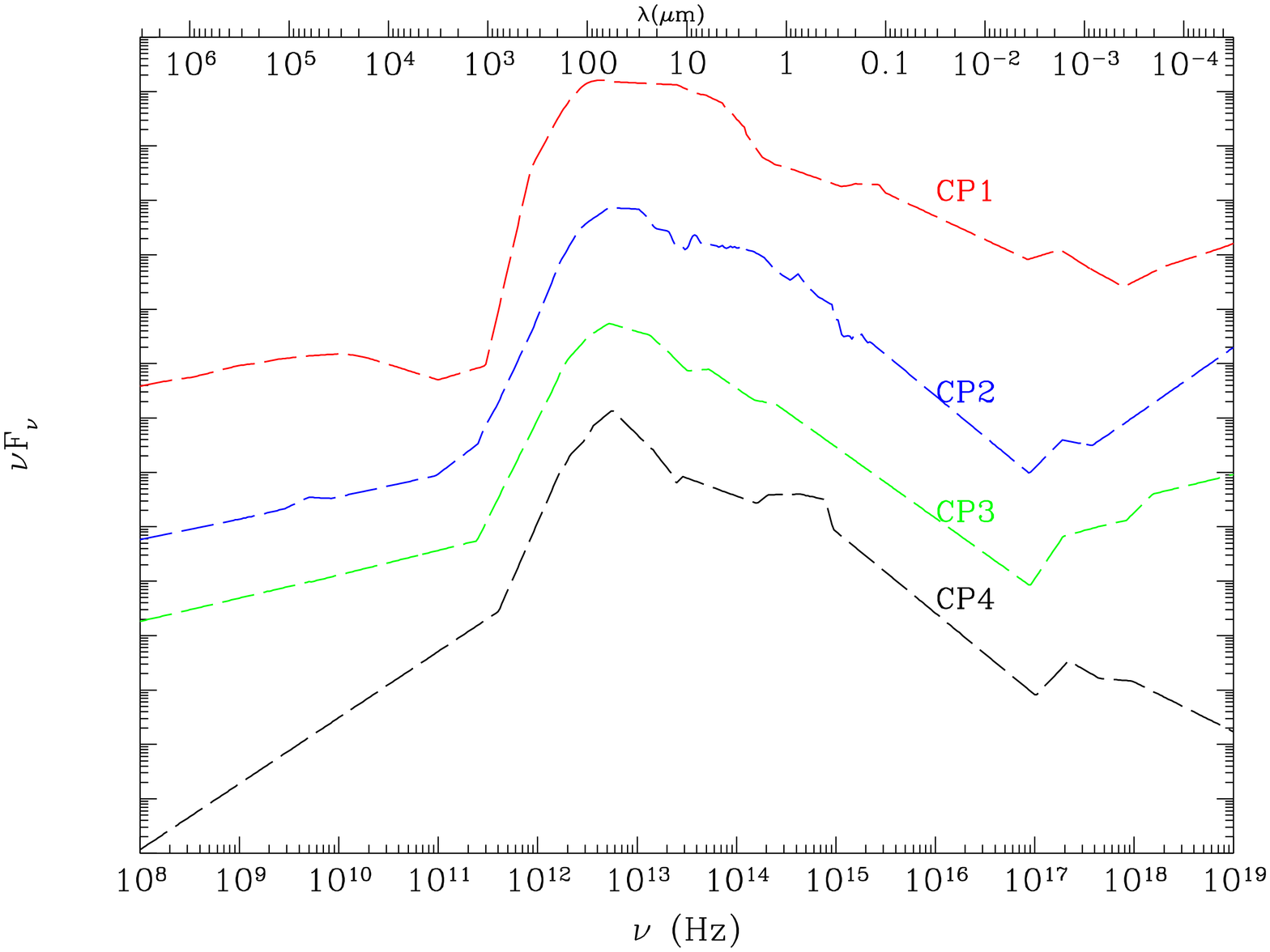}
      \label{fig:comptemp}
     }
    }

\centering
\caption{(a) Pure starburst templates. Listed downwards: NGC\,5253, NGC\,7714, M82, IRAS\,12112+0305.
         (b) Composite templates (AGN + SB). Listed downwards: NGC\,1068, Mrk\,231, IRAS\,19254-7245, IRAS\,22491-1808.
         The SED fluxes are shifted for clarity. See Sect.~\ref{sec:temp} for details.}
\end{figure*}

To represent the SB contribution we have chosen a set of four starburst galaxies well observed 
in the full spectral range (Fig.~\ref{fig:sbtemp}). We have tried to cover a broad range of burst ages,
dust contents and SFR. These physical properties are reflected in the SEDs showing 
different levels of obscuration, width and wavelength peaks.
\begin{enumerate}
 \item NGC\,5253 is a low-metallicity star-forming dwarf galaxy. Its nucleus is the site of 
  a highly obscured and extremely young ($< 10$ Myr) burst of star formation \citep{Beck96},
  with a SFR$\sim8~M_{\sun}~\mathrm{yr}^{-1}$.
 \item NGC\,7714 is a young unobscured SB \citep{Brandl04} with SFR$\sim6~M_{\sun}~\mathrm{yr}^{-1}$ 
 and a burst age between 3-5 Myr \citep{Gonzalez95}.
 \item M82 is an evolved pure SB galaxy with SFR$\sim10~M_{\sun}~\mathrm{yr}^{-1}$ \citep{Strickland04}.
 \item IRAS\,12112+0305 is a bright ULIRG powered by SB and with severe limits to any AGN 
 contribution \citep{Imanishi07,Nardini08}. The estimated SFR for this object
 is $\sim600~M_{\sun}~\mathrm{yr}^{-1}$ \citep{Fran03}.
\end{enumerate}

All of them show two bumps, peaking in the FIR and in the NIR-optical. The main difference
between the templates is the relative height between these bumps and their widths.

We included four SED templates built from sources which they harbour both an AGN 
and a SB (Fig.~\ref{fig:comptemp}):
\begin{enumerate}
 \item NGC\,1068 is a Sy2 galaxy with a composite nature, i.e. it harbours a
 heavily buried AGN \citep[$N_\mathrm{H}>10^{25}~\mathrm{cm}^{-2}$,][]{Matt97} and also an
 intense SB \citep{Telesco84}. The bolometric luminosity of this object is
 roughly evenly divided between the two component. The SB emission
 dominates longward of 30~$\mu$m and the AGN dominates shortward of 20-10~$\mu$m.

 \item Mrk 231 is an ULIRG ($L_\mathrm{IR}=3.2\times10^{12}$) optically classified as a Broad
 Absorption Line QSO \citep{Berta05} with a massive young nuclear SB which is 
 responsible for 25\%-40\% of the nuclear bolometric luminosity \citep{Davies04}.

 \item IRAS\,19254-7245, the ``Superantennae'', is a double-nucleated ULIRG
 optically classified as a Sy2 galaxy, with intense star formation. The AGN contribution
 to the total output is $\sim40-50\%$ \citep{Berta03}.

 \item IRAS\,22491-1808 is a Sy2 ULIRG \citep{Berta05} where the AGN emission is
 $\sim70\%$ of the bolometric luminosity \citep{Farrah03}.
\end{enumerate}

We fitted these composite templates to those HLIRGs where the initial AGN+SB model was
insufficient to reproduce the data (see Sect.~\ref{sec:classBfit}).

We extracted the photometric data for the templates using
VOSED\footnote{http://sdc.laeff.inta.es/vosed} and
VOSpec\footnote{http://esavo.esa.int/vospec} software. These utilities use Virtual
Observatory \citep{virobs} tools to extract photometric and spectral data from
several astronomical archives and catalogues. The templates were improved with data from NED
database in wavelength ranges where VOSED and VOSpec provided no data. These
objects are well observed at all the frequency ranges, particularly in the NIR
and optical bands. We rejected some redundant data and we tried to extract only the
nuclear emission to avoid as much contamination from the host galaxy as
possible. To this end we have chosen only those data with a roughly constant aperture
within the nucleus of the galaxy.

\section{Results}
\label{sec:results}
Figures \ref{fig:sedsplotsA} and \ref{fig:sedsplotsB} show the SED\footnote{Several 
photometric points are upper limits. The most conservative approach was chosen for 
the fit. We set the point to zero and the upper error bar to the upper limit value.} 
and the best fit model selected for each object, and Table~\ref{tab:fits} summarizes 
the results of our analysis. See Sect.~\ref{sec:notes} for comments on some particular
sources.

\subsection{Class A HLIRGs}
\label{sec:classAfit}                                                            
We have shown that our simple two-component SED model is a fair approximation
for most of these HLIRGs (see Fig.~\ref{fig:classAseds}). We found that all
class A HLIRGs but one (IRAS\,14026+4341, see Sect.~\ref{sec:notes} below) are
well fitted with type I AGN templates, consistent with their optical
classification, and an additional SB component is required in three objects. The
AGN component dominates the bolometric output in four out of these six sources,
while two objects present a powerful SB component, with 60\%-70\% contribution
of the bolometric luminosity.

We have then 3 sources with a SB-dominated SED (IRAS\,F12509+3122,
IRAS\,14026+4341 and IRAS\,F14218+3845), one AGN-dominated source with an
important SB contribution (IRAS\,18216+6418), and 3 objects which seem to be
extremely luminous quasars with no particular differences from the local ones,
judging from their SEDs and X-ray spectra \citep{Ruiz07}.

A noticeable result for the class A HLIRGs is that the AGN1 template
over-predicts the X-ray flux of these sources, as found in our previous X-ray
analysis. These discrepancies in the X-ray band can not be related to
variability effects, since the OM data, simultaneous to the X-ray observations,
match well with other optical and UV data obtained in different epochs. When we
modelled these objects with the luminosity-dependent AGN1-L SED template, we
found a significant improvement in the fit in terms of $\chi^2$ for most sources
(4 out of 6) and the X-ray emission is better predicted. This result is
consistent with the known $\alpha_\mathrm{OX}$ luminosity relationship
\citep{Strateva05,Steffen06,Kelly08}.

We must also note that the IR-to-bolometric ratio of these sources is within
$\sim40-70\%$, so an important fraction of their bolometric output is not
emitted in the IR range. Hence, strictly speaking, they should not be considered
as HLIRGs, particularly those with a completely AGN-dominated SED, where less
than ~50\% of their bolometric luminosity is in the IR. This ``contamination''
can be expected given the selection criteria of the \citet{Rowan00} parent
sample, which simply selected those known sources with
$L_\mathrm{IR}\gtrsim10^{13}~L_{\sun}$.

\subsection{Class B HLIRGs}
\label{sec:classBfit}
We found that these sources are fitted with a dominant SB component and,
in most cases, a minor AGN contribution ($<10\%$). However our model 
present some problems for class B HLIRGs that we did not find 
in class A objects (see Fig.~\ref{fig:sedsplotsB}):
\begin{enumerate}
 \item The level of obscuration in the observed X-ray spectra is higher  
 than the one expected from the AGN templates.

 \item Most sources show an excess in the MIR-NIR band not modelled by
 these templates, i.e. the width of the IR bump seems to be broader than the
 bumps in the starburst templates.

 \item The peak of the template does not match the IR peak of the SEDs
 in several sources.
\end{enumerate}

In order to improve the fit quality for the class B sources, we repeated
the SED fitting using a set of templates from composite sources (see
Sect.~\ref{sec:temp}), where both AGN and SB emission are significant. By
using these composite templates, we found that the statistical quality of our fits
was significantly improved for all but one case (IRAS\,F15307+3252, see
Sect.~\ref{sec:notes} below). For most objects, the $\chi^2$ obtained with
any of the composite templates is significantly lower than the $\chi^2$ obtained
with any combination of pure AGN and pure SB templates. CP1 is the best fit template for 4 out of 6
sources, consistent with their spectral classification (type 2 AGN) and X-ray
obscuration level (Compton-Thick). Two sources are best fitted with the CP2 template.

IRAS\,F00235+1024 is the only source that still shows a significant IR excess,
which suggest that the SB contribution may be larger in this source that in the CP1
template ($\sim50\%$).

\subsection{Fitting without X-ray data}
\label{sec:noxray}
In order to check how much X-ray data influence the SED fitting results, we
exclude X-ray data from the SED fitting procedure. Class A sources are still
well represented by the same models (see Table~\ref{tab:fits}, columns labeled
as ``no X-rays''), while class B galaxies are preferentially fitted with an AGN3
template (Compton thin model) and a SB component. Moreover, the AGN
contribution grows significantly in most sources, particularly in the class B
sources. When X-rays are included, a severe limit is imposed and the AGN
contribution decreases dramatically. This shows that X-rays are important to obtain
an accurate model with our technique and, hence, a better estimation of the
contribution of each component to the total output.

\subsection{Notes on particular sources}
\label{sec:notes}

\subsubsection*{IRAS 14026+4341}
This source is optically classified as a type I AGN \citep{Rowan00}, in
agreement with the SDSS classification, and recent MIR \Spitzer data also
suggest the presence of an AGN in this object (Ruiz et al., in preparation), but
our best fit model is obtained by using the SB2 template. The X-ray data impose a severe
constraint, rejecting the AGN templates that predict a higher emission in the
X-ray band. If we fit again this SED using no X-ray data (see
Sect.~\ref{sec:noxray}) we find that the best fit is obtained by AGN1+SB2.

The X-ray emission of this source seems to be affected by absorption
(see Fig.~\ref{fig:iras14026}): it is not detected in the soft X-ray band
(0.5-2~keV) and its 2XMMi hardness ratio
($HR3\sim-0.2$)\footnote{$HR3=\frac{CR(2.0-4.5~\mathrm{keV})-CR(1.0-2.0~\mathrm{
keV})}{ CR(2.0-4.5~\mathrm{keV})+CR(1.0-2.0~\mathrm{keV})}$, where CR is the
count rate in the given energy band.} is consistent with an X-ray absorbed
AGN \citep{DellaCeca04}. This indicates IRAS 14026+4341 as an X-ray absorbed QSO.
These objects are often embedded in ultraluminous starburst galaxies
\citep{Page07}, and they have been pointed out as a transitional phase in an
evolutionary sequence relating the growth of massive black holes to the
formation of galaxies \citep{Stevens05,Page07}.

Under these circumstances, we have selected as best fit the model
resulting from fitting the SED without X-ray data. We must note, however, that
both models (pure SB or AGN+SB) poorly fit the data between
$\sim1-100~\mu\mathrm{m}$. The observed IR excess, may be related to the X-ray emission
absorbed and reprocessed in the IR, can not be reproduced by AGN1 template (an unabsorbed
template).

\subsubsection*{IRAS F15307+3252}
This object has been optically classified as a QSO 2 \citep{Rowan00} and there is strong evidence
in X-rays favouring the presence of a heavily obscured AGN \citep{Iwasawa05}. However we have
found that its SED best fit, in terms of $\chi^2$, is obtained with a SB template 
with minor AGN1 contribution. The CP1 template is also a fair fit, but with a slightly worse $\chi^2$.

Previous analyses of the IR emission of this HLIRG \citep{Deane01,Verma02} suggest that the SB
contribution is considerably lower than what we found using a pure SB template. Hence, we have selected
the CP1 as ``best fit'', which is also consistent with its optical classification, to estimate 
the AGN and SB contribution to the bolometric luminosity.

\section{Comparison with previous results}
\label{sec:comp}

\subsection{X-ray emission}
We can estimate the expected X-ray luminosity of the AGN and SB components for each source in our
sample using the parameters obtained in our SED analysis, and compare with the X-ray luminosities 
calculated through \XMM observations.

We have seen that the AGN SED of these sources is better modelled with a luminosity-dependent 
template. Hence, we have employed the relation obtained by Sani et al. (private communication)\footnote{This 
ratio is obtained from the \citet{Steffen06} relation between X-ray and 2500\,\AA~ 
luminosities and then linking the 2500\,\AA~luminosity with the bolometric one through 
the \citet{Elvis94} SED.} to estimate the intrinsic 2-10 keV luminosity for a given AGN bolometric 
luminosity:

\begin{equation}
\label{eq:sani}
\frac{L_{2-10~\mathrm{keV}}}{L_\mathrm{BOL}}=0.043\left(\frac{L_\mathrm{BOL}}{10^{45}}\right)^{-0.357}
\end{equation} 

Figure~\ref{fig:AGNxrays} shows those sources detected in X-rays and with an AGN
component in their SED model. We plotted the bolometric luminosity of the AGN
component versus the intrinsic (absorption corrected) 2-10 keV luminosity (see
Table~\ref{tab:luminosities}), as calculated in \citet{Ruiz07}.

Most sources are scattered roughly following the Eq.~\ref{eq:sani} estimate.
This scatter is probably related to the intrinsic dispersion in X-ray
luminosities of AGN, i.e. for a given bolometric luminosity, there is a broad
range of possible X-ray luminosities \citep{Steffen06}. 

There are, however, three sources (PG\,1206+459, IRAS\,F12509+3122 and
IRAS\,14026+4341) with X-ray luminosities much lower than the estimated by
Eq.~\ref{eq:sani}. The X-ray luminosity of IRAS\,14026+4341 was
calculated using the 2XMMi X-ray fluxes so it is not corrected by absorption.
Hence, this large discrepancy between the prediction and the observed luminosity
is likely another sign of X-ray absorption (see Sect.~\ref{sec:notes}).

For the other two sources \citet{Ruiz07} did not find any sign of X-ray
absorption. This effect could be, in principle, due to an overestimate of the
AGN contribution to the bolometric luminosity. If we assume that the difference
between the bolometric luminosity calculated using the SED fitting and 
that estimated using Eq.~\ref{eq:sani} is completely originated due to star
formation, we find that the SB contribution to the total output should be larger
than 90\% in these two sources. Such a powerful SB must be clearly reflected in
the SED shape, but we did not find this kind of deviation in the SED analysis of
these sources. The X-ray weakness of these HLIRGs can not therefore be related
to the underestimate of the SB contribution to the bolometric output, or due 
to X-ray absorption. They seem to be intrinsically weak X-ray sources \citep{Leighly01,Leighly07}.

\subsection{IR SED: comparison with previous work}
The IR (1--1000\,$\mu$m) SED of our sources has been previously studied:
\citet{Rowan00}, \citet{Farrah02submm} and \citet{Verma02} modelled it using
RTM. We estimated the IR luminosities of our models, integrating between
1--1000\,$\mu$m, and compared their results with ours (see
Table~\ref{tab:luminosities}).

The IR luminosities estimated through our SED fitting and that estimated using
RTM match fairly well (see Fig.~\ref{fig:IRlumin}) for most sources. For three
objects, our luminosity estimation is almost an order of magnitude greater than
the RTM estimation, probably because our best-fit models overestimate the
FIR-submm emission (see Figs.~\ref{fig:pg1247}, \ref{fig:iras16347} and
\ref{fig:irasF15307}). This spectral emission is problably better recovered by
using RTM. Nevertheless, in spite of this large disagreement in luminosities,
our AGN contribution estimates are consistent with those obtained through RTM,
as Fig.~\ref{fig:IRratios} shows.

The latter plot shows that our AGN contribution estimates for most sources are
roughly consistent with those obtained through RTM. We can conclude that our
simple model based on templates is a fair method to obtain a first estimate of
the AGN and SB relative contribution to the IR output.

\section{Discussion}
\label{sec:discuss}
The broad band SEDs of the HLIRGs presented in this work can be roughly well
fitted using templates, and their best fits are consistent with the optical
classification of most sources (9 out of 13). Among class A sources we found
three objects fitted with pure type 1 AGN templates. They seem to be very
luminous quasars and, since most of their bolometric output is not emitted in
the IR band, should not be considered as proper HLIRGs. Four out of seven class
A HLIRGs require, in addition to a type 1 AGN template, a SB component which is,
in three cases, dominant with respect to the AGN. The AGN emission in four
sources is consistent with a luminosity-dependent SED.

On the other hand, we have found that class B sources can not be fitted with
simple combination of pure AGN and pure SB templates: a composite template is
needed, where AGN and SB phenomena are both significant. This suggests that
there should be some feedback between accretion process and star formation that
changes the shape of the SED in a way that can not be imitated just by combining
a pure SB and a pure AGN components. The main observational imprint of this
feedback seems to be an excess in the SED around $\sim10~\mu$m with respect to
the predicted emission of pure AGN and pure SB combined model.

Our division between class A and class B sources is based on the optical
spectral classification, and since all objects show a significant AGN emission,
it seems that the SED shape differences between the two groups could be an
inclination effect as in the unified model of AGNs \citep{Antonucci85}: those
HLIRGs where we have a direct view of the nucleus are luminous QSO and show a
class A SED, while those HLIRGs seen through the molecular torus and/or other
obscuring material show a class B SED. The comparable mean SB contribution of
class A (excluding pure AGN sources) and class B sources is consistent with this
hypothesis. Within this scenario, all types of HLIRGs belong to the same class
of sources, seen at different inclination angles.

\citet{Farrah02submm} proposed, however, that HLIRG population is
comprised of (1) mergers between gas rich galaxies, as found in the ULIRG
population, and (2) young active galaxies going through their maximal star
formation periods whilst harbouring an AGN.

The $N_\mathrm{H}$ distribution we found in the X-ray study seems to favour the
two population hypothesis. In a pure inclination scenario we would expect a
broad range of X-ray absorption, from no absorbed to heavily absorbed sources.
However we found only objects with no significant intrinsic absorption (all
but one class A sources) or CT absorbed objects (all class B sources). Since
AGNs observed in ULIRGs usually show heavy absorption in X-rays
\citep{Fran03,Ptak03,Teng05}, in principle class B sources could represent the
high luminosity tail of ULIRG population, while the strong SB found in class A HLIRGs 
could represent young active galaxies experiencing their maximal star formation, without 
being in interacting systems (i.e. with little connection with a recent major merger).

The study of the host galaxy morphology and environment of HLIRGs also support
the two population hypothesis. \citet{Farrah02hst} found in a sample of nine
HLIRGs observed by HST both strongly interactive systems and objects with no
clear signs of ongoing interactions. Five sources of this sample are also
included in ours: IRAS\,F00235+1024 and IRAS\,F15307+3252 (class B objects) show
signs of strong interactions, while IRAS\,F12509+3122, IRAS\,F14218+3845 and
IRAS\,16347+7037 (class A objects) are isolated systems. This result favours our
suggestion that class B HLIRGs could be objects in the extreme bright end of the
ULIRG population distribution.

Hence, while class B HLIRGs share common properties with ULIRGs (high levels of
X-ray obscuration, strong star formation, signs of mergers and interactions),
class A HLIRGs seem to be a different class of objects. Excluding the 3 pure AGN
sources, class A objects could be among the young active galaxies proposed by
\citet{Farrah02submm}. The powerful SB we found in these sources, and the large
amounts of gas available to fuel the star formation (as calculated by
\citealt{Farrah02submm}), along with the non detection of mergers or
interactions in these systems, support this idea. Moreover, the SB emission of
the bona fide class A HLIRGs is modelled with young SB templates (SB1 and SB2)
in all but one object (IRAS\,18216+6418), which is modelled with an old SB
(SB3). This source could be a more evolved object.

Therefore, sources in our sample likely belong to three different populations:
\begin{enumerate}
 \item Very luminous QSO with minor star formation activity.

 \item Young, isolated active galaxies undergoing their first episode of major 
 star formation with little connection with a recent major merger.

 \item Galaxies which have recently experienced a merger/disturbance that
 brought lots of gas and dust in the inner regions. This event trigger both the
 star formation and the AGN activity in a heavily obscured environment. These
 objects suit well as the high luminosity tail of the ULIRG population.
\end{enumerate}

Nevertheless, our sample of HLIRGs is not complete in any sense and we cannot
derive further conclusions about the global properties of the HLIRG population.
Further studies established on larger and complete samples of HLIRGs are needed
to conclude if the division between class A and class B objects is just due to
an inclination effect, or is based on intrinsic differences of their physical
properties.

\section{Conclusions}
\label{sec:conc}
In this paper we have built and analysed the multi-wavelength SED (from radio 
to X-rays) of a sample of 13 HLIRGs, previously studied in detail in X-rays
\citep{Ruiz07}. We assembled the SEDs using public data in several astronomical
databases and in the literature, and we modelled them using templates. Most
sources are roughly well fitted with this simple model and we find AGN relative
contributions consistent with those inferred by previous analyses of the IR SEDs
of HLIRGs using radiative transfer models.

We divided the HLIRGs in two groups, accordingly to their optical spectral
classification: class A (type 1 AGNs) and class B (type 2 AGNs and SB) sources.
A first look at their SED shape indicates some differences between the two
classes: class A sources show a roughly flat SED between the IR and the optical,
while class B sources have a prominent IR bump dominating the rest of the
emission.

A significant fraction (3 out of 7)  of class A HLIRGs seem to be very luminous
quasars with no particular deviations from the local quasars. Strictly speaking
these objects should not be considered HLIRGs since most of their bolometric
output is emitted outside the IR band. The SED of these QSO are consistent with
a luminosity-dependent quasar template. The remaining class A sources show
significant additional SB components, which are dominant in all but one object.
Given their strong SB activity and the lack of any sign of mergers in these
systems, these HLIRGs could be very young galaxies experiencing their first
episode of maximal star formation.

Class B HLIRGs show an IR excess that can not be modelled with any combination
of our selected pure AGN and pure SB templates. This feature can be properly
fitted using composite templates (SEDs from objects where AGN and SB emission
are both important). This suggests that a significant fraction of the emission
of this class of objects is originated in a SB. This also shows that the
feedback between accretion and star formation processes modifies the SED of
class B HLIRGs in a way that can not be replicated by just the addition of pure
AGN and pure SB independent templates. Class B HLIRGs share many properties with
ULIRGs (high X-ray absorption, strong star formation, signs of mergers and
interactions), so they could be just the high luminosity tail of this
population.

Therefore, we have found some evidence supporting the idea that bona fide HLIRGs
are composed of two populations: young active galaxies with no sign for recent
mergers most likely going through their first episode of strong star formation,
and the high-luminosity end of the ULIRG population, where both the SB and AGN
are likely triggered by a recent merger/interaction. Further observational
studies based on larger and, most importantly, complete samples of HLIRGs are
needed to obtain stronger evidence for this hypothesis. Moreover our simple
template-fitting approach should be complemented with RTMs (or other theoretical
models of AGN and SB emission), since the two approaches are complementary in
many ways and their combination may shed further light onto the relative SB-AGN
contribution and on the feedback processes that take place in the most
interesting HLIRGs, namely those that are well represented by composite
templates within our approach.

\begin{acknowledgements}
We are grateful to the referee M. Rowan-Robinson for the constructive comments
and suggestions that improved this paper. A.R. acknowledges support from a
Universidad de Cantabria fellowship. Financial support for A.R. and F.J.C. was
provided by the Spanish Ministry of Education and Science, under projects
ESP2003-00812 and ESP2006-13608-C02-01. FP acknowledges financial support under
the project ASI INAF I/08/07/0. GM thanks the Ministerio de Ciencia e
Innovaci\'on and CSIC for support through a Ram\'on y Cajal contract.

This research has made use of the NASA/IPAC Extragalactic Database (NED) which
is operated by the Jet Propulsion Laboratory, California Institute of
Technology, under contract with the National Aeronautics and Space
Administration. This paper is based also on data from the VOSED tool at LAEFF.

The 2.5m Isaac Newton Telescope and its service programme are operated on the
island of La Palma by the Isaac Newton Group in the Spanish Observatorio del
Roque de los Muchachos of the Instituto de Astrof\'isica de Canarias. 
\end{acknowledgements}
\balance

\cleardoublepage

\begin{table*}
\caption{Best fit models for the HLIRG's SEDs}
\label{tab:fits}       
\begin{center}
\begin{tabular}{lllcccc{|}ccc{|}clc}
\hline\hline\noalign{\smallskip}
Source & z & Type & CT$^a$ & \multicolumn{7}{c}{Best Fit model$^b$} & $\log L_\mathrm{BOL}$ $^c$ & AGN / SB$^d$ \\
\cline{5-11}
       &   &      &        & \multicolumn{3}{c}{all data$^e$} & \multicolumn{3}{c}{no X-rays$^f$} & composite temp.$^g$ & (erg s$^{-1}$) &  \\
       &   &      &        & \multicolumn{2}{c}{Model} & $\alpha$ & \multicolumn{2}{c}{Model} & $\alpha$ &                  &  &  \\       
\hline\noalign{\smallskip}
\multicolumn{11}{l}{Class A HLIRGs} \\
\hline
\object{PG\,1206+459}      & 1.158 & QSO   & N & AGN1-L & ... & 1   & AGN1 & ... & 1   & \textbf{AGN1-L}    & 48.4 &   1 / 0 \\
\object{PG\,1247+267}      & 2.038 & QSO   & N & AGN1-L & ... & 1   & AGN1 & ... & 1   & \textbf{AGN1-L}    & 49.2 &   1 / 0 \\
\object{IRAS\,F12509+3122} & 0.780 & QSO   & N & AGN1-L & SB1 & 0.3 & AGN1 & SB4 & 0.5 & \textbf{AGN1-L+SB1}& 47.7 & 0.3 / 0.7 \\
\object{IRAS\,14026+4341}  & 0.323 & QSO   & N & ...    & SB2 & 0   & \textbf{AGN1} & \textbf{SB2} & 0.3 & SB2 & 46.7 & 0.3 / 0.7 \\
\object{IRAS\,F14218+3845} & 1.21  & QSO   & N & AGN1   & SB1 & 0.4 & AGN1 & SB1 & 0.3 & \textbf{AGN1+SB1}  & 47.2 & 0.4 / 0.6 \\
\object{IRAS\,16347+7037}  & 1.334 & QSO   & N & AGN1-L & ... & 1   & AGN1 & ... & 1   & \textbf{AGN1-L}    & 48.9 &   1 / 0 \\
\object{IRAS\,18216+6418}  & 0.297 & QSO   & N & AGN1   & SB3 & 0.8 & AGN1 & SB3 & 0.8 & \textbf{AGN1+SB3}  & 47.4 & 0.8 / 0.2 \\
\hline
\multicolumn{11}{l}{Class B HLIRGs} \\
\hline
\object{IRAS\,F00235+1024} & 0.575 & NL-SB & Y & ...    & SB3 & 0    & ...  & SB3 & 0   & \textbf{CP1} & 46.7 & $\sim$0.5 / $\sim$0.5 \\
\object{IRAS\,07380-2342}  & 0.292 & SB    & N & AGN4   & SB1 & 0.06 & AGN3 & SB1 & 0.3 & \textbf{CP1} & 47.0 & $\sim$0.5 / $\sim$0.5 \\
\object{IRAS\,00182-7112}  & 0.327 & QSO 2 & Y & AGN3   & SB4 & 0.06 & AGN3 & SB3 & 0.3 & \textbf{CP1} & 46.6 & $\sim$0.5 / $\sim$0.5 \\
\object{IRAS\,09104+4109}  & 0.442 & QSO 2 & Y & AGN4   & SB4 & 0.09 & AGN3 & SB1 & 0.8 & \textbf{CP2} & 47.3 & $\sim$0.7 / $\sim$0.3 \\
\object{IRAS\,12514+1027}  & 0.32  & Sy2   & Y & AGN5   & SB4 & 0.06 & AGN3 & SB2 & 0.9 & \textbf{CP2} & 46.7 & $\sim$0.7 / $\sim$0.3 \\
\object{IRAS\,F15307+3252} & 0.926 & QSO 2 & Y & AGN1   & SB3 & 0.03 & AGN3 & SB1 & 0.8 & \textbf{CP1} & 47.9 & $\sim$0.5 / $\sim$0.5 \\
\noalign{\smallskip}\hline\noalign{\smallskip}
\end{tabular}
\end{center}
 $^a$ Compton Thick candidates. \\
 $^b$ The best fit adopted to estimate the bolometric luminosity and the AGN and SB fraction
      is marked in bold fonts. \\
 $^c$ Bolometric luminosity in CGS units. \\
 $^d$ Fraction of the bolometric luminosity originated in AGN and SB. Calculated through
      the parameter $\alpha$ of the best fit model. \\
 $^e$ Best fit using our original set of templates. \\
 $^f$ Best fit not using X-ray data. \\
 $^g$ Best fit including the templates of composite sources.
\end{table*}

\begin{table*}
\caption{SED templates used as models.}
\label{tab:temp}
\begin{center}
\begin{tabular}{lll}
\hline\noalign{\smallskip}
Label  & Source & Description \\
\hline\noalign{\smallskip}
AGN1   & ...                & \tiny local quasar's mean SED$^1$\\
AGN1-L & ...                & \tiny luminosity-dependent QSO SED$^2$ \\
AGN3   & NGC\,5506           & Sy2, $N_\mathrm{H}=3\times10^{22}\mathrm{cm^{-2}}$ \\
AGN4   & NGC\,4507           & Sy2, $N_\mathrm{H}=4\times10^{23}\mathrm{cm^{-2}}$ \\
AGN5   & Mnk~3              & Sy2, $N_\mathrm{H}=1.4\times10^{24}\mathrm{cm^{-2}}$ \\
AGN6   & NGC\,3393           & Sy2, $N_\mathrm{H}>1\times10^{25}\mathrm{cm^{-2}}$ \\
SB1    & NGC\,5253           & Young and dusty SB\\
SB2    & NGC\,7714           & Young and unobscured SB\\
SB3    & M82                & Old SB \\
SB4    & IRAS\,12112+0305    & ULIRG \\
CP1    & NGC1068            & \tiny Composite template: AGN: $\sim50\%$ \\
CP2    & Mnk~231            & \tiny Composite template: AGN: $\sim70\%$ \\
CP3    & IRAS\,19254-7245    & \tiny Composite template: AGN: $\sim45\%$ \\
CP4    & IRAS\,22491-1808    & \tiny Composite template: AGN: $\sim70\%$ \\
\noalign{\smallskip}\hline\noalign{\smallskip}
\end{tabular}
\linebreak
References. (1) \citealt{Richards06}; (2) \citealt{Hopkins07}.
\end{center}
\end{table*}

\begin{table*}
\caption{IR and X-ray luminosities.}
\label{tab:luminosities}
\begin{center}
\begin{tabular}{lccc|ccc|cc}
\hline\hline\noalign{\smallskip}
Source & $\log L_\mathrm{IR}^\mathrm{tot}$ $^a$ & $\log L_\mathrm{IR}^\mathrm{AGN}$ $^a$ & $\log L_\mathrm{IR}^\mathrm{SB}$ $^a$ &  $\log L_\mathrm{IR,RTM}^\mathrm{tot}$ $^b$ & $\log L_\mathrm{IR,RTM}^\mathrm{AGN}$ $^b$ & $\log L_\mathrm{IR,RTM}^\mathrm{SB}$ $^b$ & $\log L_\mathrm{X}^\mathrm{AGN}$ $^c$ & $N_\mathrm{H}$ $^d$ \\
       & (erg s$^{-1}$)           & (erg s$^{-1}$)           & (erg s$^{-1}$)          &  (erg s$^{-1}$)               & (erg s$^{-1}$)               & (erg s$^{-1}$)              & (erg s$^{-1}$)        & (cm$^{-2}$) \\
\hline\noalign{\smallskip}
\multicolumn{9}{l}{Class A HLIRGs} \\
\hline
PG\,1206+459      & 48.0 & 48.0 & 0    & 47.8 & 47.8 & $<46.7$ & $45.11^{+0.02}_{-0.04}$ & ... \\
PG\,1247+267      & 48.8 & 48.8 & 0    & 47.9 & 47.9 & $<46.8$ & $45.93^{+0.02}_{-0.03}$ & ... \\ 
IRAS\,F12509+3122 & 47.6 & 46.8 & 47.5 & 47.0 & 46.8 & 46.6    & $42.26^{+0.05}_{-0.05}$ & ... \\ 
IRAS\,14026+4341$^e$  & 46.5 & 45.8 & 46.5 & 46.5 & 46.3 & 46.1    & $42.7^{+0.2}_{-0.5}$    & ... \\
IRAS\,F14218+3845 & 47.1 & 46.5 & 46.9 & 46.9 & 46.1 & 46.8    & $44.60^{+0.03}_{-0.03}$ & ... \\ 
IRAS\,16347+7037  & 48.5 & 48.5 & 0    & 47.7 & 47.7 & $<46.8$ & $46.00^{+0.07}_{-0.09}$ & ... \\ 
IRAS\,18216+6418  & 47.1 & 46.9 & 46.6 & 46.8 & 46.6 & 46.4    & $45.6^{+0.04}_{-0.05}$  & ... \\ 
\hline
\multicolumn{9}{l}{Class B HLIRGs} \\
\hline
IRAS\,F00235+1024 & 46.7 & 46.4 & 46.4 & 46.7 & 46.4    & 46.4    & $<42.2$ & $>10^{25}$\\ 
IRAS\,07380-2342  & 47.0 & 46.7 & 46.7 & 47.0 & 46.8    & 46.5    & $<41.7$ & ... \\ 
IRAS\,00182-7112  & 46.6 & 46.3 & 46.3 & 46.7 & $<46.5$ & 46.7    & $44.82^{+0.16}_{-0.14}$ & $>10^{25}$ \\ 
IRAS\,09104+4109  & 47.3 & 47.1 & 46.8 & 46.8 & 46.8    & $<46.2$ & $45.30^{+0.36}_{-0.09}$ & $>10^{25}$ \\ 
IRAS\,12514+1027  & 46.7 & 46.5 & 46.2 & 46.5 & 46.2    & 46.2    & $43.3^{+1.4}_{-0.7}$    & $(4^{+20}_{-3})\times10^{23}$ \\ 
IRAS\,F15307+3252 & 47.9 & 47.6 & 47.6 & 46.9 & 46.6    & 46.7    & $45.49^{+0.09}_{-0.11}$ & $>10^{25}$ \\ 
\noalign{\smallskip}\hline\noalign{\smallskip}
\end{tabular}
\end{center}
a) IR luminosities (1-1000 $\mu\mathrm{m}$) estimated using our SED fitting. \\
b) IR luminosities (1-1000 $\mu\mathrm{m}$) estimated by the analysis of the IR SED using RTM \citep{Rowan00,Farrah02submm}. \\
c) Absorption corrected 2-10 keV luminosities from \citet{Ruiz07}. \\
d) Intrinsic absorption estimated using X-ray spectra \citep{Ruiz07}. \\
e) The X-ray luminosity of this source has been calculated from 2XMMi fluxes \citep{2xmmi}, and it is not corrected from absorption.
\end{table*}


\begin{figure*}
\begin{center}
    \mbox{
     \subfigure[Class A HLIRGs.]{
      \includegraphics[angle=0,width=.45\linewidth]{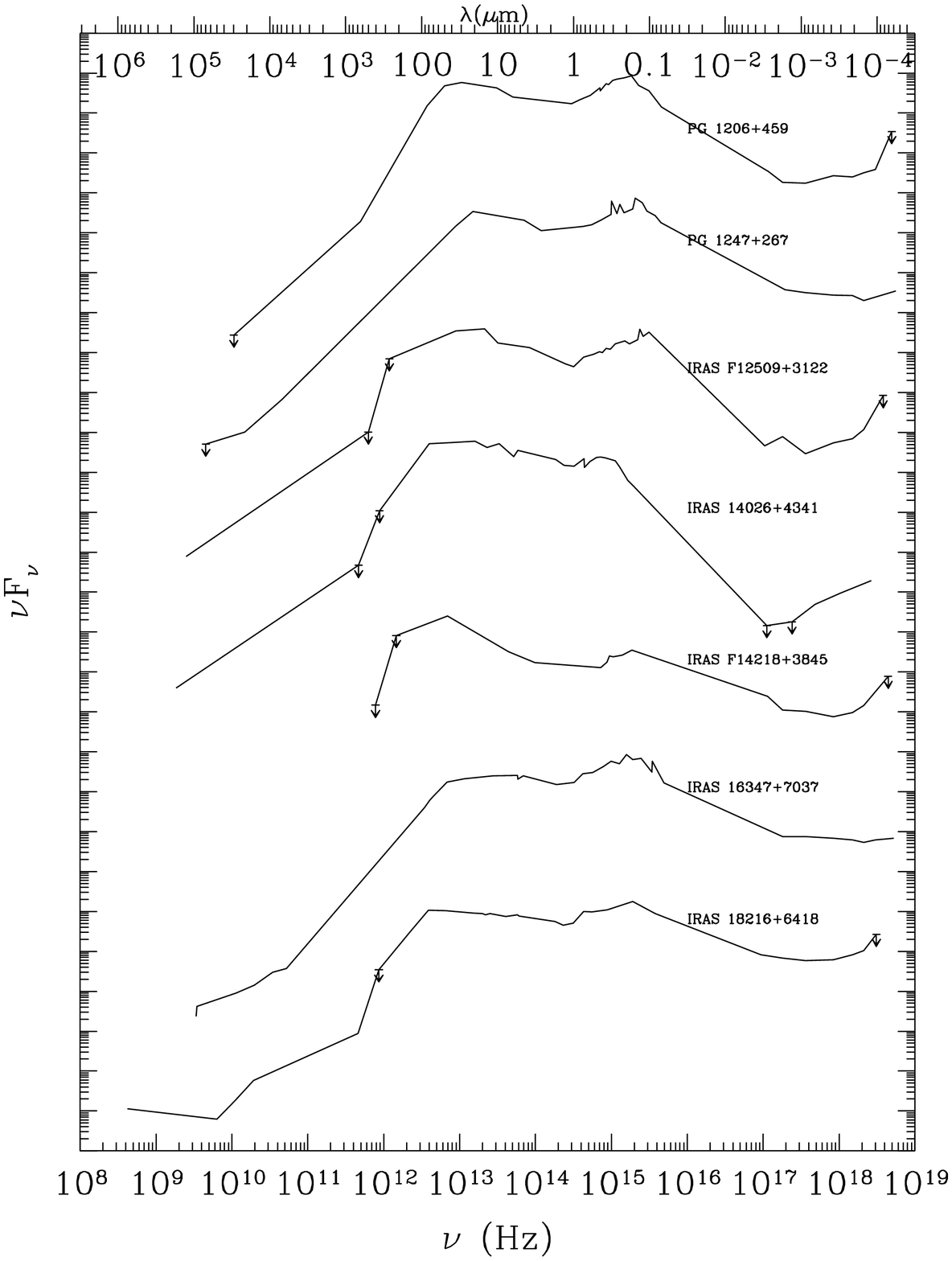}
      \label{fig:classAseds}
     }
    \quad
     \subfigure[Class B HLIRGs.]{
      \includegraphics[angle=0,width=.45\linewidth]{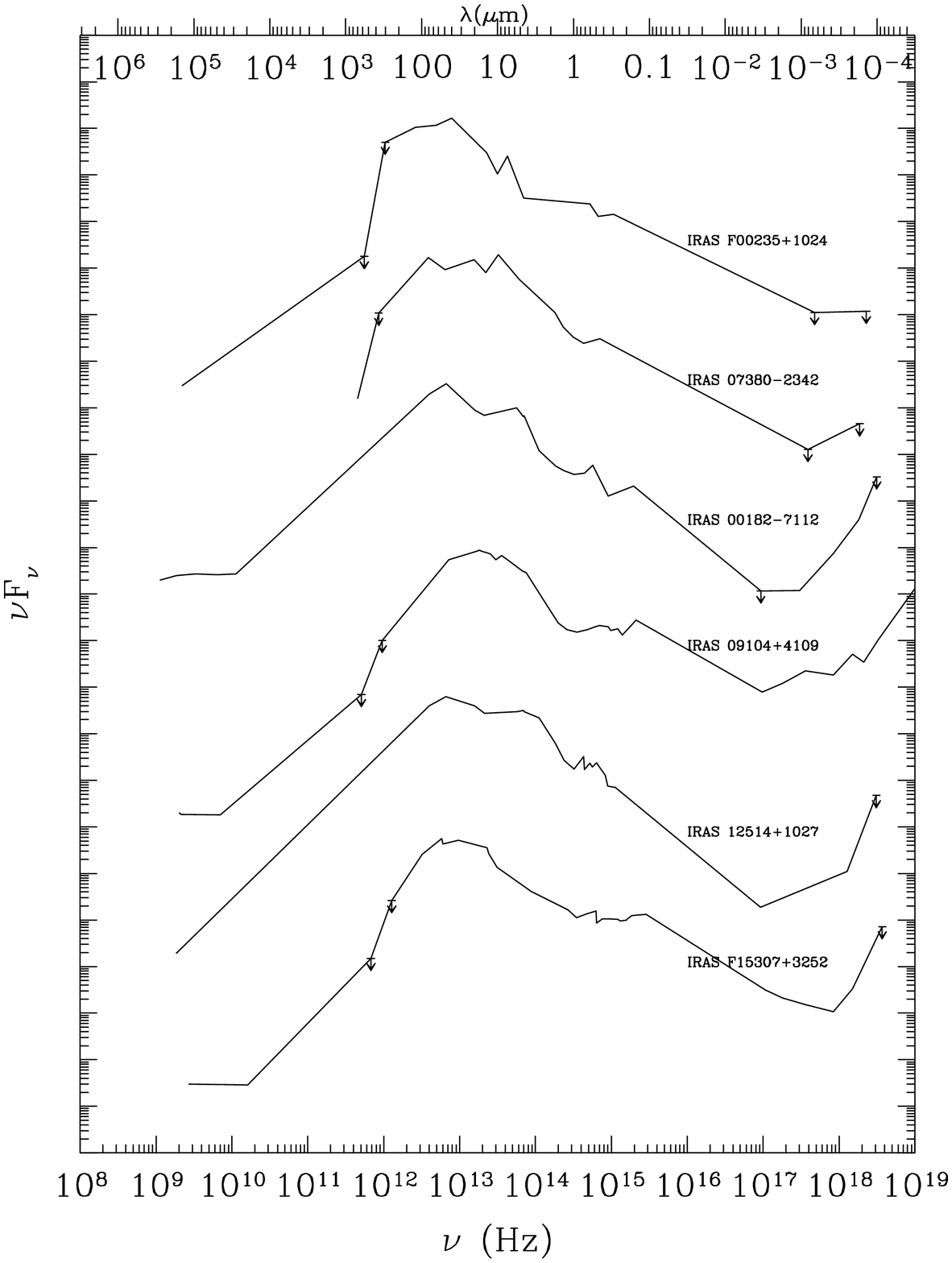}
      \label{fig:classBseds}
     }
    }

\caption{Rest-frame spectral energy distributions of the sample. Fluxes are shifted for clarity.}
\label{fig:seds}
\end{center}
\end{figure*}

\begin{figure*}
   \begin{center}
      \mbox{ 
       \subfigure[]{
       \includegraphics[angle=-90,width=.45\linewidth]{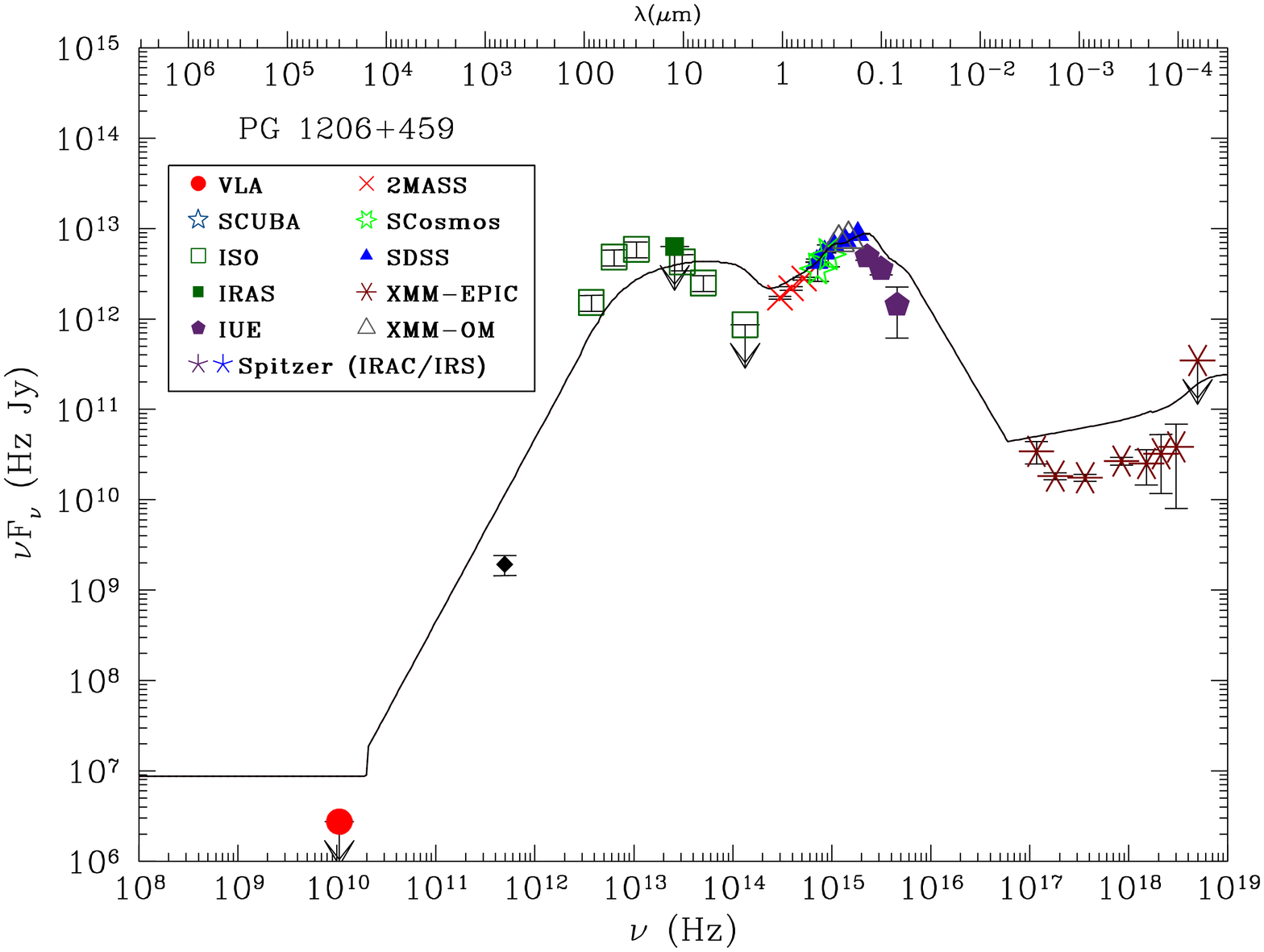}
       \label{fig:pg1206} 
      } 
      \quad 
       \subfigure[]{ 
       \includegraphics[angle=-90,width=.45\linewidth]{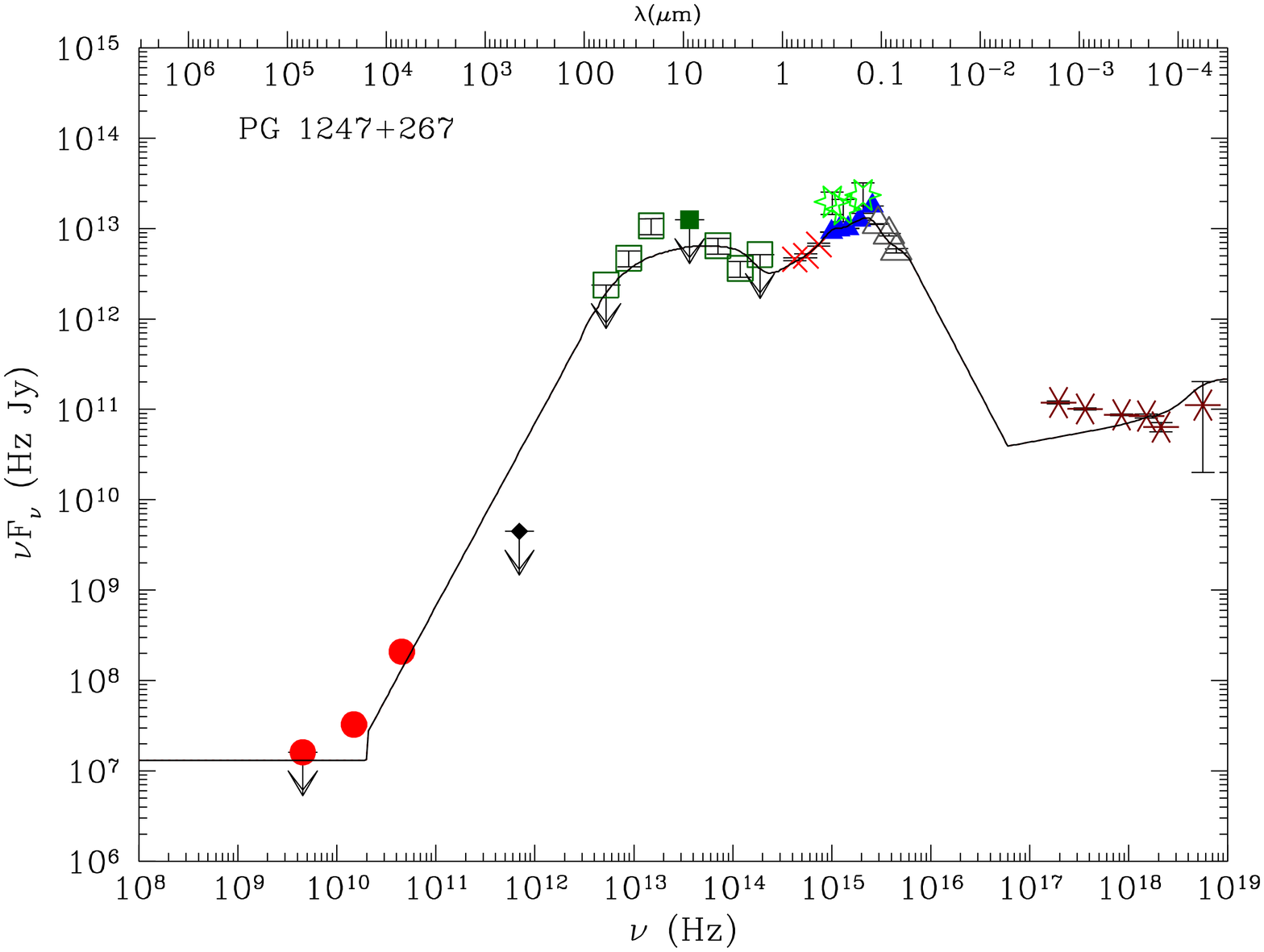}
       \label{fig:pg1247} 
       }
      }
      \mbox{ 
       \subfigure[]{
        \includegraphics[angle=-90,width=.45\linewidth]{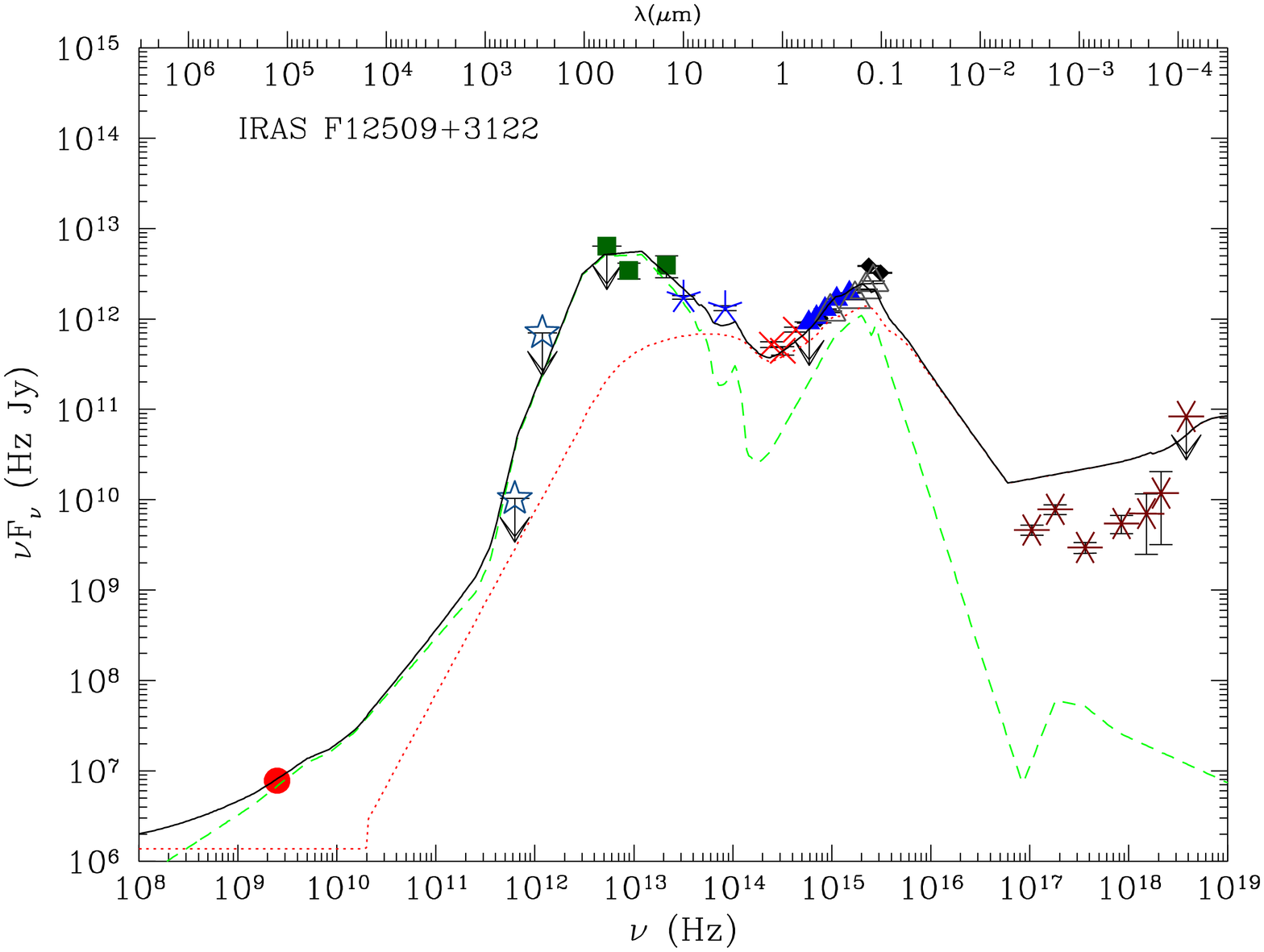}
        \label{fig:irasF12509} 
      } 
      \quad 
       \subfigure[]{
        \includegraphics[angle=-90,width=.45\linewidth]{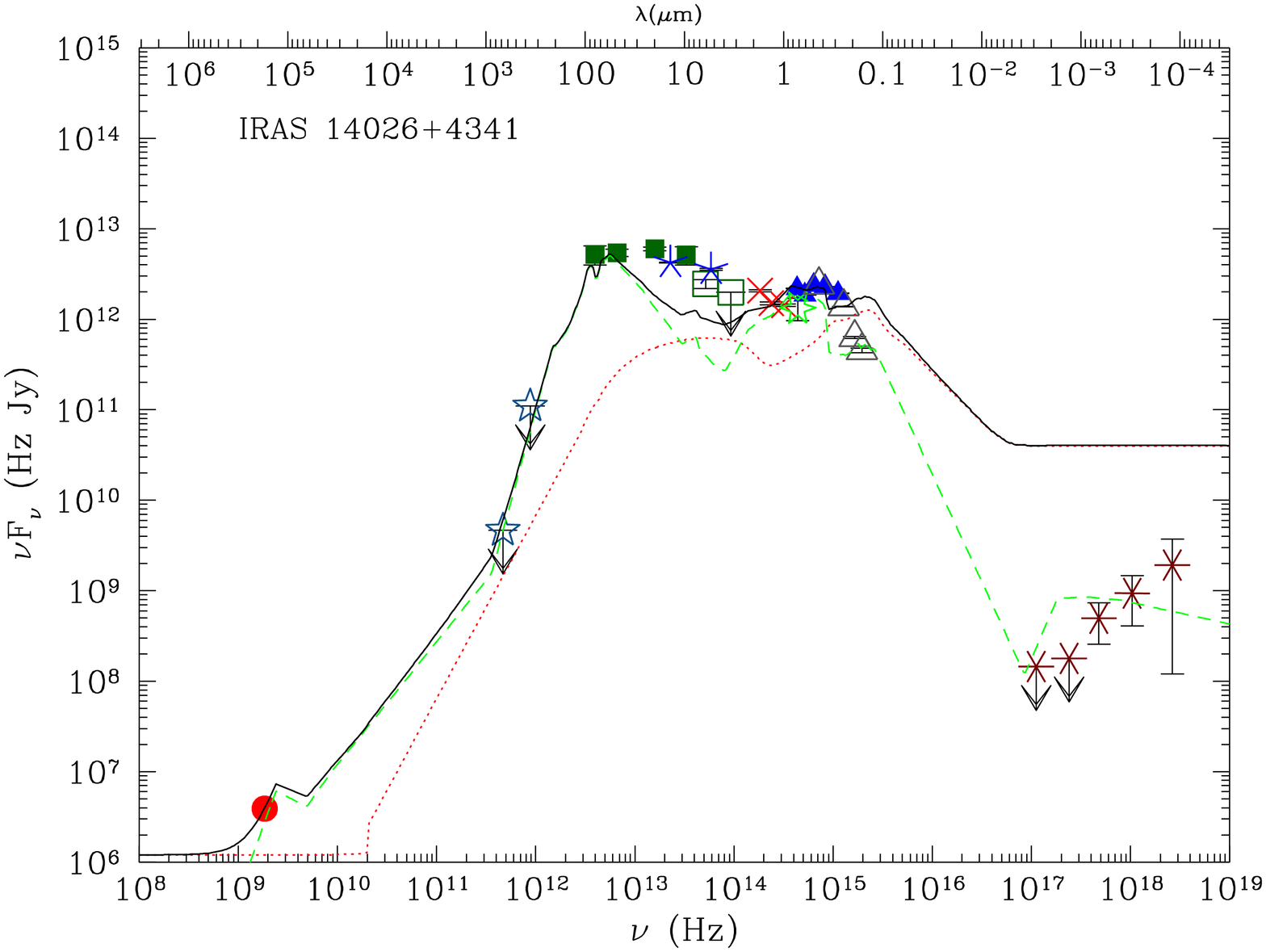}
        \label{fig:iras14026} 
       }
      }
      \mbox{
       \subfigure[]{
        \includegraphics[angle=-90,width=.45\linewidth]{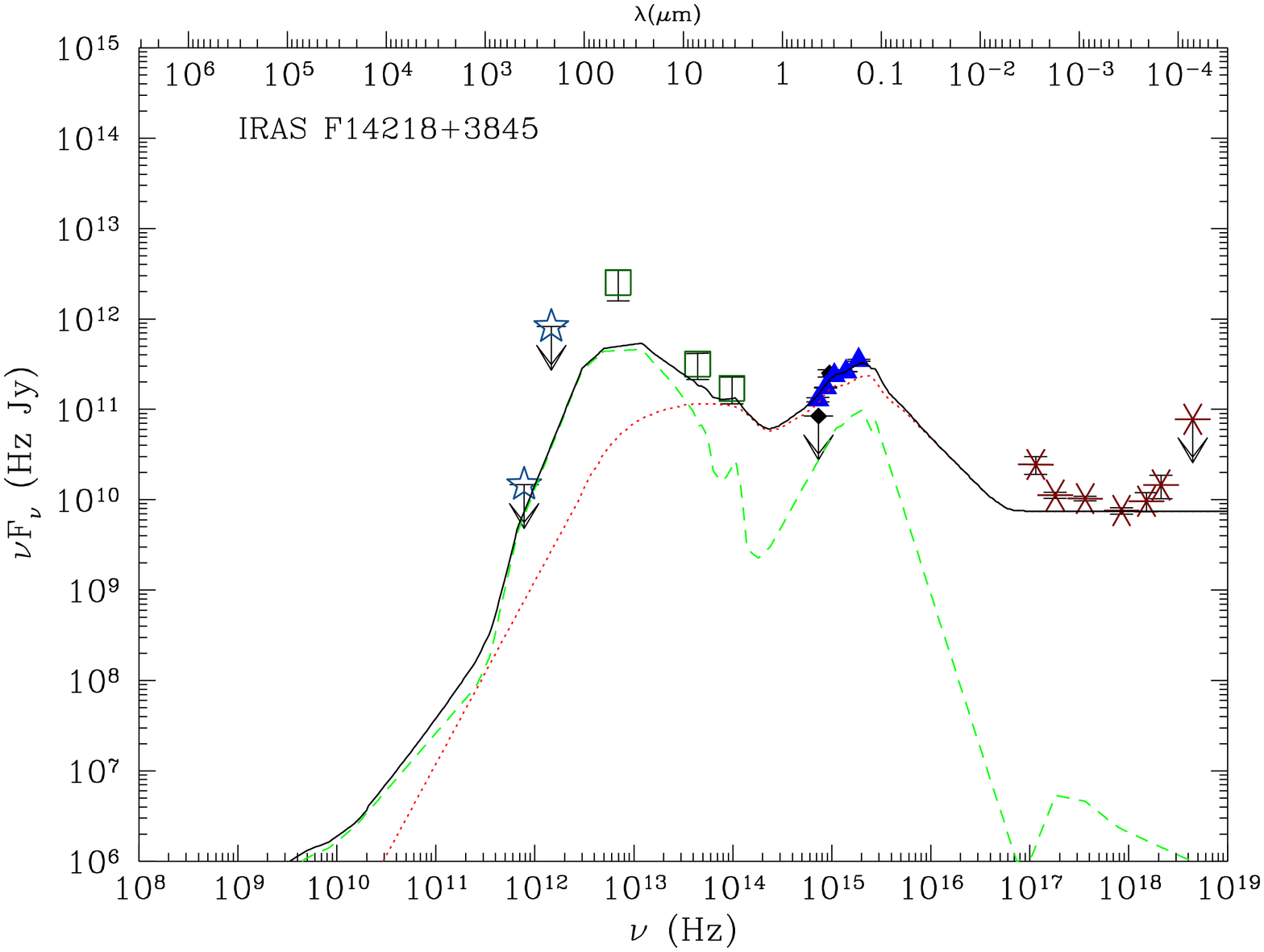}
        \label{fig:irasF14218} 
      } 
      \quad
       \subfigure[]{
        \includegraphics[angle=-90,width=.45\linewidth]{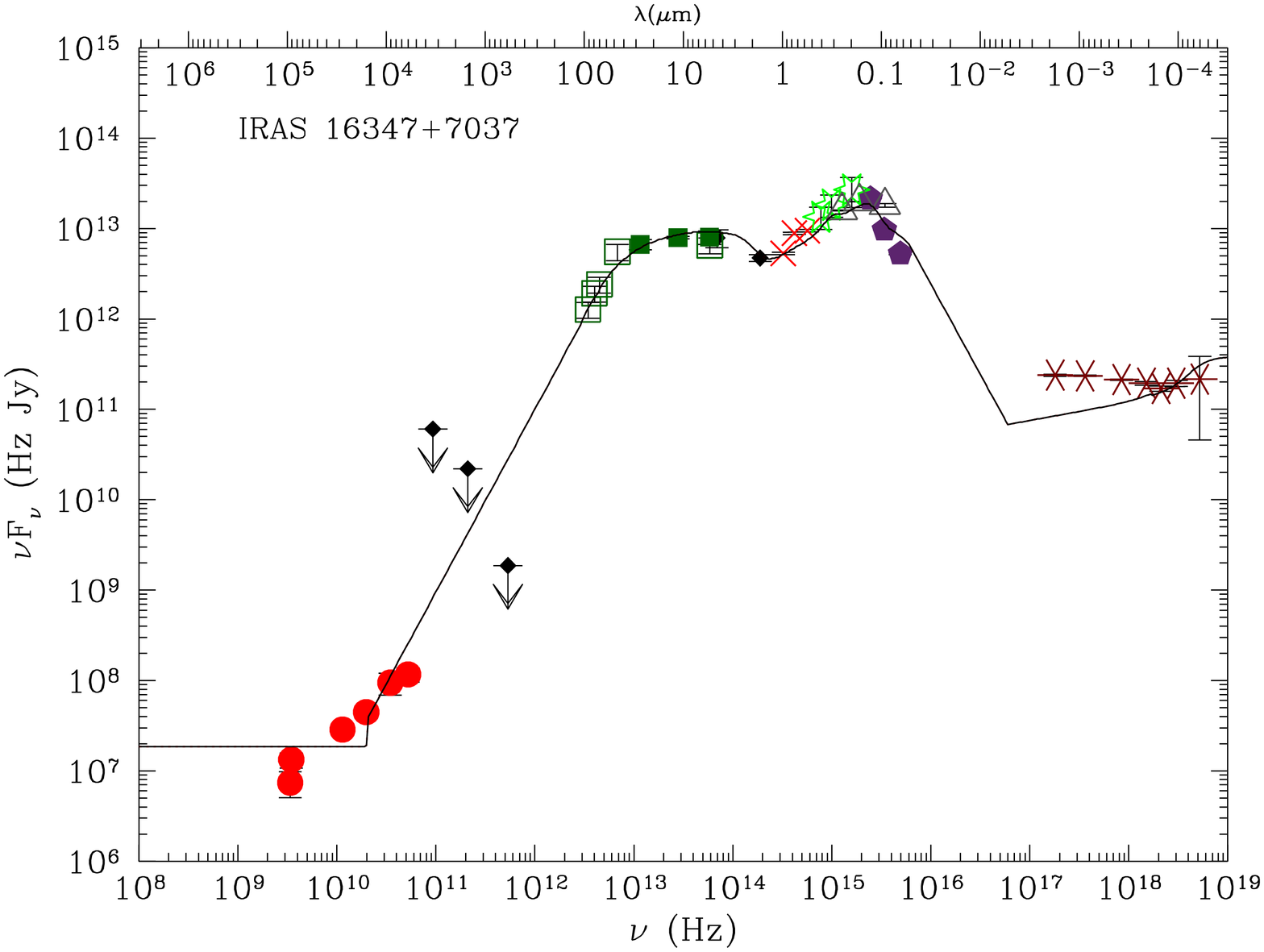}
        \label{fig:iras16347}
        }
      }
   \end{center}
\caption{Rest-frame Spectral Energy Distributions of class A HLIRGs and their best fit models. 
         Dotted lines (red in the colour version) are the AGN components and dashed lines (green 
         in the colour version) are the SB components. Black solid lines are the sum 
         of the AGN and SB components.}
\label{fig:sedsplotsA}
\end{figure*}

\addtocounter{figure}{-1}

\begin{figure*}
\centering
 \addtocounter{subfigure}{6}
      \mbox{
        \subfigure[]{
          \includegraphics[angle=-90,width=.45\linewidth]{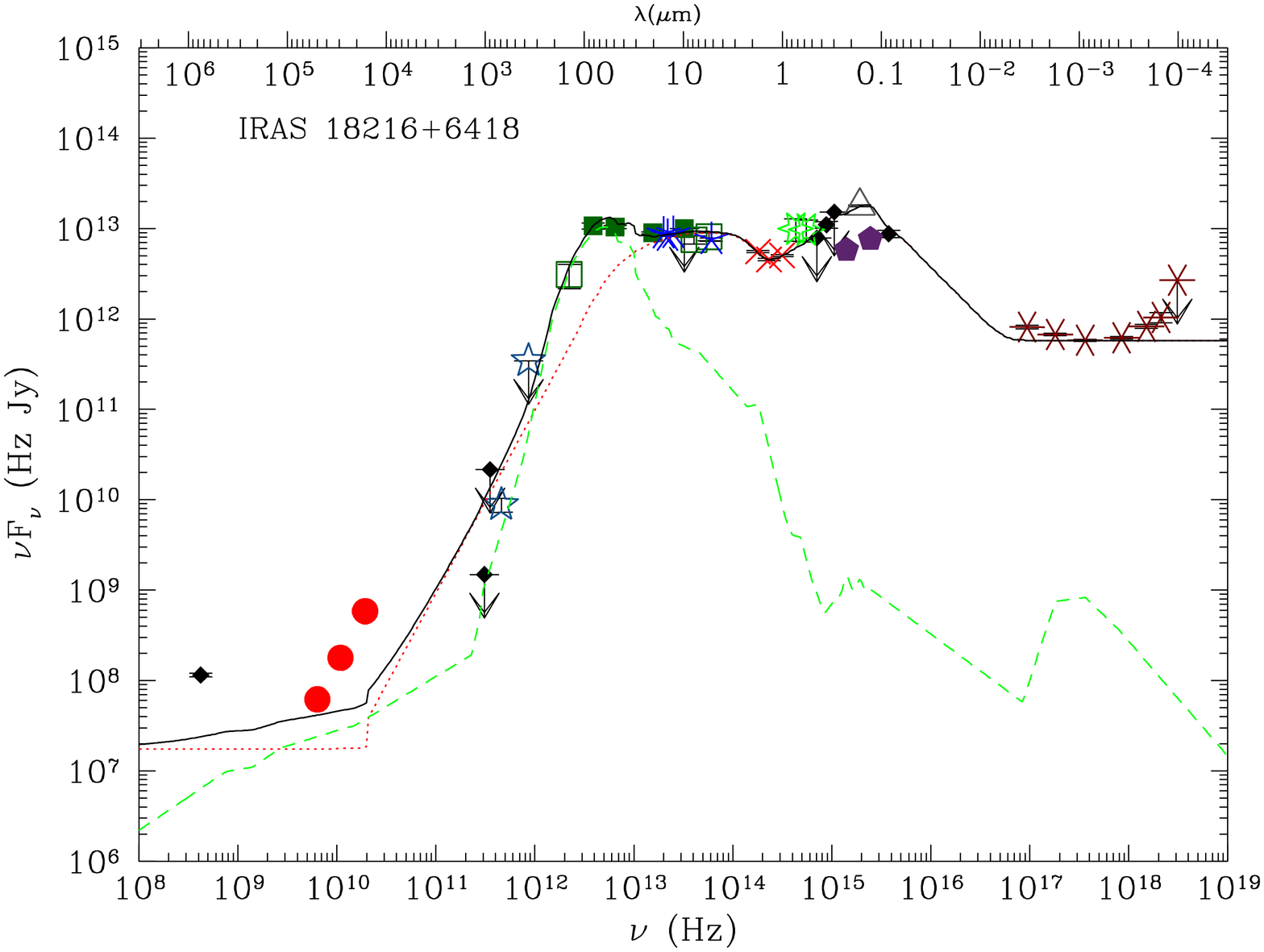} 
           \label{fig:iras18216}
        } \quad
      }
\caption{Continued.}
\label{fig:sedsplotsA:b}
\end{figure*}


\begin{figure*}
   \begin{center} 
    \mbox{ 
     \subfigure[]{
      \includegraphics[angle=-90,width=.45\linewidth]{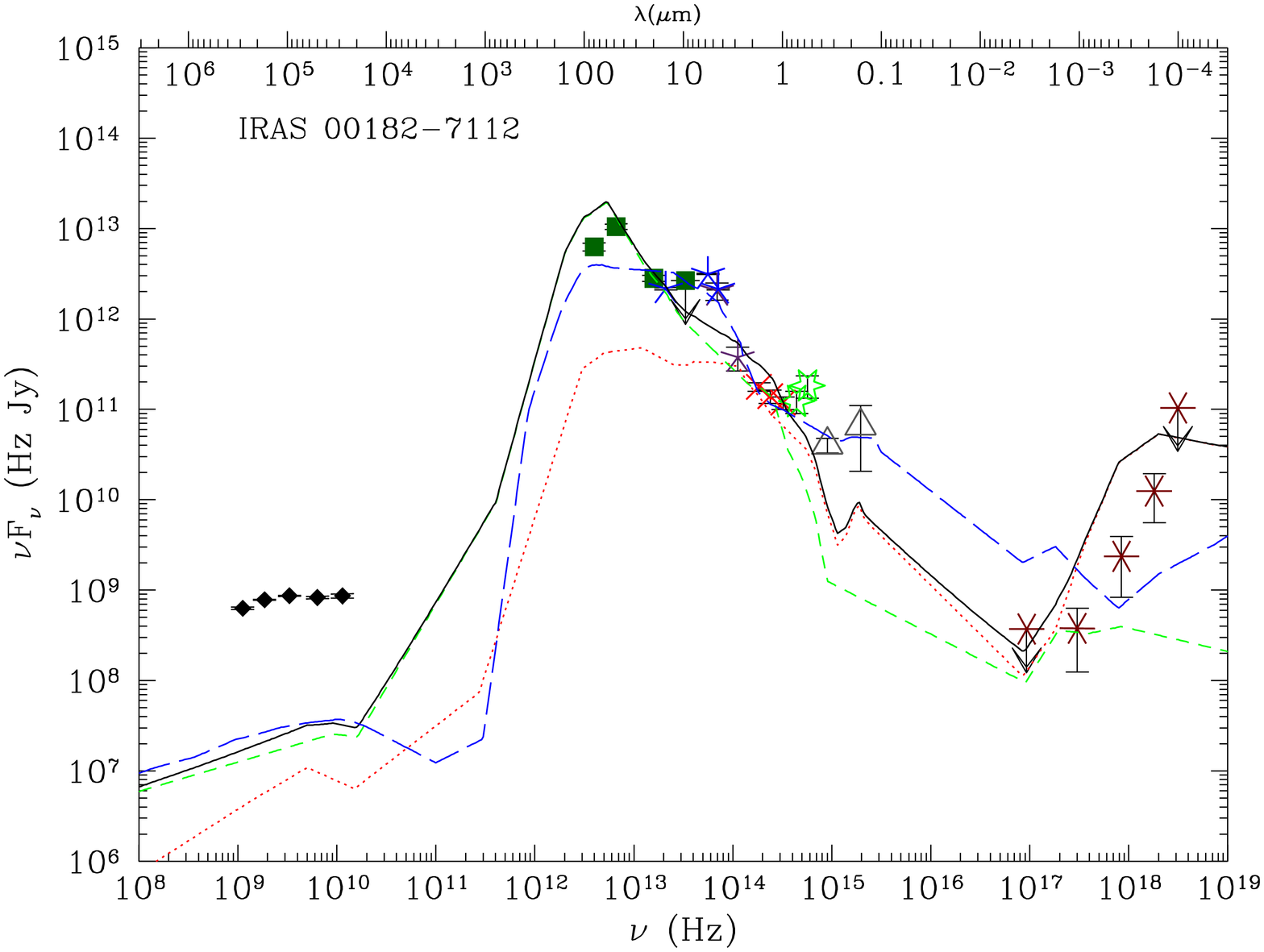}
      \label{fig:iras00182} 
     } 
    \quad 
     \subfigure[]{
      \includegraphics[angle=-90,width=.45\linewidth]{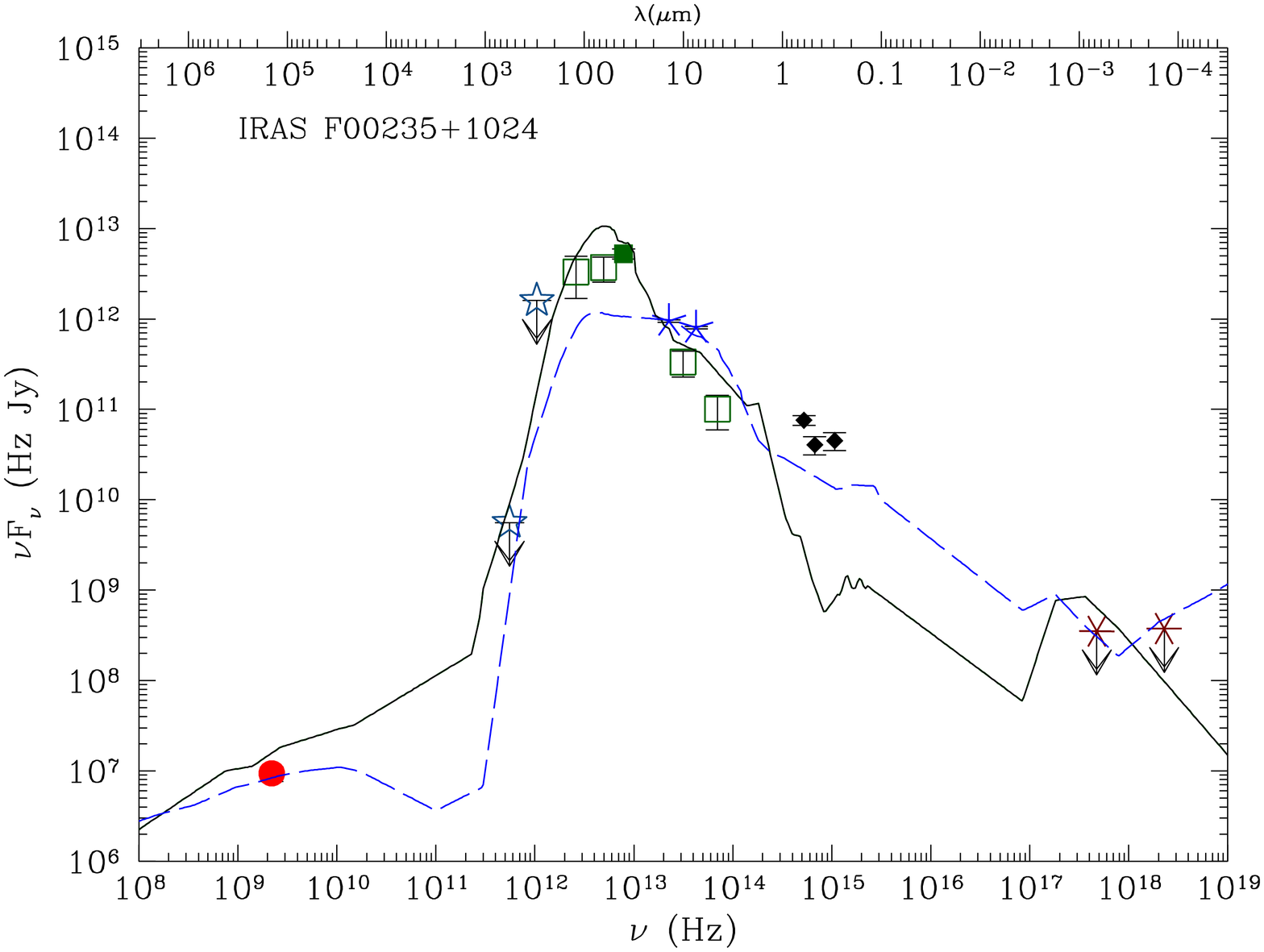}
      \label{fig:irasF00235} 
     } 
    } 
    \mbox{ 
     \subfigure[]{
      \includegraphics[angle=-90,width=.45\linewidth]{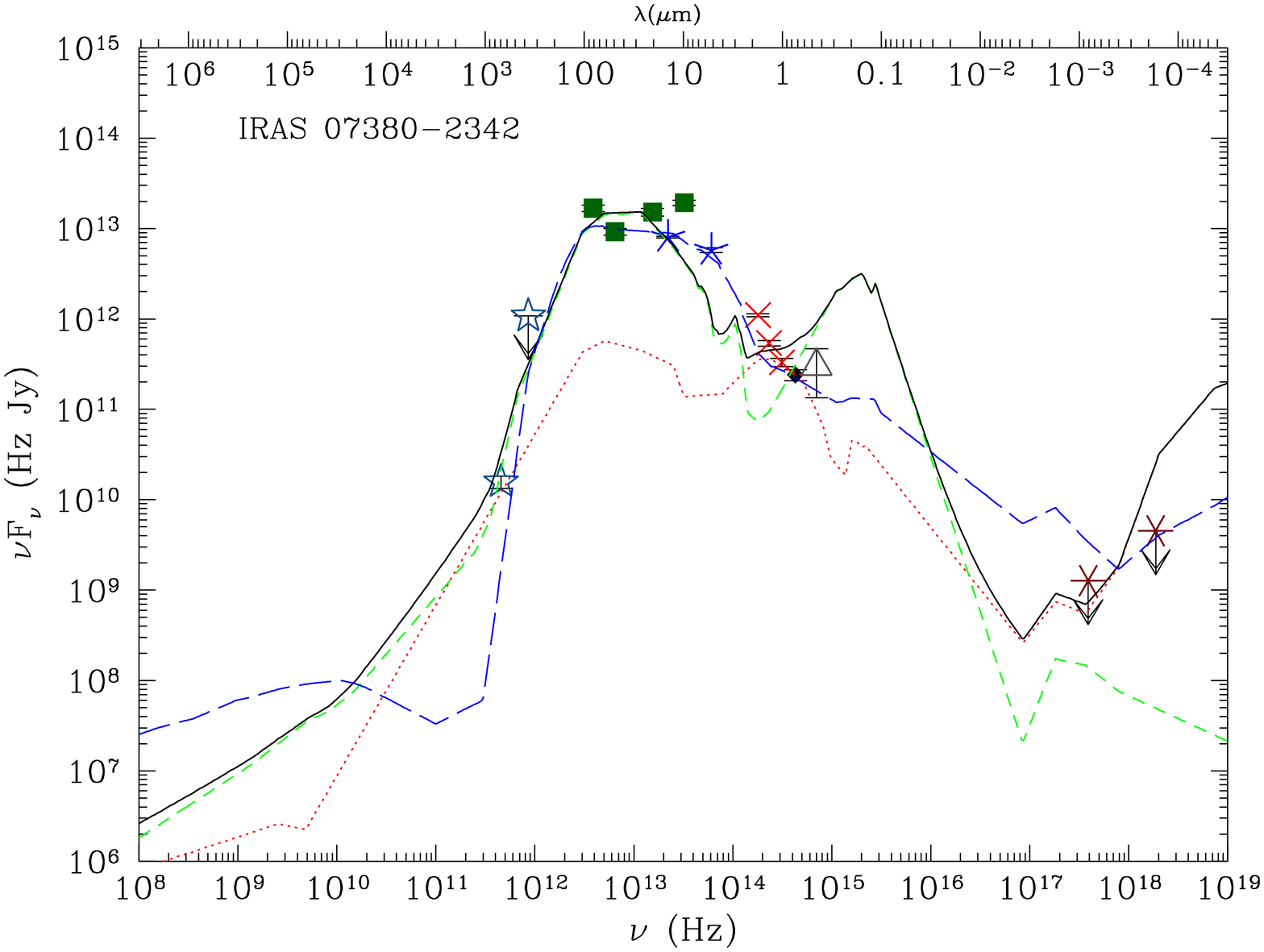}
      \label{fig:iras07380} 
     } 
    \quad 
     \subfigure[]{
      \includegraphics[angle=-90,width=.45\linewidth]{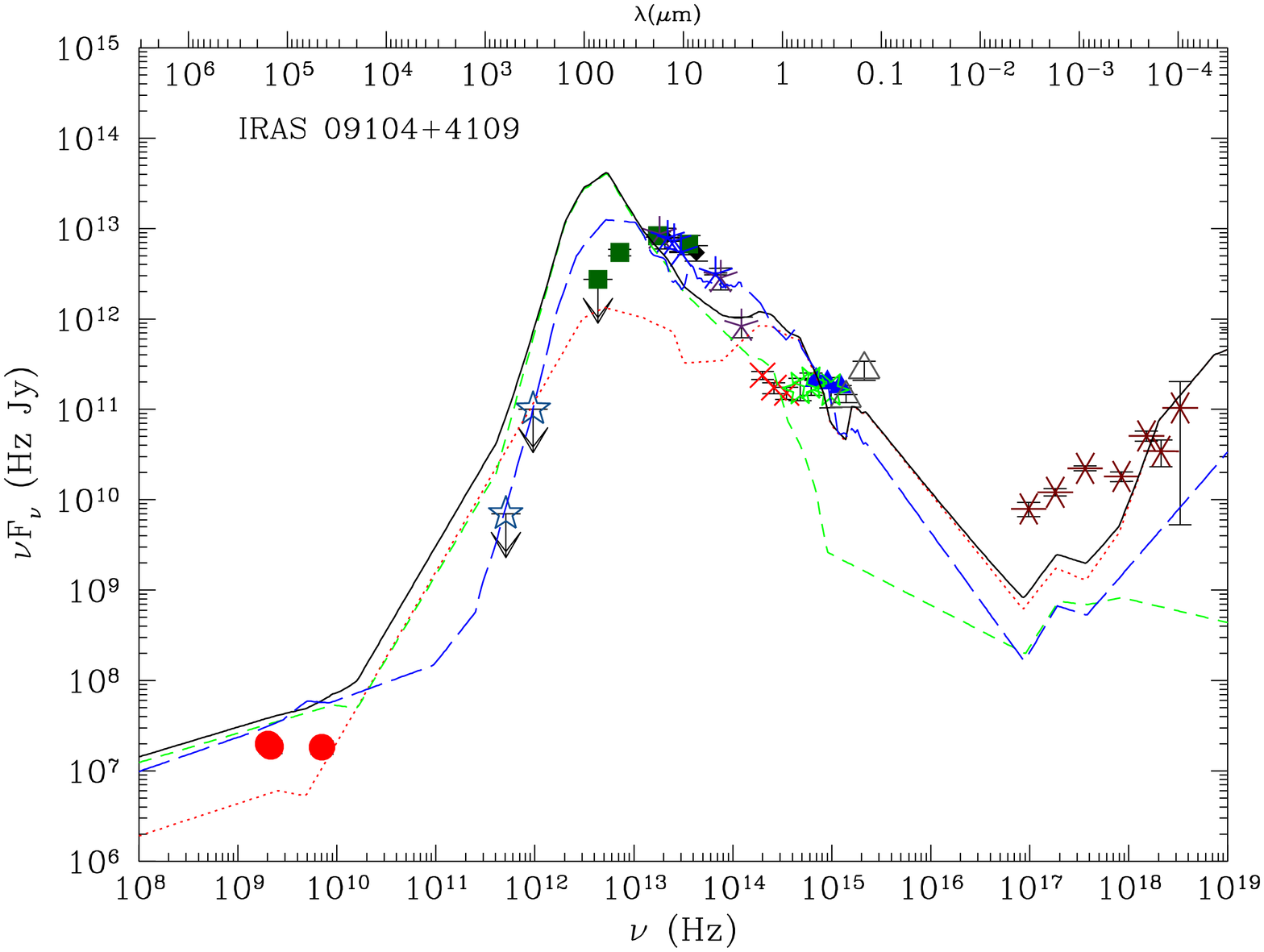}
      \label{fig:iras09104} 
     } 
    } 
    \mbox{ 
     \subfigure[]{
      \includegraphics[angle=-90,width=.45\linewidth]{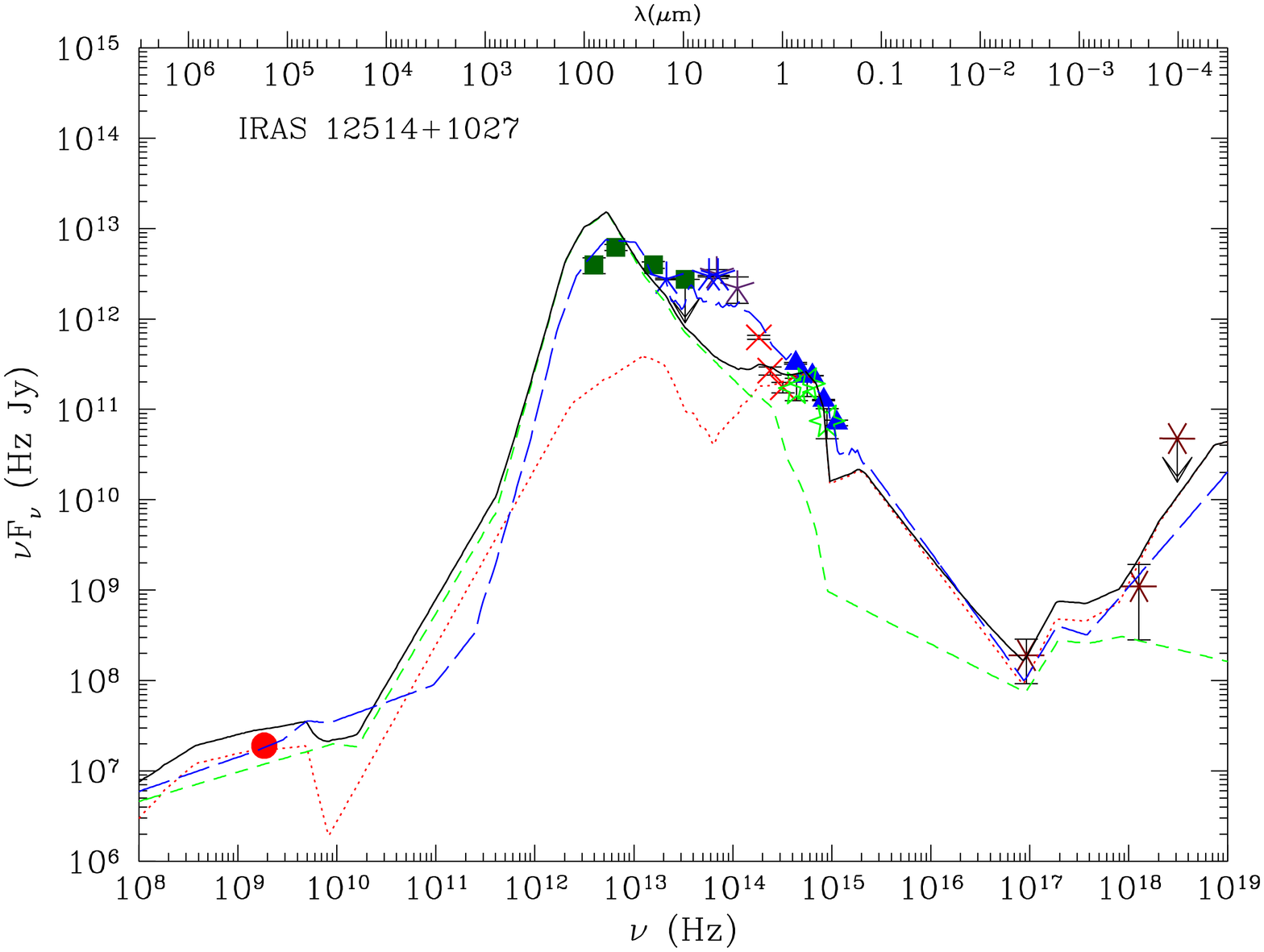}
      \label{fig:iras12514} 
     } 
    \quad 
     \subfigure[]{
      \includegraphics[angle=-90,width=.45\linewidth]{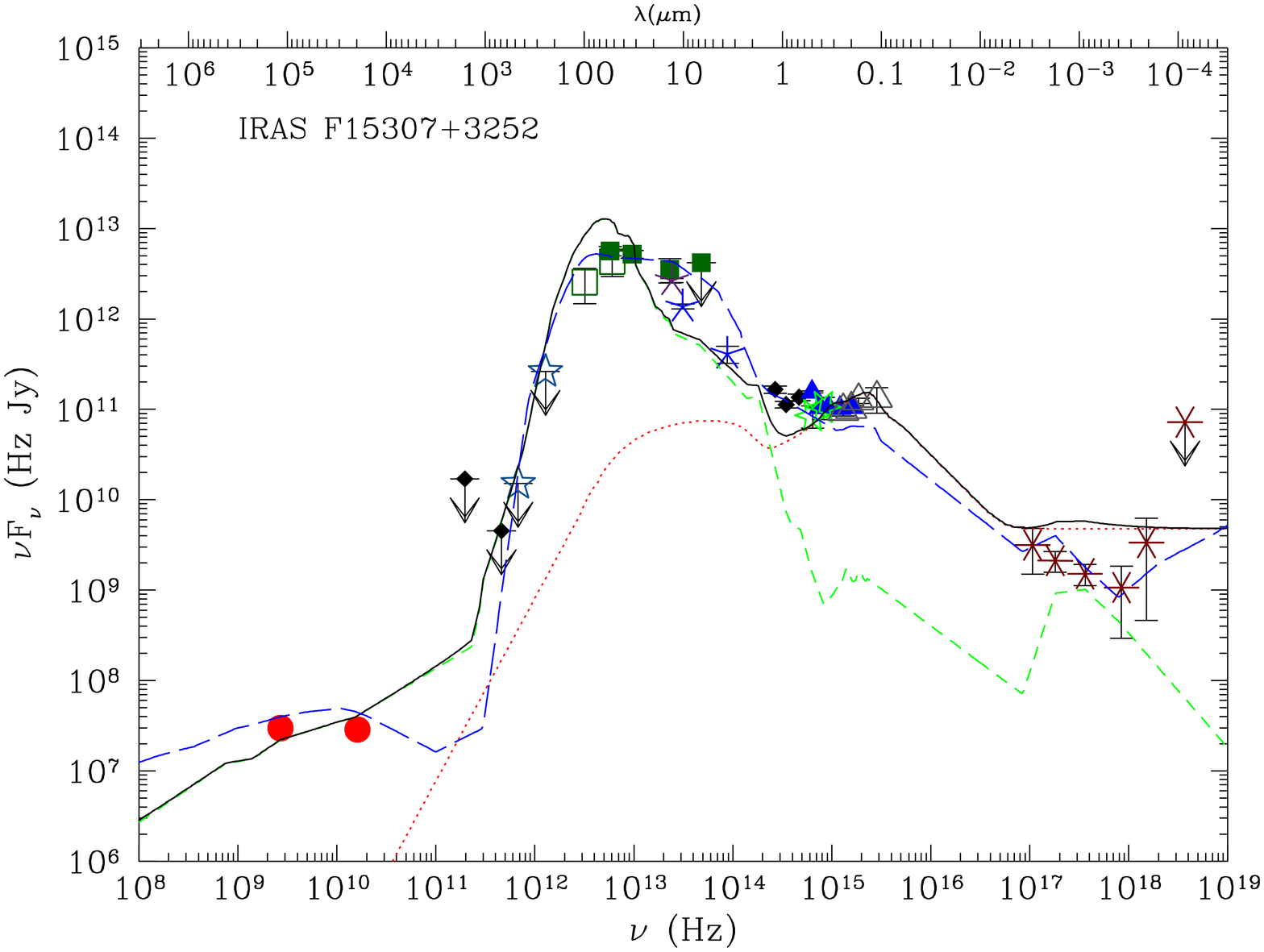}
      \label{fig:irasF15307}
     } 
    } 
   \end{center}
 \caption{Rest-frame Spectral Energy Distributions of class B HLIRGs and their best fit models.
         Symbols as in Fig.~\ref{fig:sedsplotsA}. The long-dashed lines (blue in the colour 
         version) are the best fits obtained using composite templates (see Sects.~\ref{sec:temp} 
         and \ref{sec:classBfit}).}
 \label{fig:sedsplotsB}
\end{figure*}

\begin{figure*}
\centering
\includegraphics[angle=-90, width=\linewidth]{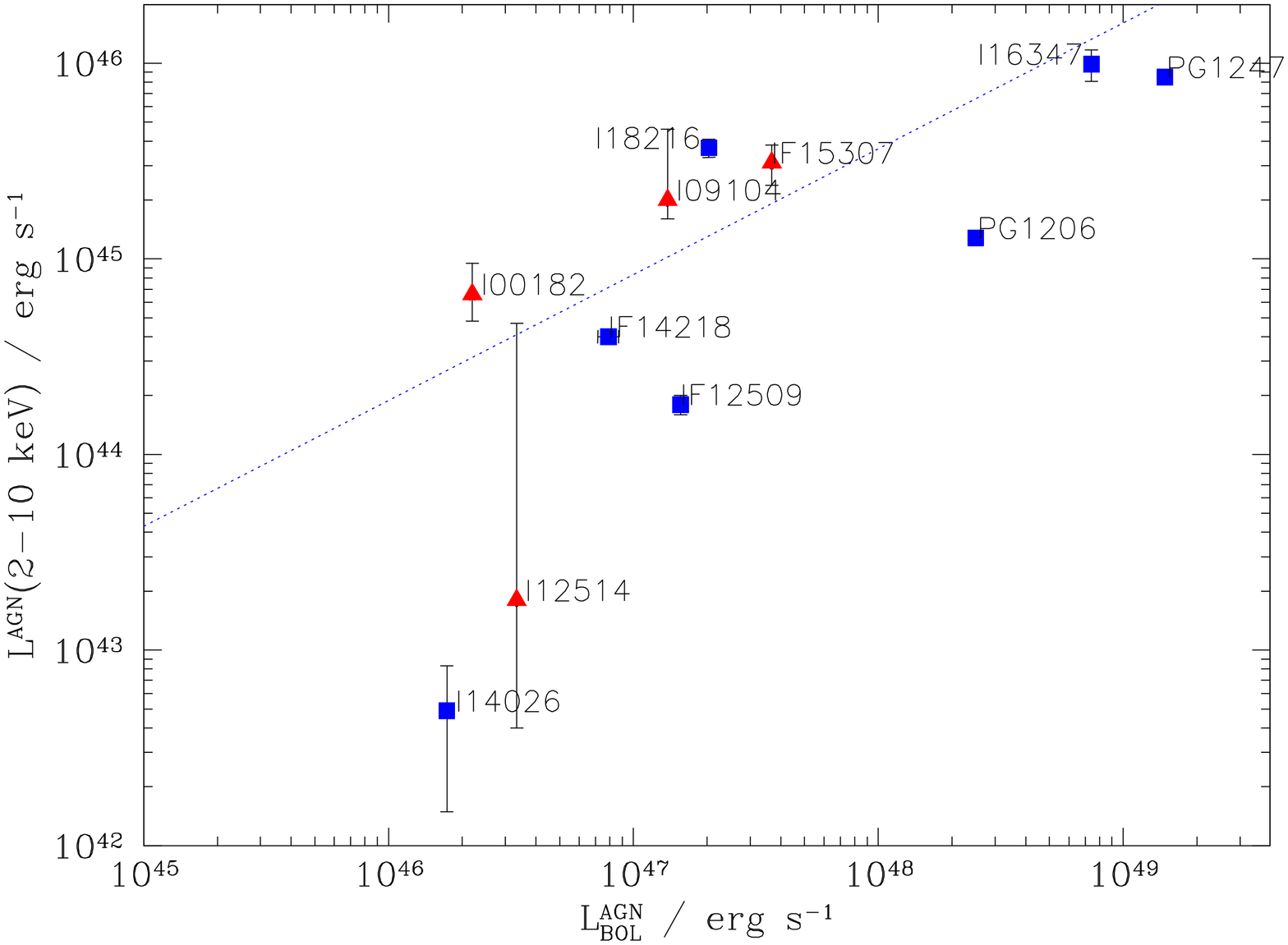}
\caption{Bolometric versus observed, absorption-corrected 2-10 keV AGN luminosities. Squares
         (blue in the colour version) are class A HLIRGs, triangles (red in the colour version)
         are class B HLIRGs. The dotted line reflects the ratio between these luminosities obtained 
         by Sani et al.}
\label{fig:AGNxrays}
\end{figure*}

\begin{figure*}
\centering
   \mbox{
     \subfigure[Total IR luminosities.]{
      \includegraphics[angle=-90,width=.45\linewidth]{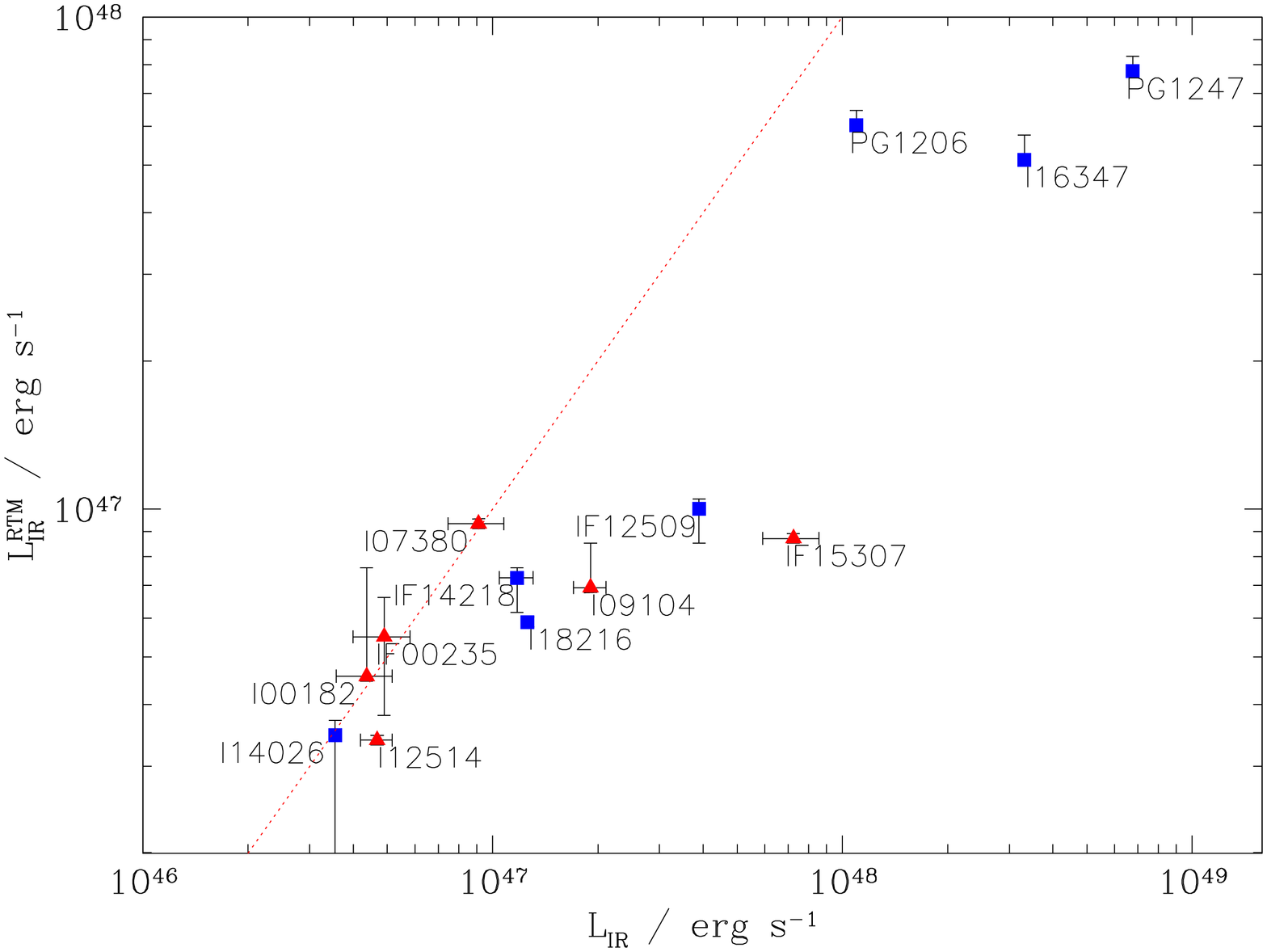}
      \label{fig:IRlumin}
     }
    \quad
     \subfigure[AGN to total IR luminosity ratios.]{
      \includegraphics[angle=-90,width=.45\linewidth]{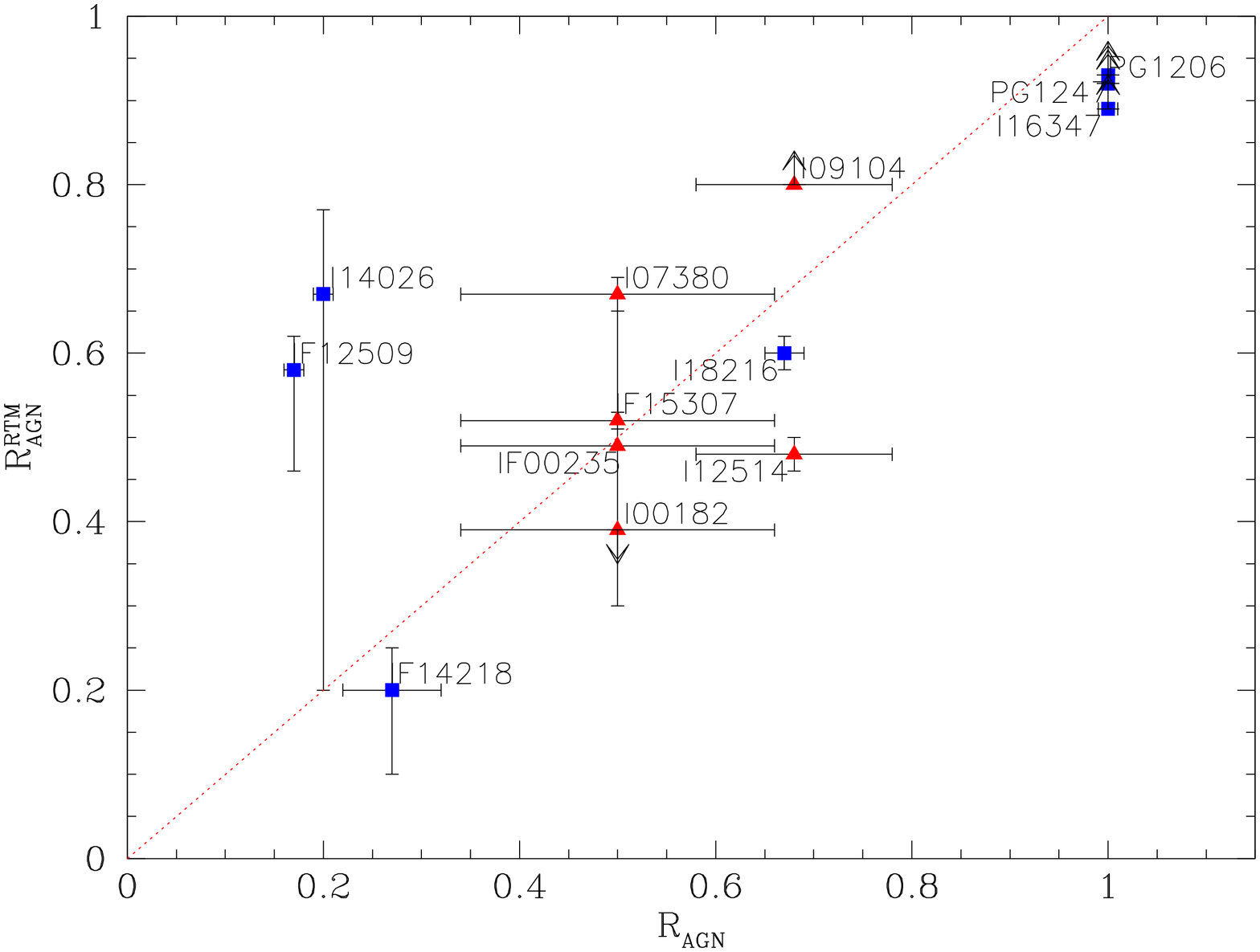}
      \label{fig:IRratios}
     }
    }

\caption{
         (a) Total IR luminosity estimated using our templates ($L_\mathrm{IR}$) and
             using radiative transfer models ($L_\mathrm{IR}^\mathrm{RTM}$).
         (b) AGN to total IR luminosity ratios estimated through our model ($R_\mathrm{AGN}$)
             and using radiative transfer models ($R_\mathrm{AGN}^\mathrm{RTM}$).
         Symbols as in Fig.~\ref{fig:AGNxrays}. The dotted lines mean equal values.
        }
\label{fig:IRcomp}
\end{figure*}

\clearpage

\begin{appendix}
\section{IRAS~13279+3401}
\label{sec:iras13279}
The object IRAS\,13279+3401 has been previously classified as a QSO, and the IR luminosity
estimated through the redshift presented in the literature \citep[z=0.36,][]{Rowan00,Farrah02submm}
exceed the HLIRG limit. However, we have now strong evidence showing that this source is a much 
closer galaxy. 

\begin{figure*}
    \mbox{ 
     \subfigure[Optical spectrum (2.5m INT).]{
      \includegraphics[angle=-90,width=.45\linewidth]{aruiz_SEDhlirgs_figA1a}
      \label{fig:optspec13279} 
     } 
    \quad 
     \subfigure[MIR spectrum (\Spitzer Infrared Spectrograph).]{
      \includegraphics[angle=-90,width=.5\linewidth]{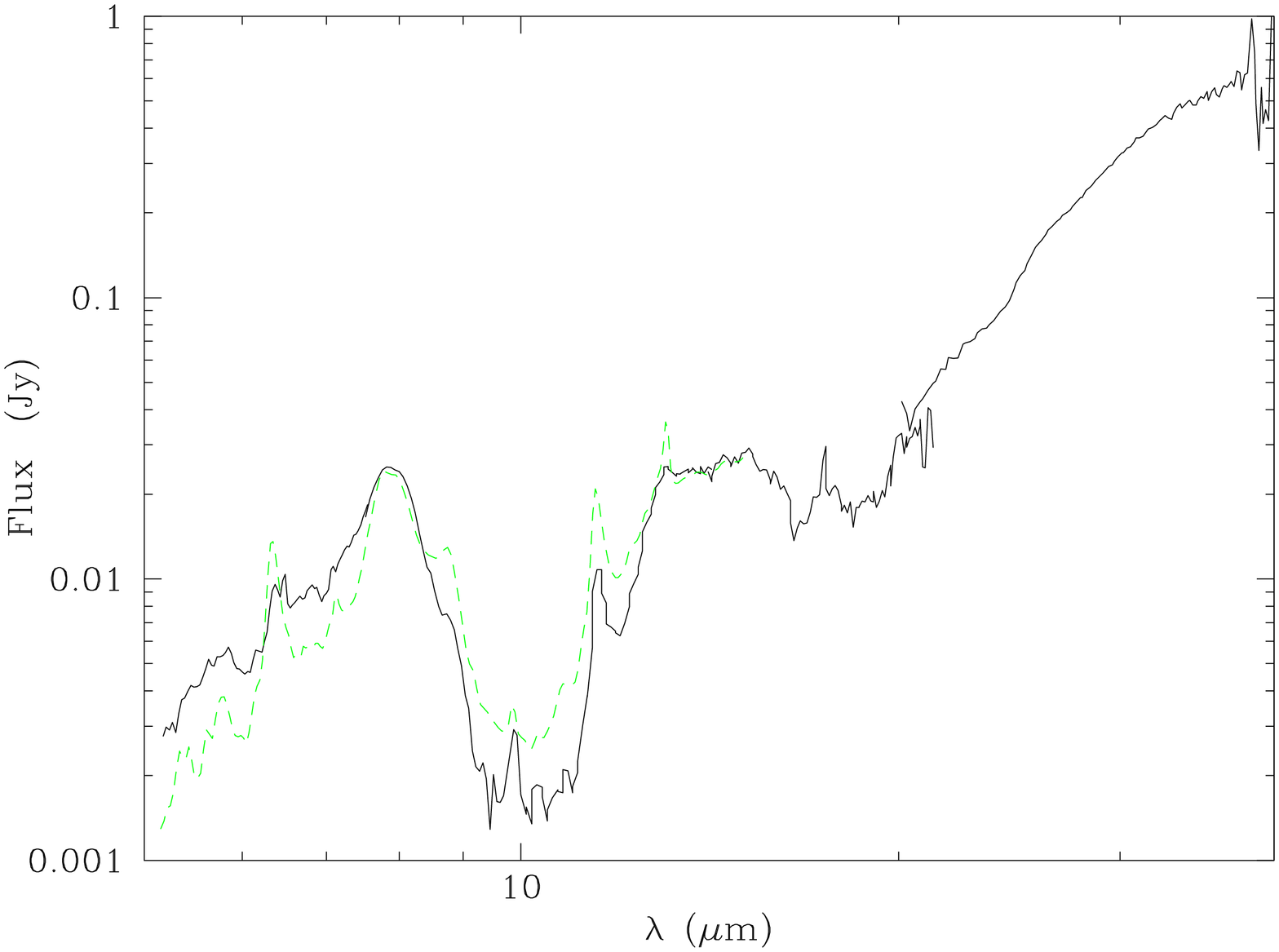}
      \label{fig:mirspec13279} 
     } 
    } 

\caption{(a) Optical and (b) MIR spectra of IRAS\ 13279+3401 in the observer frame. The 
         slashed line (green in the colour version) in the right hand side plot is the 
         SB template from \citet{Nardini08}.}
\label{fig:spec13279}
\end{figure*}

Figure~A.\ref{fig:optspec13279} shows its optical spectrum obtained by the
2.5m Isaac Newton Telescope, where we do not find any type I feature. A QSO with z=0.36 should 
present a broad H$_\beta$ emission line at $\sim6600$~\AA. We estimate z=0.023 for this spectrum,
from stellar absorption features.

We have also the MIR spectrum of this source, observed by \Spitzer (see Fig.~A.\ref{fig:mirspec13279}).
We have estimated the redshift of the source using a SB template from \citet{Nardini08}. 
We redshifted the template matching the most important spectral features and we found $z\sim0.02$, 
which is consistent with our estimate using the optical spectrum. The IR luminosity derived 
from this redshift is $\sim3\times10^{10}~L_{\sun}$, well below the HLIRG limit and even below
the LIRG.
\end{appendix}

\Online
\begin{appendix}
\section{Tables of data}
\label{sec:tables}
\nobalance
Along this appendix a table is presented for each object with the fluxes employed
to build the SEDs and the origin of each data. The re-binned spectra from \XMM and
\Spitzer are not included in these tables. Fluxes shown with no errors are upper limits.

\begin{table}[ht]
\caption{Photometric data for IRAS~00182-7112.}
\label{tab:iras00182data}
\begin{center}
\begin{tabular}{llll}
\hline\noalign{\smallskip}
$\nu$ (Hz) & $F_{\nu}$ (Jy) & Error (Jy) & Ref. \\
\hline\noalign{\smallskip}
$8.43\times10^{8}$ & $4.23\times10^{-1}$ & $1.28\times10^{-2}$ & \scriptsize{SUMSS (NED)} \\ 
$1.40\times10^{9}$ & $3.17\times10^{-1}$ & $3.00\times10^{-3}$ & \scriptsize{ATCA \citep{Drake04}} \\ 
$2.50\times10^{9}$ & $1.97\times10^{-1}$ & $3.00\times10^{-3}$ & \scriptsize{ATCA \citep{Drake04}} \\ 
$4.80\times10^{9}$ & $9.80\times10^{-2}$ & $3.00\times10^{-3}$ & \scriptsize{ATCA \citep{Drake04}} \\ 
$8.60\times10^{9}$ & $5.70\times10^{-2}$ & $3.00\times10^{-3}$ & \scriptsize{ATCA \citep{Drake04}} \\ 
$3.00\times10^{12}$ & $1.19\times10^{0}$ & $1.19\times10^{-1}$ & \scriptsize{IRAS (NED)} \\ 
$5.00\times10^{12}$ & $1.20\times10^{0}$ & $8.37\times10^{-2}$ & \scriptsize{IRAS (NED)} \\ 
$1.20\times10^{13}$ & $1.33\times10^{-1}$ & $1.02\times10^{-2}$ & \scriptsize{IRAS (NED)} \\ 
$2.50\times10^{13}$ & $6.02\times10^{-2}$ & ... & \scriptsize{IRAS (NED)} \\ 
$5.23\times10^{13}$ & $2.23\times10^{-2}$ & $4.78\times10^{-3}$ & \scriptsize{\Spitzer-IRAC} \\ 
$8.44\times10^{13}$ & $2.53\times10^{-3}$ & $7.47\times10^{-4}$ & \scriptsize{\Spitzer-IRAC} \\ 
$1.39\times10^{14}$ & $7.22\times10^{-4}$ & $7.83\times10^{-5}$ & \scriptsize{2MASS} \\ 
$1.80\times10^{14}$ & $4.37\times10^{-4}$ & $7.34\times10^{-5}$ & \scriptsize{2MASS} \\ 
$2.43\times10^{14}$ & $2.68\times10^{-4}$ & $4.34\times10^{-5}$ & \scriptsize{2MASS} \\ 
$3.33\times10^{14}$ & $2.02\times10^{-4}$ & $5.57\times10^{-5}$ & \scriptsize{SSS} \\ 
$4.28\times10^{14}$ & $2.28\times10^{-4}$ & $6.30\times10^{-5}$ & \scriptsize{SSS} \\ 
$6.81\times10^{14}$ & $2.97\times10^{-5}$ & $5.40\times10^{-6}$ & \scriptsize{\XMM-OM} \\ 
$1.48\times10^{15}$ & $1.90\times10^{-5}$ & $1.30\times10^{-5}$ & \scriptsize{\XMM-OM} \\ 
\noalign{\smallskip}\hline\noalign{\smallskip}
\end{tabular}
\end{center}
\end{table}

\begin{table}[ht]
\caption{Photometric data for IRAS~F00235+1024.}
\label{tab:irasF00235data}
\begin{center}
\begin{tabular}{llll}
\hline\noalign{\smallskip}
$\nu$ (Hz) & $F_{\nu}$ (Jy) & Error (Jy) & Ref. \\
\hline\noalign{\smallskip}
$1.40\times10^{9}$ & $2.70\times10^{-3}$ & $5.00\times10^{-4}$ & \scriptsize{VLA (NED)} \\ 
$3.53\times10^{11}$ & $6.38\times10^{-3}$ & ... & \scriptsize{SCUBA \citep{Farrah02submm}} \\ 
$6.66\times10^{11}$ & $9.67\times10^{-1}$ & ... & \scriptsize{SCUBA \citep{Farrah02submm}} \\ 
$1.66\times10^{12}$ & $8.05\times10^{-1}$ & $3.93\times10^{-1}$ & \scriptsize{ISO (NED)} \\ 
$3.15\times10^{12}$ & $4.78\times10^{-1}$ & $1.48\times10^{-1}$ & \scriptsize{ISO (NED)} \\ 
$5.00\times10^{12}$ & $4.28\times10^{-1}$ & $5.56\times10^{-2}$ & \scriptsize{IRAS (NED)} \\ 
$2.00\times10^{13}$ & $6.75\times10^{-3}$ & $2.14\times10^{-3}$ & \scriptsize{ISO (NED)} \\ 
$4.44\times10^{13}$ & $9.20\times10^{-4}$ & $3.80\times10^{-4}$ & \scriptsize{ISO (NED)} \\ 
$1.39\times10^{14}$ & $1.01\times10^{-4}$ & $5.00\times10^{-6}$ & \scriptsize{\citep{Farrah02submm}} \\ 
$3.33\times10^{14}$ & $8.00\times10^{-5}$ & $1.00\times10^{-5}$ & \scriptsize{\citep{Farrah02submm}} \\ 
$4.28\times10^{14}$ & $3.10\times10^{-5}$ & $7.00\times10^{-6}$ & \scriptsize{USNO-B1.0} \\ 
$6.81\times10^{14}$ & $1.80\times10^{-5}$ & $4.00\times10^{-6}$ & \scriptsize{USNO-B1.0} \\ 
$3.02\times10^{17}$ & $4.68\times10^{-10}$ & ... & \scriptsize{\XMM-EPIC} \\ 
$1.45\times10^{18}$ & $1.04\times10^{-10}$ & ... & \scriptsize{\XMM-EPIC} \\ 
\noalign{\smallskip}\hline\noalign{\smallskip}
\end{tabular}
\end{center}
\end{table}

\begin{table}[ht]
\caption{Photometric data for IRAS~07380-2342.}
\label{tab:iras07380data}
\begin{center}
\begin{tabular}{llll}
\hline\noalign{\smallskip}
$\nu$ (Hz) & $F_{\nu}$ (Jy) & Error (Jy) & Ref. \\
\hline\noalign{\smallskip}
$3.53\times10^{11}$ & $2.68\times10^{-2}$ & $4.20\times10^{-3}$ & \scriptsize{SCUBA \citep{Farrah02submm}} \\ 
$6.66\times10^{11}$ & $9.67\times10^{-1}$ & ... & \scriptsize{SCUBA \citep{Farrah02submm}} \\ 
$3.00\times10^{12}$ & $3.55\times10^{0}$ & $2.84\times10^{-1}$ & \scriptsize{IRAS (NED)} \\ 
$5.00\times10^{12}$ & $1.17\times10^{0}$ & $9.36\times10^{-2}$ & \scriptsize{IRAS (NED)} \\ 
$1.20\times10^{13}$ & $8.00\times10^{-1}$ & $8.00\times10^{-2}$ & \scriptsize{IRAS (NED)} \\ 
$2.50\times10^{13}$ & $4.84\times10^{-1}$ & $3.39\times10^{-2}$ & \scriptsize{IRAS (NED)} \\ 
$1.39\times10^{14}$ & $4.08\times10^{-3}$ & $1.49\times10^{-4}$ & \scriptsize{2MASS} \\ 
$1.80\times10^{14}$ & $1.35\times10^{-3}$ & $9.71\times10^{-5}$ & \scriptsize{2MASS} \\ 
$2.43\times10^{14}$ & $4.74\times10^{-4}$ & $4.79\times10^{-5}$ & \scriptsize{2MASS} \\ 
$3.33\times10^{14}$ & $1.62\times10^{-4}$ & $2.23\times10^{-5}$ & \scriptsize{DENIS} \\ 
$5.45\times10^{14}$ & $3.80\times10^{-5}$ & $2.10\times10^{-5}$ & \scriptsize{\XMM-OM} \\ 
$3.02\times10^{17}$ & $2.51\times10^{-9}$ & ... & \scriptsize{\XMM-EPIC} \\ 
$1.45\times10^{18}$ & $1.87\times10^{-9}$ & ... & \scriptsize{\XMM-EPIC} \\ 
\noalign{\smallskip}\hline\noalign{\smallskip}
\end{tabular}
\end{center}
\end{table}

\begin{table}[ht]
\caption{Photometric data for IRAS~09104+4109.}
\label{tab:iras09104data}
\begin{center}
\begin{tabular}{llll}
\hline\noalign{\smallskip}
$\nu$ (Hz) & $F_{\nu}$ (Jy) & Error (Jy) & Ref. \\
\hline\noalign{\smallskip}
$1.40\times10^{9}$ & $6.88\times10^{-3}$ & $1.31\times10^{-4}$ & \scriptsize{VLA (NED)} \\ 
$1.49\times10^{9}$ & $6.00\times10^{-3}$ & $1.00\times10^{-3}$ & \scriptsize{VLA (NED)} \\ 
$4.90\times10^{9}$ & $1.80\times10^{-3}$ & $3.00\times10^{-4}$ & \scriptsize{VLA (NED)} \\ 
$3.53\times10^{11}$ & $9.54\times10^{-3}$ & ... & \scriptsize{SCUBA \citep{Farrah02submm}} \\ 
$6.66\times10^{11}$ & $7.28\times10^{-2}$ & ... & \scriptsize{SCUBA \citep{Farrah02submm}} \\ 
$3.00\times10^{12}$ & $4.38\times10^{-1}$ & ... & \scriptsize{IRAS (NED)} \\ 
$5.00\times10^{12}$ & $5.25\times10^{-1}$ & $4.20\times10^{-2}$ & \scriptsize{IRAS (NED)} \\ 
$1.20\times10^{13}$ & $3.34\times10^{-1}$ & $1.30\times10^{-2}$ & \scriptsize{IRAS (NED)} \\ 
$1.26\times10^{13}$ & $3.33\times10^{-1}$ & $1.70\times10^{-2}$ & \scriptsize{\Spitzer-IRAC} \\ 
$1.43\times10^{13}$ & $2.70\times10^{-1}$ & $7.00\times10^{-2}$ & \scriptsize{IRTF \citep{Klein88}} \\ 
$2.50\times10^{13}$ & $1.30\times10^{-1}$ & $3.11\times10^{-2}$ & \scriptsize{IRAS (NED)} \\ 
$2.97\times10^{13}$ & $8.80\times10^{-2}$ & $1.70\times10^{-2}$ & \scriptsize{IRTF \citep{Klein88}} \\ 
$5.23\times10^{13}$ & $2.64\times10^{-2}$ & $7.11\times10^{-3}$ & \scriptsize{\Spitzer-IRAC} \\ 
$8.44\times10^{13}$ & $4.74\times10^{-3}$ & $1.21\times10^{-3}$ & \scriptsize{\Spitzer-IRAC} \\ 
$1.39\times10^{14}$ & $8.21\times10^{-4}$ & $8.61\times10^{-5}$ & \scriptsize{2MASS} \\ 
$1.80\times10^{14}$ & $4.59\times10^{-4}$ & $6.24\times10^{-5}$ & \scriptsize{2MASS} \\ 
$2.43\times10^{14}$ & $2.97\times10^{-4}$ & $4.64\times10^{-5}$ & \scriptsize{2MASS} \\ 
$3.33\times10^{14}$ & $2.42\times10^{-4}$ & $6.69\times10^{-5}$ & \scriptsize{SSS} \\ 
$3.28\times10^{14}$ & $5.06\times10^{-4}$ & $7.45\times10^{-6}$ & \scriptsize{SDSS-DR5} \\ 
$3.93\times10^{14}$ & $7.04\times10^{-4}$ & $3.42\times10^{-6}$ & \scriptsize{SDSS-DR5} \\ 
$4.28\times10^{14}$ & $2.12\times10^{-4}$ & $5.85\times10^{-5}$ & \scriptsize{SSS} \\ 
$4.81\times10^{14}$ & $2.11\times10^{-4}$ & $1.69\times10^{-6}$ & \scriptsize{SDSS-DR5} \\ 
$6.28\times10^{14}$ & $1.51\times10^{-4}$ & $1.26\times10^{-6}$ & \scriptsize{SDSS-DR5} \\ 
$6.81\times10^{14}$ & $1.08\times10^{-4}$ & $3.97\times10^{-5}$ & \scriptsize{SSS} \\ 
$8.47\times10^{14}$ & $1.01\times10^{-4}$ & $3.16\times10^{-6}$ & \scriptsize{SDSS-DR5} \\ 
$9.67\times10^{14}$ & $5.90\times10^{-5}$ & $6.00\times10^{-6}$ & \scriptsize{\XMM-OM} \\ 
$1.48\times10^{15}$ & $7.50\times10^{-5}$ & $1.80\times10^{-5}$ & \scriptsize{\XMM-OM} \\ 
$7.86\times10^{18}$ & $1.06\times10^{-7}$ & $5.91\times10^{-8}$ & \scriptsize{BeppoSAX (NED)} \\ 
\noalign{\smallskip}\hline\noalign{\smallskip}
\end{tabular}
\end{center}
\end{table}

\begin{table}[ht]
\caption{Photometric data for PG~1206+459.}
\label{tab:pg1206data}
\begin{center}
\begin{tabular}{llll}
\hline\noalign{\smallskip}
$\nu$ (Hz) & $F_{\nu}$ (Jy) & Error (Jy) & Ref. \\
\hline\noalign{\smallskip}
$4.90\times10^{9}$ & $1.20\times10^{-4}$ & ... & \scriptsize{VLA (NED)} \\ 
$2.30\times10^{11}$ & $1.80\times10^{-3}$ & $4.50\times10^{-4}$ & \scriptsize{IRAM (NED)} \\ 
$1.72\times10^{12}$ & $1.89\times10^{-1}$ & $3.78\times10^{-2}$ & \scriptsize{ISO (NED)} \\ 
$2.93\times10^{12}$ & $3.53\times10^{-1}$ & $7.06\times10^{-2}$ & \scriptsize{ISO (NED)} \\ 
$4.93\times10^{12}$ & $2.57\times10^{-1}$ & $5.14\times10^{-2}$ & \scriptsize{ISO (NED)} \\ 
$1.20\times10^{13}$ & $1.13\times10^{-1}$ & ... & \scriptsize{IRAS (NED)} \\ 
$1.43\times10^{13}$ & $6.40\times10^{-2}$ & $1.28\times10^{-2}$ & \scriptsize{ISO (NED)} \\ 
$2.34\times10^{13}$ & $2.30\times10^{-2}$ & $4.60\times10^{-3}$ & \scriptsize{ISO (NED)} \\ 
$2.50\times10^{13}$ & $2.07\times10^{-1}$ & $3.60\times10^{-2}$ & \scriptsize{IRAS (NED)} \\ 
$6.17\times10^{13}$ & $3.00\times10^{-3}$ & ... & \scriptsize{ISO (NED)} \\ 
$1.39\times10^{14}$ & $2.64\times10^{-3}$ & $9.46\times10^{-5}$ & \scriptsize{2MASS} \\ 
$1.80\times10^{14}$ & $2.57\times10^{-3}$ & $1.22\times10^{-4}$ & \scriptsize{2MASS} \\ 
$2.43\times10^{14}$ & $2.43\times10^{-3}$ & $7.76\times10^{-5}$ & \scriptsize{2MASS} \\ 
$3.33\times10^{14}$ & $2.27\times10^{-3}$ & $6.26\times10^{-4}$ & \scriptsize{SSS} \\ 
$3.28\times10^{14}$ & $2.78\times10^{-3}$ & $1.26\times10^{-5}$ & \scriptsize{SDSS-DR5} \\ 
$3.93\times10^{14}$ & $2.94\times10^{-3}$ & $9.76\times10^{-6}$ & \scriptsize{SDSS-DR5} \\ 
$4.28\times10^{14}$ & $2.51\times10^{-3}$ & $6.93\times10^{-4}$ & \scriptsize{SSS} \\ 
$4.81\times10^{14}$ & $2.92\times10^{-3}$ & $9.45\times10^{-6}$ & \scriptsize{SDSS-DR5} \\ 
$5.45\times10^{14}$ & $2.62\times10^{-3}$ & $3.00\times10^{-5}$ & \scriptsize{\XMM-OM} \\ 
$6.28\times10^{14}$ & $2.53\times10^{-3}$ & $7.01\times10^{-6}$ & \scriptsize{SDSS-DR5} \\ 
$6.81\times10^{14}$ & $2.25\times10^{-3}$ & $1.30\times10^{-5}$ & \scriptsize{\XMM-OM} \\ 
$8.47\times10^{14}$ & $2.20\times10^{-3}$ & $9.06\times10^{-6}$ & \scriptsize{SDSS-DR5} \\ 
$1.05\times10^{15}$ & $1.00\times10^{-3}$ & $9.21\times10^{-5}$ & \scriptsize{IUE} \\ 
$1.46\times10^{15}$ & $5.25\times10^{-4}$ & $7.25\times10^{-5}$ & \scriptsize{IUE} \\ 
$2.11\times10^{15}$ & $1.45\times10^{-4}$ & $8.32\times10^{-5}$ & \scriptsize{IUE} \\ 
\noalign{\smallskip}\hline\noalign{\smallskip}
\end{tabular}
\end{center}
\end{table}

\begin{table}[ht]
\caption{Photometric data for PG~1247+267.}
\label{tab:pg1247data}
\begin{center}
\begin{tabular}{llll}
\hline\noalign{\smallskip}
$\nu$ (Hz) & $F_{\nu}$ (Jy) & Error (Jy) & Ref. \\
\hline\noalign{\smallskip}
$1.49\times10^{9}$ & $1.17\times10^{-3}$ & ... & \scriptsize{VLA (NED)} \\ 
$4.90\times10^{9}$ & $7.20\times10^{-4}$ & $8.00\times10^{-5}$ & \scriptsize{VLA (NED)} \\ 
$1.49\times10^{10}$ & $1.51\times10^{-3}$ & $2.20\times10^{-4}$ & \scriptsize{VLA (NED)} \\ 
$2.30\times10^{11}$ & $2.10\times10^{-3}$ & ... & \scriptsize{IRAM (NED)} \\ 
$1.72\times10^{12}$ & $1.50\times10^{-1}$ & ... & \scriptsize{ISO (NED)} \\ 
$2.93\times10^{12}$ & $1.74\times10^{-1}$ & $3.48\times10^{-2}$ & \scriptsize{ISO (NED)} \\ 
$4.93\times10^{12}$ & $2.36\times10^{-1}$ & $4.72\times10^{-2}$ & \scriptsize{ISO (NED)} \\ 
$1.20\times10^{13}$ & $1.13\times10^{-1}$ & ... & \scriptsize{IRAS (NED)} \\ 
$2.34\times10^{13}$ & $3.00\times10^{-2}$ & $6.00\times10^{-3}$ & \scriptsize{ISO (NED)} \\ 
$3.91\times10^{13}$ & $1.00\times10^{-2}$ & $2.00\times10^{-3}$ & \scriptsize{ISO (NED)} \\ 
$6.17\times10^{13}$ & $9.00\times10^{-3}$ & ... & \scriptsize{ISO (NED)} \\ 
$1.39\times10^{14}$ & $3.55\times10^{-3}$ & $1.36\times10^{-4}$ & \scriptsize{2MASS} \\ 
$1.80\times10^{14}$ & $2.99\times10^{-3}$ & $1.45\times10^{-4}$ & \scriptsize{2MASS} \\ 
$2.43\times10^{14}$ & $2.93\times10^{-3}$ & $1.10\times10^{-4}$ & \scriptsize{2MASS} \\ 
$3.33\times10^{14}$ & $6.31\times10^{-3}$ & $1.74\times10^{-3}$ & \scriptsize{SSS} \\ 
$3.28\times10^{14}$ & $3.01\times10^{-3}$ & $1.37\times10^{-5}$ & \scriptsize{SDSS-DR5} \\ 
$3.93\times10^{14}$ & $2.66\times10^{-3}$ & $9.69\times10^{-6}$ & \scriptsize{SDSS-DR5} \\ 
$4.28\times10^{14}$ & $4.03\times10^{-3}$ & $1.11\times10^{-3}$ & \scriptsize{SSS} \\ 
$4.81\times10^{14}$ & $2.26\times10^{-3}$ & $8.02\times10^{-6}$ & \scriptsize{SDSS-DR5} \\ 
$6.28\times10^{14}$ & $2.13\times10^{-3}$ & $6.42\times10^{-6}$ & \scriptsize{SDSS-DR5} \\ 
$6.81\times10^{14}$ & $3.54\times10^{-3}$ & $1.30\times10^{-3}$ & \scriptsize{SSS} \\ 
$8.47\times10^{14}$ & $2.27\times10^{-3}$ & $9.92\times10^{-6}$ & \scriptsize{SDSS-DR5} \\ 
$9.67\times10^{14}$ & $1.17\times10^{-3}$ & $8.00\times10^{-6}$ & \scriptsize{\XMM-OM} \\ 
$1.25\times10^{15}$ & $6.70\times10^{-4}$ & $2.00\times10^{-5}$ & \scriptsize{\XMM-OM} \\ 
$1.48\times10^{15}$ & $3.70\times10^{-4}$ & $2.00\times10^{-5}$ & \scriptsize{\XMM-OM} \\ 
\noalign{\smallskip}\hline\noalign{\smallskip}
\end{tabular}
\end{center}
\end{table}

\begin{table}[ht]
\caption{Photometric data for IRAS~F12509+3122.}
\label{tab:irasF12509data}
\begin{center}
\begin{tabular}{llll}
\hline\noalign{\smallskip}
$\nu$ (Hz) & $F_{\nu}$ (Jy) & Error (Jy) & Ref. \\
\hline\noalign{\smallskip}
$1.40\times10^{9}$ & $1.76\times10^{-3}$ & $1.25\times10^{-4}$ & \scriptsize{VLA (NED)} \\ 
$3.53\times10^{11}$ & $9.23\times10^{-3}$ & ... & \scriptsize{SCUBA \citep{Farrah02submm}} \\ 
$6.66\times10^{11}$ & $3.33\times10^{-1}$ & ... & \scriptsize{SCUBA \citep{Farrah02submm}} \\ 
$3.00\times10^{12}$ & $6.75\times10^{-1}$ & ... & \scriptsize{IRAS (NED)} \\ 
$5.00\times10^{12}$ & $2.18\times10^{-1}$ & $4.36\times10^{-2}$ & \scriptsize{IRAS (NED)} \\ 
$1.20\times10^{13}$ & $1.03\times10^{-1}$ & $2.75\times10^{-2}$ & \scriptsize{IRAS (NED)} \\ 
$1.39\times10^{14}$ & $1.18\times10^{-3}$ & $8.78\times10^{-5}$ & \scriptsize{2MASS} \\ 
$1.80\times10^{14}$ & $7.75\times10^{-4}$ & $8.07\times10^{-5}$ & \scriptsize{2MASS} \\ 
$2.43\times10^{14}$ & $9.77\times10^{-4}$ & $6.01\times10^{-5}$ & \scriptsize{2MASS} \\ 
$3.33\times10^{14}$ & $8.60\times10^{-4}$ & ... & \scriptsize{\citep{Farrah02submm}} \\ 
$3.28\times10^{14}$ & $8.73\times10^{-4}$ & $7.57\times10^{-6}$ & \scriptsize{SDSS-DR5} \\ 
$3.93\times10^{14}$ & $8.39\times10^{-4}$ & $3.56\times10^{-6}$ & \scriptsize{SDSS-DR5} \\ 
$4.28\times10^{14}$ & $7.20\times10^{-4}$ & $7.00\times10^{-5}$ & \scriptsize{APM} \\ 
$4.81\times10^{14}$ & $8.39\times10^{-4}$ & $3.12\times10^{-6}$ & \scriptsize{SDSS-DR5} \\ 
$5.45\times10^{14}$ & $6.60\times10^{-4}$ & $1.60\times10^{-5}$ & \scriptsize{\XMM-OM} \\ 
$6.28\times10^{14}$ & $8.32\times10^{-4}$ & $2.89\times10^{-6}$ & \scriptsize{SDSS-DR5} \\ 
$8.47\times10^{14}$ & $7.22\times10^{-4}$ & $4.89\times10^{-6}$ & \scriptsize{SDSS-DR5} \\ 
$9.67\times10^{14}$ & $4.94\times10^{-4}$ & $6.00\times10^{-6}$ & \scriptsize{\XMM-OM} \\ 
$1.25\times10^{15}$ & $4.68\times10^{-4}$ & $7.00\times10^{-6}$ & \scriptsize{\XMM-OM} \\ 
$1.33\times10^{15}$ & $7.83\times10^{-4}$ & $9.94\times10^{-7}$ & \scriptsize{HST - FOS} \\ 
$1.48\times10^{15}$ & $4.69\times10^{-4}$ & $1.60\times10^{-5}$ & \scriptsize{\XMM-OM} \\ 
$1.75\times10^{15}$ & $5.16\times10^{-4}$ & $1.38\times10^{-6}$ & \scriptsize{HST - FOS} \\ 
\noalign{\smallskip}\hline\noalign{\smallskip}
\end{tabular}
\end{center}
\end{table}

\begin{table}[ht]
\caption{Photometric data for IRAS~12514+1027.}
\label{tab:iras12514data}
\begin{center}
\begin{tabular}{llll}
\hline\noalign{\smallskip}
$\nu$ (Hz) & $F_{\nu}$ (Jy) & Error (Jy) & Ref. \\
\hline\noalign{\smallskip}
$1.40\times10^{9}$ & $7.78\times10^{-3}$ & $1.46\times10^{-4}$ & \scriptsize{VLA (NED)} \\ 
$3.00\times10^{12}$ & $7.55\times10^{-1}$ & $1.51\times10^{-1}$ & \scriptsize{IRAS (NED)} \\ 
$5.00\times10^{12}$ & $7.12\times10^{-1}$ & $5.70\times10^{-2}$ & \scriptsize{IRAS (NED)} \\ 
$1.20\times10^{13}$ & $1.90\times10^{-1}$ & $1.58\times10^{-2}$ & \scriptsize{IRAS (NED)} \\ 
$2.50\times10^{13}$ & $6.32\times10^{-2}$ & ... & \scriptsize{IRAS (NED)} \\ 
$5.23\times10^{13}$ & $3.46\times10^{-2}$ & $4.03\times10^{-3}$ & \scriptsize{\Spitzer-IRAC} \\ 
$8.44\times10^{13}$ & $1.49\times10^{-2}$ & $4.82\times10^{-3}$ & \scriptsize{\Spitzer-IRAC} \\ 
$1.39\times10^{14}$ & $2.57\times10^{-3}$ & $1.26\times10^{-4}$ & \scriptsize{2MASS} \\ 
$1.80\times10^{14}$ & $8.45\times10^{-4}$ & $8.57\times10^{-5}$ & \scriptsize{2MASS} \\ 
$2.43\times10^{14}$ & $4.07\times10^{-4}$ & $5.40\times10^{-5}$ & \scriptsize{2MASS} \\ 
$3.33\times10^{14}$ & $2.88\times10^{-4}$ & $7.96\times10^{-5}$ & \scriptsize{SSS} \\ 
$3.28\times10^{14}$ & $5.66\times10^{-4}$ & $1.15\times10^{-5}$ & \scriptsize{SDSS-DR5} \\ 
$3.93\times10^{14}$ & $3.39\times10^{-4}$ & $2.96\times10^{-6}$ & \scriptsize{SDSS-DR5} \\ 
$4.28\times10^{14}$ & $2.46\times10^{-4}$ & $6.81\times10^{-5}$ & \scriptsize{SSS} \\ 
$4.81\times10^{14}$ & $2.82\times10^{-4}$ & $2.30\times10^{-6}$ & \scriptsize{SDSS-DR5} \\ 
$6.28\times10^{14}$ & $1.16\times10^{-4}$ & $1.50\times10^{-6}$ & \scriptsize{SDSS-DR5} \\ 
$6.81\times10^{14}$ & $5.78\times10^{-5}$ & $2.13\times10^{-5}$ & \scriptsize{SSS} \\ 
$8.47\times10^{14}$ & $4.79\times10^{-5}$ & $3.53\times10^{-6}$ & \scriptsize{SDSS-DR5} \\ 
\noalign{\smallskip}\hline\noalign{\smallskip}
\end{tabular}
\end{center}
\end{table}

\begin{table}[ht]
\caption{Photometric data for IRAS~14026+4341.}
\label{tab:iras14026data}
\begin{center}
\begin{tabular}{llll}
\hline\noalign{\smallskip}
$\nu$ (Hz) & $F_{\nu}$ (Jy) & Error (Jy) & Ref. \\
\hline\noalign{\smallskip}
$1.40\times10^{9}$ & $1.59\times10^{-3}$ & $1.39\times10^{-4}$ & \scriptsize{VLA (NED)} \\ 
$3.53\times10^{11}$ & $7.53\times10^{-3}$ & ... & \scriptsize{SCUBA \citep{Farrah02submm}} \\ 
$6.66\times10^{11}$ & $9.40\times10^{-2}$ & ... & \scriptsize{SCUBA \citep{Farrah02submm}} \\ 
$3.00\times10^{12}$ & $9.94\times10^{-1}$ & $2.39\times10^{-1}$ & \scriptsize{IRAS (NED)} \\ 
$5.00\times10^{12}$ & $6.22\times10^{-1}$ & $5.60\times10^{-2}$ & \scriptsize{IRAS (NED)} \\ 
$1.20\times10^{13}$ & $2.85\times10^{-1}$ & $1.41\times10^{-2}$ & \scriptsize{IRAS (NED)} \\ 
$2.50\times10^{13}$ & $1.18\times10^{-1}$ & $2.71\times10^{-2}$ & \scriptsize{IRAS (NED)} \\ 
$3.91\times10^{13}$ & $3.60\times10^{-2}$ & $4.00\times10^{-3}$ & \scriptsize{ISO (NED)} \\ 
$7.05\times10^{13}$ & $1.60\times10^{-2}$ & ... & \scriptsize{ISO (NED)} \\ 
$1.39\times10^{14}$ & $8.47\times10^{-3}$ & $2.29\times10^{-4}$ & \scriptsize{2MASS} \\ 
$1.80\times10^{14}$ & $4.75\times10^{-3}$ & $1.57\times10^{-4}$ & \scriptsize{2MASS} \\ 
$2.43\times10^{14}$ & $3.33\times10^{-3}$ & $1.01\times10^{-4}$ & \scriptsize{2MASS} \\ 
$3.33\times10^{14}$ & $2.26\times10^{-3}$ & $6.25\times10^{-4}$ & \scriptsize{SSS} \\ 
$3.28\times10^{14}$ & $3.83\times10^{-3}$ & $2.05\times10^{-5}$ & \scriptsize{SDSS-DR5} \\ 
$3.93\times10^{14}$ & $2.70\times10^{-3}$ & $1.23\times10^{-5}$ & \scriptsize{SDSS-DR5} \\ 
$4.81\times10^{14}$ & $2.82\times10^{-3}$ & $1.17\times10^{-5}$ & \scriptsize{SDSS-DR5} \\ 
$5.45\times10^{14}$ & $2.44\times10^{-3}$ & $1.90\times10^{-5}$ & \scriptsize{\XMM-OM} \\ 
$6.28\times10^{14}$ & $2.09\times10^{-3}$ & $9.05\times10^{-6}$ & \scriptsize{SDSS-DR5} \\ 
$8.47\times10^{14}$ & $1.30\times10^{-3}$ & $7.19\times10^{-6}$ & \scriptsize{SDSS-DR5} \\ 
$9.67\times10^{14}$ & $7.70\times10^{-4}$ & $4.00\times10^{-6}$ & \scriptsize{\XMM-OM} \\ 
$1.25\times10^{15}$ & $2.65\times10^{-4}$ & $7.00\times10^{-6}$ & \scriptsize{\XMM-OM} \\ 
$1.48\times10^{15}$ & $1.59\times10^{-4}$ & $9.00\times10^{-6}$ & \scriptsize{\XMM-OM} \\ 
$8.46\times10^{16}$ & $1.42\times10^{-10}$ & $8.39\times10^{-10}$ & \scriptsize{2XMMi} \\ 
$1.81\times10^{17}$ & $2.14\times10^{-10}$ & $3.50\times10^{-10}$ & \scriptsize{2XMMi} \\ 
$3.63\times10^{17}$ & $7.82\times10^{-10}$ & $3.76\times10^{-10}$ & \scriptsize{2XMMi} \\ 
$7.86\times10^{17}$ & $6.81\times10^{-10}$ & $3.85\times10^{-10}$ & \scriptsize{2XMMi} \\ 
$1.99\times10^{18}$ & $5.49\times10^{-10}$ & $5.17\times10^{-10}$ & \scriptsize{2XMMi} \\ 
\noalign{\smallskip}\hline\noalign{\smallskip}
\end{tabular}
\end{center}
\end{table}

\begin{table}[ht]
\caption{Photometric data for IRAS~F14218+3845.}
\label{tab:irasF14218data}
\begin{center}
\begin{tabular}{llll}
\hline\noalign{\smallskip}
$\nu$ (Hz) & $F_{\nu}$ (Jy) & Error (Jy) & Ref. \\
\hline\noalign{\smallskip}
$3.53\times10^{11}$ & $8.55\times10^{-3}$ & ... & \scriptsize{SCUBA \citep{Farrah02submm}} \\ 
$6.66\times10^{11}$ & $2.53\times10^{-1}$ & ... & \scriptsize{SCUBA \citep{Farrah02submm}} \\ 
$3.15\times10^{12}$ & $1.64\times10^{-1}$ & $6.10\times10^{-2}$ & \scriptsize{ISO (NED)} \\ 
$2.00\times10^{13}$ & $3.23\times10^{-3}$ & $1.04\times10^{-3}$ & \scriptsize{ISO (NED)} \\ 
$4.44\times10^{13}$ & $7.90\times10^{-4}$ & $2.60\times10^{-4}$ & \scriptsize{ISO (NED)} \\ 
$3.33\times10^{14}$ & $5.00\times10^{-5}$ & ... & \scriptsize{\citep{Farrah02submm}} \\ 
$3.28\times10^{14}$ & $7.97\times10^{-5}$ & $4.34\times10^{-6}$ & \scriptsize{SDSS-DR5} \\ 
$3.93\times10^{14}$ & $9.09\times10^{-5}$ & $1.26\times10^{-6}$ & \scriptsize{SDSS-DR5} \\ 
$4.28\times10^{14}$ & $1.14\times10^{-4}$ & $1.05\times10^{-5}$ & \scriptsize{APM} \\ 
$4.81\times10^{14}$ & $1.01\times10^{-4}$ & $1.02\times10^{-6}$ & \scriptsize{SDSS-DR5} \\ 
$6.28\times10^{14}$ & $8.52\times10^{-5}$ & $7.85\times10^{-7}$ & \scriptsize{SDSS-DR5} \\ 
$8.47\times10^{14}$ & $8.47\times10^{-5}$ & $2.26\times10^{-6}$ & \scriptsize{SDSS-DR5} \\ 
\noalign{\smallskip}\hline\noalign{\smallskip}
\end{tabular}
\end{center}
\end{table}

\begin{table}[ht]
\caption{Photometric data for IRAS~F15307+3252.}
\label{tab:irasF15307data}
\begin{center}
\begin{tabular}{llll}
\hline\noalign{\smallskip}
$\nu$ (Hz) & $F_{\nu}$ (Jy) & Error (Jy) & Ref. \\
\hline\noalign{\smallskip}
$1.40\times10^{9}$ & $5.71\times10^{-3}$ & $1.09\times10^{-4}$ & \scriptsize{VLA (NED)} \\ 
$8.42\times10^{9}$ & $9.20\times10^{-4}$ & $4.00\times10^{-5}$ & \scriptsize{VLA (NED)} \\ 
$1.02\times10^{11}$ & $4.50\times10^{-2}$ & ... & \scriptsize{OVMA (NED)} \\ 
$2.39\times10^{11}$ & $5.10\times10^{-3}$ & ... & \scriptsize{OVMA (NED)} \\ 
$3.53\times10^{11}$ & $1.15\times10^{-2}$ & ... & \scriptsize{SCUBA \citep{Farrah02submm}} \\ 
$6.66\times10^{11}$ & $1.06\times10^{-1}$ & ... & \scriptsize{SCUBA \citep{Farrah02submm}} \\ 
$1.66\times10^{12}$ & $4.14\times10^{-1}$ & $1.76\times10^{-1}$ & \scriptsize{ISO (NED)} \\ 
$3.00\times10^{12}$ & $5.10\times10^{-1}$ & $6.20\times10^{-2}$ & \scriptsize{IRAS (NED)} \\ 
$3.15\times10^{12}$ & $3.68\times10^{-1}$ & $1.16\times10^{-1}$ & \scriptsize{ISO (NED)} \\ 
$5.00\times10^{12}$ & $2.80\times10^{-1}$ & $2.70\times10^{-2}$ & \scriptsize{IRAS (NED)} \\ 
$1.20\times10^{13}$ & $8.00\times10^{-2}$ & $2.40\times10^{-2}$ & \scriptsize{IRAS (NED)} \\ 
$1.26\times10^{13}$ & $5.70\times10^{-2}$ & $3.00\times10^{-3}$ & \scriptsize{\Spitzer-IRAC} \\ 
$2.50\times10^{13}$ & $4.50\times10^{-2}$ & ... & \scriptsize{IRAS (NED)} \\ 
$1.39\times10^{14}$ & $3.20\times10^{-4}$ & $3.00\times10^{-5}$ & \scriptsize{MMT (NED)} \\ 
$1.80\times10^{14}$ & $1.68\times10^{-4}$ & $1.40\times10^{-5}$ & \scriptsize{MMT (NED)} \\ 
$2.43\times10^{14}$ & $1.48\times10^{-4}$ & $1.20\times10^{-5}$ & \scriptsize{MMT (NED)} \\ 
$3.33\times10^{14}$ & $6.70\times10^{-5}$ & $1.85\times10^{-5}$ & \scriptsize{SSS} \\ 
$3.28\times10^{14}$ & $1.28\times10^{-4}$ & $3.91\times10^{-6}$ & \scriptsize{SDSS-DR5} \\ 
$3.93\times10^{14}$ & $7.17\times10^{-5}$ & $1.06\times10^{-6}$ & \scriptsize{SDSS-DR5} \\ 
$4.28\times10^{14}$ & $6.32\times10^{-5}$ & $1.75\times10^{-5}$ & \scriptsize{SSS} \\ 
$4.81\times10^{14}$ & $5.90\times10^{-5}$ & $8.70\times10^{-7}$ & \scriptsize{SDSS-DR5} \\ 
$6.28\times10^{14}$ & $4.40\times10^{-5}$ & $6.89\times10^{-7}$ & \scriptsize{SDSS-DR5} \\ 
$6.81\times10^{14}$ & $3.40\times10^{-5}$ & $4.00\times10^{-6}$ & \scriptsize{\XMM-OM} \\ 
$8.21\times10^{14}$ & $2.90\times10^{-5}$ & $2.00\times10^{-6}$ & \scriptsize{\XMM-OM} \\ 
$8.47\times10^{14}$ & $3.34\times10^{-5}$ & $1.69\times10^{-6}$ & \scriptsize{SDSS-DR5} \\ 
$9.67\times10^{14}$ & $3.00\times10^{-5}$ & $2.00\times10^{-6}$ & \scriptsize{\XMM-OM} \\ 
$1.48\times10^{15}$ & $1.90\times10^{-5}$ & $6.00\times10^{-6}$ & \scriptsize{\XMM-OM} \\ 
\noalign{\smallskip}\hline\noalign{\smallskip}
\end{tabular}
\end{center}
\end{table}

\begin{table}[ht]
\caption{Photometric data for IRAS~16347+7037.}
\label{tab:iras16347data}
\begin{center}
\begin{tabular}{llll}
\hline\noalign{\smallskip}
$\nu$ (Hz) & $F_{\nu}$ (Jy) & Error (Jy) & Ref. \\
\hline\noalign{\smallskip}
$1.45\times10^{9}$ & $9.40\times10^{-4}$ & $3.00\times10^{-4}$ & \scriptsize{VLA (NED)} \\ 
$1.49\times10^{9}$ & $1.65\times10^{-3}$ & $3.10\times10^{-4}$ & \scriptsize{VLA (NED)} \\ 
$4.90\times10^{9}$ & $1.08\times10^{-3}$ & $1.10\times10^{-4}$ & \scriptsize{VLA (NED)} \\ 
$8.48\times10^{9}$ & $9.70\times10^{-4}$ & $1.50\times10^{-4}$ & \scriptsize{VLA (NED)} \\ 
$1.49\times10^{10}$ & $1.17\times10^{-3}$ & $3.20\times10^{-4}$ & \scriptsize{VLA (NED)} \\ 
$2.25\times10^{10}$ & $9.60\times10^{-4}$ & $1.70\times10^{-4}$ & \scriptsize{VLA (NED)} \\ 
$4.00\times10^{10}$ & $2.78\times10^{-1}$ & ... & \scriptsize{FCRAO (NED)} \\ 
$9.00\times10^{10}$ & $4.50\times10^{-2}$ & ... & \scriptsize{NRAO-12m (NED)} \\ 
$2.30\times10^{11}$ & $1.50\times10^{-3}$ & ... & \scriptsize{IRAM (NED)} \\ 
$1.48\times10^{12}$ & $1.58\times10^{-1}$ & $3.16\times10^{-2}$ & \scriptsize{ISO (NED)} \\ 
$1.72\times10^{12}$ & $2.06\times10^{-1}$ & $4.12\times10^{-2}$ & \scriptsize{ISO (NED)} \\ 
$1.93\times10^{12}$ & $2.30\times10^{-1}$ & $4.60\times10^{-2}$ & \scriptsize{ISO (NED)} \\ 
$2.93\times10^{12}$ & $3.49\times10^{-1}$ & $6.98\times10^{-2}$ & \scriptsize{ISO (NED)} \\ 
$5.00\times10^{12}$ & $2.46\times10^{-1}$ & $3.20\times10^{-2}$ & \scriptsize{IRAS (NED)} \\ 
$1.20\times10^{13}$ & $1.22\times10^{-1}$ & $4.15\times10^{-3}$ & \scriptsize{IRAS (NED)} \\ 
$2.50\times10^{13}$ & $5.93\times10^{-2}$ & $1.01\times10^{-2}$ & \scriptsize{IRAS (NED)} \\ 
$2.52\times10^{13}$ & $4.80\times10^{-2}$ & $9.60\times10^{-3}$ & \scriptsize{ISO (NED)} \\ 
$2.97\times10^{13}$ & $4.90\times10^{-2}$ & $1.10\times10^{-2}$ & \scriptsize{Hale-5m \citep{Neugebauer87}} \\ 
$8.10\times10^{13}$ & $1.07\times10^{-2}$ & $9.48\times10^{-4}$ & \scriptsize{Hale-5m \citep{Neugebauer87}} \\ 
$1.39\times10^{14}$ & $6.99\times10^{-3}$ & $2.13\times10^{-4}$ & \scriptsize{2MASS} \\ 
$1.80\times10^{14}$ & $8.91\times10^{-3}$ & $2.75\times10^{-4}$ & \scriptsize{2MASS} \\ 
$2.43\times10^{14}$ & $6.96\times10^{-3}$ & $2.06\times10^{-4}$ & \scriptsize{2MASS} \\ 
$3.33\times10^{14}$ & $7.07\times10^{-3}$ & $1.95\times10^{-3}$ & \scriptsize{SSS} \\ 
$4.28\times10^{14}$ & $7.27\times10^{-3}$ & $2.01\times10^{-3}$ & \scriptsize{SSS} \\ 
$5.45\times10^{14}$ & $4.76\times10^{-3}$ & $4.00\times10^{-5}$ & \scriptsize{\XMM-OM} \\ 
$6.81\times10^{14}$ & $6.20\times10^{-3}$ & $2.28\times10^{-3}$ & \scriptsize{SSS} \\ 
$8.21\times10^{14}$ & $3.71\times10^{-3}$ & $1.60\times10^{-5}$ & \scriptsize{\XMM-OM} \\ 
$1.05\times10^{15}$ & $3.77\times10^{-3}$ & $4.97\times10^{-6}$ & \scriptsize{IUE} \\ 
$1.46\times10^{15}$ & $1.22\times10^{-3}$ & $4.71\times10^{-6}$ & \scriptsize{IUE} \\ 
$1.48\times10^{15}$ & $1.59\times10^{-3}$ & $6.00\times10^{-5}$ & \scriptsize{\XMM-OM} \\ 
$2.11\times10^{15}$ & $4.60\times10^{-4}$ & $8.92\times10^{-6}$ & \scriptsize{IUE} \\ 
\noalign{\smallskip}\hline\noalign{\smallskip}
\end{tabular}
\end{center}
\end{table}

\begin{table}[ht]
\caption{Photometric data for IRAS~18216+6418.}
\label{tab:iras18216data}
\begin{center}
\begin{tabular}{llll}
\hline\noalign{\smallskip}
$\nu$ (Hz) & $F_{\nu}$ (Jy) & Error (Jy) & Ref. \\
\hline\noalign{\smallskip}
$3.25\times10^{8}$ & $2.11\times10^{-1}$ & $9.65\times10^{-3}$ & \scriptsize{WENSS (NED)} \\ 
$4.90\times10^{9}$ & $7.50\times10^{-3}$ & $4.00\times10^{-4}$ & \scriptsize{VLA (NED)} \\ 
$8.42\times10^{9}$ & $1.26\times10^{-2}$ & $6.00\times10^{-4}$ & \scriptsize{VLA (NED)} \\ 
$1.49\times10^{10}$ & $2.33\times10^{-2}$ & $1.10\times10^{-3}$ & \scriptsize{VLA (NED)} \\ 
$2.39\times10^{11}$ & $3.70\times10^{-3}$ & ... & \scriptsize{\citep{Farrah02submm}} \\ 
$2.73\times10^{11}$ & $4.70\times10^{-2}$ & ... & \scriptsize{\citep{Farrah02submm}} \\ 
$3.53\times10^{11}$ & $1.48\times10^{-2}$ & $2.60\times10^{-3}$ & \scriptsize{SCUBA \citep{Farrah02submm}} \\ 
$6.66\times10^{11}$ & $3.08\times10^{-1}$ & ... & \scriptsize{SCUBA \citep{Farrah02submm}} \\ 
$1.72\times10^{12}$ & $1.07\times10^{0}$ & $3.21\times10^{-1}$ & \scriptsize{ISO (NED)} \\ 
$3.00\times10^{12}$ & $2.13\times10^{0}$ & $1.70\times10^{-1}$ & \scriptsize{IRAS (NED)} \\ 
$5.00\times10^{12}$ & $1.24\times10^{0}$ & $4.96\times10^{-2}$ & \scriptsize{IRAS (NED)} \\ 
$1.20\times10^{13}$ & $4.45\times10^{-1}$ & $1.19\times10^{-2}$ & \scriptsize{IRAS (NED)} \\ 
$2.50\times10^{13}$ & $2.38\times10^{-1}$ & ... & \scriptsize{IRAS (NED)} \\ 
$3.12\times10^{13}$ & $1.43\times10^{-1}$ & $1.43\times10^{-2}$ & \scriptsize{ISO (NED)} \\ 
$4.44\times10^{13}$ & $1.09\times10^{-1}$ & $1.10\times10^{-2}$ & \scriptsize{ISO (NED)} \\ 
$1.39\times10^{14}$ & $2.35\times10^{-2}$ & $6.04\times10^{-4}$ & \scriptsize{2MASS} \\ 
$1.80\times10^{14}$ & $1.47\times10^{-2}$ & $4.25\times10^{-4}$ & \scriptsize{2MASS} \\ 
$2.43\times10^{14}$ & $1.19\times10^{-2}$ & $3.18\times10^{-4}$ & \scriptsize{2MASS} \\ 
$3.33\times10^{14}$ & $1.69\times10^{-2}$ & $4.68\times10^{-3}$ & \scriptsize{SSS} \\ 
$4.28\times10^{14}$ & $1.24\times10^{-2}$ & $3.42\times10^{-3}$ & \scriptsize{SSS} \\ 
$5.45\times10^{14}$ & $7.55\times10^{-3}$ & ... & \scriptsize{\citep{Farrah02submm}} \\ 
$6.81\times10^{14}$ & $8.10\times10^{-3}$ & $7.00\times10^{-4}$ & \scriptsize{APM} \\ 
$8.21\times10^{14}$ & $9.04\times10^{-3}$ & ... & \scriptsize{\citep{Farrah02submm}} \\ 
$1.48\times10^{15}$ & $4.94\times10^{-3}$ & $6.00\times10^{-5}$ & \scriptsize{\XMM-OM} \\ 
\noalign{\smallskip}\hline\noalign{\smallskip}
\end{tabular}
\end{center}
\end{table}

\end{appendix}
 

\begin{thebibliography}{71}
\expandafter\ifx\csname natexlab\endcsname\relax\def\natexlab#1{#1}\fi

\bibitem[{{Adelman-McCarthy} {et~al.}(2007){Adelman-McCarthy}, {Ag{\"u}eros},
  {Allam}, {Anderson}, {Anderson}, {Annis}, {Bahcall}, {Bailer-Jones},
  {Baldry}, {Barentine}, {Beers}, {Belokurov}, {Berlind}, {Bernardi},
  {Blanton}, {Bochanski}, {Boroski}, {Bramich}, {Brewington}, {Brinchmann},
  {Brinkmann}, {Brunner}, {Budav{\'a}ri}, {Carey}, {Carliles}, {Carr},
  {Castander}, {Connolly}, {Cool}, {Cunha}, {Csabai}, {Dalcanton}, {Doi},
  {Eisenstein}, {Evans}, {Evans}, {Fan}, {Finkbeiner}, {Friedman}, {Frieman},
  {Fukugita}, {Gillespie}, {Gilmore}, {Glazebrook}, {Gray}, {Grebel}, {Gunn},
  {de Haas}, {Hall}, {Harvanek}, {Hawley}, {Hayes}, {Heckman}, {Hendry},
  {Hennessy}, {Hindsley}, {Hirata}, {Hogan}, {Hogg}, {Holtzman}, {Ichikawa},
  {Ichikawa}, {Ivezi{\'c}}, {Jester}, {Johnston}, {Jorgensen}, {Juri{\'c}},
  {Kauffmann}, {Kent}, {Kleinman}, {Knapp}, {Kniazev}, {Kron}, {Krzesinski},
  {Kuropatkin}, {Lamb}, {Lampeitl}, {Lee}, {Leger}, {Lima}, {Lin}, {Long},
  {Loveday}, {Lupton}, {Mandelbaum}, {Margon}, {Mart{\'{\i}}nez-Delgado},
  {Matsubara}, {McGehee}, {McKay}, {Meiksin}, {Munn}, {Nakajima}, {Nash},
  {Neilsen}, {Newberg}, {Nichol}, {Nieto-Santisteban}, {Nitta}, {Oyaizu},
  {Okamura}, {Ostriker}, {Padmanabhan}, {Park}, {Peoples}, {Pier}, {Pope},
  {Pourbaix}, {Quinn}, {Raddick}, {Re Fiorentin}, {Richards}, {Richmond},
  {Rix}, {Rockosi}, {Schlegel}, {Schneider}, {Scranton}, {Seljak}, {Sheldon},
  {Shimasaku}, {Silvestri}, {Smith}, {Smol{\v c}i{\'c}}, {Snedden}, {Stebbins},
  {Stoughton}, {Strauss}, {SubbaRao}, {Suto}, {Szalay}, {Szapudi}, {Szkody},
  {Tegmark}, {Thakar}, {Tremonti}, {Tucker}, {Uomoto}, {Vanden Berk},
  {Vandenberg}, {Vidrih}, {Vogeley}, {Voges}, {Vogt}, {Weinberg}, {West},
  {White}, {Wilhite}, {Yanny}, {Yocum}, {York}, {Zehavi}, {Zibetti}, \&
  {Zucker}}]{sloan5}
{Adelman-McCarthy}, J.~K., {Ag{\"u}eros}, M.~A., {Allam}, S.~S., {et~al.} 2007,
  \apjs, 172, 634

\bibitem[{{Antonucci} \& {Miller}(1985)}]{Antonucci85}
{Antonucci}, R.~R.~J. \& {Miller}, J.~S. 1985, \apj, 297, 621

\bibitem[{{Beck} {et~al.}(1996){Beck}, {Turner}, {Ho}, {Lacy}, \&
  {Kelly}}]{Beck96}
{Beck}, S.~C., {Turner}, J.~L., {Ho}, P.~T.~P., {Lacy}, J.~H., \& {Kelly},
  D.~M. 1996, \apj, 457, 610

\bibitem[{{Berta}(2005)}]{Berta05}
{Berta}, S. 2005, PhD thesis, AA(Dipartimento di Astronomia, Univ.~di Padova,
  Vicolo dell'Osservatorio 2, I-35122, Padova, Italy)

\bibitem[{{Berta} {et~al.}(2003){Berta}, {Fritz}, {Franceschini}, {Bressan}, \&
  {Pernechele}}]{Berta03}
{Berta}, S., {Fritz}, J., {Franceschini}, A., {Bressan}, A., \& {Pernechele},
  C. 2003, \aap, 403, 119

\bibitem[{{Bianchi} {et~al.}(2006){Bianchi}, {Guainazzi}, \&
  {Chiaberge}}]{Bianchi06}
{Bianchi}, S., {Guainazzi}, M., \& {Chiaberge}, M. 2006, \aap, 448, 499

\bibitem[{{Brandl} {et~al.}(2004){Brandl}, {Devost}, {Higdon}, {Charmandaris},
  {Weedman}, {Spoon}, {Herter}, {Hao}, {Bernard-Salas}, {Houck}, {Armus},
  {Soifer}, {Grillmair}, \& {Appleton}}]{Brandl04}
{Brandl}, B.~R., {Devost}, D., {Higdon}, S.~J.~U., {et~al.} 2004, \apjs, 154,
  188

\bibitem[{{Cutri} {et~al.}(2003){Cutri}, {Skrutskie}, {van Dyk}, {Beichman},
  {Carpenter}, {Chester}, {Cambresy}, {Evans}, {Fowler}, {Gizis}, {Howard},
  {Huchra}, {Jarrett}, {Kopan}, {Kirkpatrick}, {Light}, {Marsh}, {McCallon},
  {Schneider}, {Stiening}, {Sykes}, {Weinberg}, {Wheaton}, {Wheelock}, \&
  {Zacarias}}]{2MASS}
{Cutri}, R.~M., {Skrutskie}, M.~F., {van Dyk}, S., {et~al.} 2003, {2MASS All
  Sky Catalog of point sources.} (The IRSA 2MASS All-Sky Point Source Catalog,
  NASA/IPAC Infrared Science
  Archive.~http://irsa.ipac.caltech.edu/applications/Gator/)

\bibitem[{{Davies} {et~al.}(2004){Davies}, {Tacconi}, \& {Genzel}}]{Davies04}
{Davies}, R.~I., {Tacconi}, L.~J., \& {Genzel}, R. 2004, \apj, 613, 781

\bibitem[{{Davis} {et~al.}(2007){Davis}, {Guhathakurta}, {Konidaris}, {Newman},
  {Ashby}, {Biggs}, {Barmby}, {Bundy}, {Chapman}, {Coil}, {Conselice},
  {Cooper}, {Croton}, {Eisenhardt}, {Ellis}, {Faber}, {Fang}, {Fazio},
  {Georgakakis}, {Gerke}, {Goss}, {Gwyn}, {Harker}, {Hopkins}, {Huang},
  {Ivison}, {Kassin}, {Kirby}, {Koekemoer}, {Koo}, {Laird}, {Le Floc'h}, {Lin},
  {Lotz}, {Marshall}, {Martin}, {Metevier}, {Moustakas}, {Nandra}, {Noeske},
  {Papovich}, {Phillips}, {Rich}, {Rieke}, {Rigopoulou}, {Salim},
  {Schiminovich}, {Simard}, {Smail}, {Small}, {Weiner}, {Willmer}, {Willner},
  {Wilson}, {Wright}, \& {Yan}}]{Davis07}
{Davis}, M., {Guhathakurta}, P., {Konidaris}, N.~P., {et~al.} 2007, \apjl, 660,
  L1

\bibitem[{{Deane} \& {Trentham}(2001)}]{Deane01}
{Deane}, J.~R. \& {Trentham}, N. 2001, \mnras, 326, 1467

\bibitem[{{Della Ceca} {et~al.}(2004){Della Ceca}, {Maccacaro}, {Caccianiga},
  {Severgnini}, {Braito}, {Barcons}, {Carrera}, {Watson}, {Tedds}, {Brunner},
  {Lehmann}, {Page}, {Lamer}, \& {Schwope}}]{DellaCeca04}
{Della Ceca}, R., {Maccacaro}, T., {Caccianiga}, A., {et~al.} 2004, \aap, 428,
  383

\bibitem[{{Dickey} \& {Lockman}(1990)}]{nh}
{Dickey}, J.~M. \& {Lockman}, F.~J. 1990, \araa, 28, 215

\bibitem[{{Dickinson} {et~al.}(2003){Dickinson}, {Giavalisco}, \& {The Goods
  Team}}]{Dickinson03}
{Dickinson}, M., {Giavalisco}, M., \& {The Goods Team}. 2003, in The Mass of
  Galaxies at Low and High Redshift, ed. R.~{Bender} \& A.~{Renzini}, 324

\bibitem[{{Drake} {et~al.}(2004){Drake}, {Bicknell}, {McGregor}, \&
  {Dopita}}]{Drake04}
{Drake}, C.~L., {Bicknell}, G.~V., {McGregor}, P.~J., \& {Dopita}, M.~A. 2004,
  \aj, 128, 969

\bibitem[{{Efstathiou} \& {Rowan-Robinson}(1995)}]{Efstathiou95}
{Efstathiou}, A. \& {Rowan-Robinson}, M. 1995, \mnras, 273, 649

\bibitem[{{Efstathiou} {et~al.}(2000){Efstathiou}, {Rowan-Robinson}, \&
  {Siebenmorgen}}]{Efstathiou00}
{Efstathiou}, A., {Rowan-Robinson}, M., \& {Siebenmorgen}, R. 2000, \mnras,
  313, 734

\bibitem[{{Elvis} {et~al.}(1994){Elvis}, {Wilkes}, {McDowell}, {Green},
  {Bechtold}, {Willner}, {Oey}, {Polomski}, \& {Cutri}}]{Elvis94}
{Elvis}, M., {Wilkes}, B.~J., {McDowell}, J.~C., {et~al.} 1994, \apjs, 95, 1

\bibitem[{{Fabian} \& {Iwasawa}(1999)}]{Fabian99}
{Fabian}, A.~C. \& {Iwasawa}, K. 1999, \mnras, 303, L34

\bibitem[{{Farrah} {et~al.}(2003){Farrah}, {Afonso}, {Efstathiou},
  {Rowan-Robinson}, {Fox}, \& {Clements}}]{Farrah03}
{Farrah}, D., {Afonso}, J., {Efstathiou}, A., {et~al.} 2003, \mnras, 343, 585

\bibitem[{{Farrah} {et~al.}(2001){Farrah}, {Rowan-Robinson}, {Oliver},
  {Serjeant}, {Borne}, {Lawrence}, {Lucas}, {Bushouse}, \& {Colina}}]{Farrah01}
{Farrah}, D., {Rowan-Robinson}, M., {Oliver}, S., {et~al.} 2001, \mnras, 326,
  1333

\bibitem[{{Farrah} {et~al.}(2002{\natexlab{a}}){Farrah}, {Serjeant},
  {Efstathiou}, {Rowan-Robinson}, \& {Verma}}]{Farrah02submm}
{Farrah}, D., {Serjeant}, S., {Efstathiou}, A., {Rowan-Robinson}, M., \&
  {Verma}, A. 2002{\natexlab{a}}, \mnras, 335, 1163

\bibitem[{{Farrah} {et~al.}(2002{\natexlab{b}}){Farrah}, {Verma}, {Oliver},
  {Rowan-Robinson}, \& {McMahon}}]{Farrah02hst}
{Farrah}, D., {Verma}, A., {Oliver}, S., {Rowan-Robinson}, M., \& {McMahon}, R.
  2002{\natexlab{b}}, \mnras, 329, 605

\bibitem[{{Franceschini} {et~al.}(2003){Franceschini}, {Braito}, {Persic},
  {Della Ceca}, {Bassani}, {Cappi}, {Malaguti}, {Palumbo}, {Risaliti},
  {Salvati}, \& {Severgnini}}]{Fran03}
{Franceschini}, A., {Braito}, V., {Persic}, M., {et~al.} 2003, \mnras, 343,
  1181

\bibitem[{{Franceschini} {et~al.}(1999){Franceschini}, {Hasinger}, {Miyaji}, \&
  {Malquori}}]{Fran99}
{Franceschini}, A., {Hasinger}, G., {Miyaji}, T., \& {Malquori}, D. 1999,
  \mnras, 310, L5

\bibitem[{{Freeman} {et~al.}(2001){Freeman}, {Doe}, \&
  {Siemiginowska}}]{sherpa}
{Freeman}, P., {Doe}, S., \& {Siemiginowska}, A. 2001, in Presented at the
  Society of Photo-Optical Instrumentation Engineers (SPIE) Conference, Vol.
  4477, Society of Photo-Optical Instrumentation Engineers (SPIE) Conference
  Series, ed. {J.-L.~Starck \& F.~D.~Murtagh}, 76--87

\bibitem[{{Gilli} {et~al.}(2007){Gilli}, {Comastri}, \& {Hasinger}}]{Gilli07}
{Gilli}, R., {Comastri}, A., \& {Hasinger}, G. 2007, \aap, 463, 79

\bibitem[{{Gonzalez-Delgado} {et~al.}(1995){Gonzalez-Delgado}, {Perez}, {Diaz},
  {Garcia-Vargas}, {Terlevich}, \& {Vilchez}}]{Gonzalez95}
{Gonzalez-Delgado}, R.~M., {Perez}, E., {Diaz}, A.~I., {et~al.} 1995, \apj,
  439, 604

\bibitem[{{Hopkins} {et~al.}(2007){Hopkins}, {Richards}, \&
  {Hernquist}}]{Hopkins07}
{Hopkins}, P.~F., {Richards}, G.~T., \& {Hernquist}, L. 2007, \apj, 654, 731

\bibitem[{{Imanishi} {et~al.}(2007){Imanishi}, {Dudley}, {Maiolino}, {Maloney},
  {Nakagawa}, \& {Risaliti}}]{Imanishi07}
{Imanishi}, M., {Dudley}, C.~C., {Maiolino}, R., {et~al.} 2007, \apjs, 171, 72

\bibitem[{{Iwasawa} {et~al.}(2005){Iwasawa}, {Crawford}, {Fabian}, \&
  {Wilman}}]{Iwasawa05}
{Iwasawa}, K., {Crawford}, C.~S., {Fabian}, A.~C., \& {Wilman}, R.~J. 2005,
  \mnras, 362, L20

\bibitem[{{Joint IRAS Science Working Group}(1988)}]{irasPSC}
{Joint IRAS Science Working Group}. 1988, in IRAS Point Source Catalog (1988)

\bibitem[{{Kelly} {et~al.}(2008){Kelly}, {Bechtold}, {Trump}, {Vestergaard}, \&
  {Siemiginowska}}]{Kelly08}
{Kelly}, B.~C., {Bechtold}, J., {Trump}, J.~R., {Vestergaard}, M., \&
  {Siemiginowska}, A. 2008, \apjs, 176, 355

\bibitem[{{Kleinmann} {et~al.}(1988){Kleinmann}, {Hamilton}, {Keel},
  {Wynn-Williams}, {Eales}, {Becklin}, \& {Kuntz}}]{Klein88}
{Kleinmann}, S.~G., {Hamilton}, D., {Keel}, W.~C., {et~al.} 1988, \apj, 328,
  161

\bibitem[{{Kormendy} \& {Gebhardt}(2001)}]{Kormendy01}
{Kormendy}, J. \& {Gebhardt}, K. 2001, in American Institute of Physics
  Conference Series, Vol. 586, 20th Texas Symposium on relativistic
  astrophysics, ed. J.~C. {Wheeler} \& H.~{Martel}, 363

\bibitem[{{Leighly} {et~al.}(2001){Leighly}, {Halpern}, {Helfand}, {Becker}, \&
  {Impey}}]{Leighly01}
{Leighly}, K.~M., {Halpern}, J.~P., {Helfand}, D.~J., {Becker}, R.~H., \&
  {Impey}, C.~D. 2001, \aj, 121, 2889

\bibitem[{{Leighly} {et~al.}(2007){Leighly}, {Halpern}, {Jenkins}, {Grupe},
  {Choi}, \& {Prescott}}]{Leighly07}
{Leighly}, K.~M., {Halpern}, J.~P., {Jenkins}, E.~B., {et~al.} 2007, \apj, 663,
  103

\bibitem[{{Lonsdale} {et~al.}(2006){Lonsdale}, {Farrah}, \&
  {Smith}}]{Lonsdale06}
{Lonsdale}, C.~J., {Farrah}, D., \& {Smith}, H.~E. 2006, {Ultraluminous
  Infrared Galaxies}, ed. J.~W. Mason (Springer Verlag), 285

\bibitem[{{Magdziarz} \& {Zdziarski}(1995)}]{pexrav}
{Magdziarz}, P. \& {Zdziarski}, A.~A. 1995, \mnras, 273, 837

\bibitem[{{Magorrian} {et~al.}(1998){Magorrian}, {Tremaine}, {Richstone},
  {Bender}, {Bower}, {Dressler}, {Faber}, {Gebhardt}, {Green}, {Grillmair},
  {Kormendy}, \& {Lauer}}]{Magorrian98}
{Magorrian}, J., {Tremaine}, S., {Richstone}, D., {et~al.} 1998, \aj, 115, 2285

\bibitem[{{Matt} {et~al.}(1997){Matt}, {Guainazzi}, {Frontera}, {Bassani},
  {Brandt}, {Fabian}, {Fiore}, {Haardt}, {Iwasawa}, {Maiolino}, {Malaguti},
  {Marconi}, {Matteuzzi}, {Molendi}, {Perola}, {Piraino}, \& {Piro}}]{Matt97}
{Matt}, G., {Guainazzi}, M., {Frontera}, F., {et~al.} 1997, \aap, 325, L13

\bibitem[{{McLure} \& {Dunlop}(2002)}]{McLure02}
{McLure}, R.~J. \& {Dunlop}, J.~S. 2002, \mnras, 331, 795

\bibitem[{{Moshir} {et~al.}(1990){Moshir}, {Kopan}, {Conrow}, {McCallon},
  {Hacking}, {Gregorich}, {Rohrback}, {Melnyk}, {Rice}, {Fullmer}, {White}, \&
  {Chester}}]{irasFSC}
{Moshir}, M., {Kopan}, G., {Conrow}, T., {et~al.} 1990, in IRAS Faint Source
  Catalogue, version 2.0 (1990)

\bibitem[{{Nardini} {et~al.}(2008){Nardini}, {Risaliti}, {Salvati}, {Sani},
  {Imanishi}, {Marconi}, \& {Maiolino}}]{Nardini08}
{Nardini}, E., {Risaliti}, G., {Salvati}, M., {et~al.} 2008, \mnras, 385, L130

\bibitem[{{Neugebauer} {et~al.}(1987){Neugebauer}, {Green}, {Matthews},
  {Schmidt}, {Soifer}, \& {Bennett}}]{Neugebauer87}
{Neugebauer}, G., {Green}, R.~F., {Matthews}, K., {et~al.} 1987, \apjs, 63, 615

\bibitem[{{Page} {et~al.}(2007){Page}, {Carrera}, {Ebrero}, {Stevens}, \&
  {Ivison}}]{Page07}
{Page}, M.~J., {Carrera}, F.~J., {Ebrero}, J., {Stevens}, J.~A., \& {Ivison},
  R.~J. 2007, in Studying Galaxy Evolution with Spitzer and Herschel, ed.
  V.~{Charmandaris}, D.~{Rigopoulou}, \& N.~{Kylafis}

\bibitem[{{Persic} \& {Rephaeli}(2002)}]{Persic02}
{Persic}, M. \& {Rephaeli}, Y. 2002, \aap, 382, 843

\bibitem[{{Ptak} {et~al.}(2003){Ptak}, {Heckman}, {Levenson}, {Weaver}, \&
  {Strickland}}]{Ptak03}
{Ptak}, A., {Heckman}, T., {Levenson}, N.~A., {Weaver}, K., \& {Strickland}, D.
  2003, \apj, 592, 782

\bibitem[{{Quinn} {et~al.}(2004){Quinn}, {Barnes}, {Csabai}, {Cui}, {Genova},
  {Hanisch}, {Kembhavi}, {Kim}, {Lawrence}, {Malkov}, {Ohishi}, {Pasian},
  {Schade}, \& {Voges}}]{virobs}
{Quinn}, P.~J., {Barnes}, D.~G., {Csabai}, I., {et~al.} 2004, in Presented at
  the Society of Photo-Optical Instrumentation Engineers (SPIE) Conference,
  Vol. 5493, Society of Photo-Optical Instrumentation Engineers (SPIE)
  Conference Series, ed. {P.~J.~Quinn \& A.~Bridger}, 137--145

\bibitem[{{Richards} {et~al.}(2006){Richards}, {Lacy}, {Storrie-Lombardi},
  {Hall}, {Gallagher}, {Hines}, {Fan}, {Papovich}, {Vanden Berk}, {Trammell},
  {Schneider}, {Vestergaard}, {York}, {Jester}, {Anderson}, {Budav{\'a}ri}, \&
  {Szalay}}]{Richards06}
{Richards}, G.~T., {Lacy}, M., {Storrie-Lombardi}, L.~J., {et~al.} 2006, \apjs,
  166, 470

\bibitem[{{Risaliti} \& {Elvis}(2004)}]{Risaliti04}
{Risaliti}, G. \& {Elvis}, M. 2004, {A Panchromatic View of AGN} (ASSL
  Vol.~308: Supermassive Black Holes in the Distant Universe), 187

\bibitem[{{Rowan-Robinson}(1995)}]{Rowan95}
{Rowan-Robinson}, M. 1995, \mnras, 272, 737

\bibitem[{{Rowan-Robinson}(2000)}]{Rowan00}
{Rowan-Robinson}, M. 2000, \mnras, 316, 885

\bibitem[{{Ruiz} {et~al.}(2007){Ruiz}, {Carrera}, \& {Panessa}}]{Ruiz07}
{Ruiz}, A., {Carrera}, F.~J., \& {Panessa}, F. 2007, \aap, 471, 775

\bibitem[{{Scoville} {et~al.}(2007){Scoville}, {Aussel}, {Brusa}, {Capak},
  {Carollo}, {Elvis}, {Giavalisco}, {Guzzo}, {Hasinger}, {Impey}, {Kneib},
  {LeFevre}, {Lilly}, {Mobasher}, {Renzini}, {Rich}, {Sanders}, {Schinnerer},
  {Schminovich}, {Shopbell}, {Taniguchi}, \& {Tyson}}]{Scoville07}
{Scoville}, N., {Aussel}, H., {Brusa}, M., {et~al.} 2007, \apjs, 172, 1

\bibitem[{{Silverman} {et~al.}(2005){Silverman}, {Green}, {Barkhouse},
  {Cameron}, {Foltz}, {Jannuzi}, {Kim}, {Kim}, {Mossman}, {Tananbaum},
  {Wilkes}, {Smith}, {Smith}, \& {Smith}}]{Silverman05}
{Silverman}, J.~D., {Green}, P.~J., {Barkhouse}, W.~A., {et~al.} 2005, \apj,
  624, 630

\bibitem[{{Spergel} {et~al.}(2003){Spergel}, {Verde}, {Peiris}, {Komatsu},
  {Nolta}, {Bennett}, {Halpern}, {Hinshaw}, {Jarosik}, {Kogut}, {Limon},
  {Meyer}, {Page}, {Tucker}, {Weiland}, {Wollack}, \& {Wright}}]{Spergel03}
{Spergel}, D.~N., {Verde}, L., {Peiris}, H.~V., {et~al.} 2003, \apjs, 148, 175

\bibitem[{{Steffen} {et~al.}(2006){Steffen}, {Strateva}, {Brandt}, {Alexander},
  {Koekemoer}, {Lehmer}, {Schneider}, \& {Vignali}}]{Steffen06}
{Steffen}, A.~T., {Strateva}, I., {Brandt}, W.~N., {et~al.} 2006, \aj, 131,
  2826

\bibitem[{{Stevens} {et~al.}(2005){Stevens}, {Page}, {Ivison}, {Carrera},
  {Mittaz}, {Smail}, \& {McHardy}}]{Stevens05}
{Stevens}, J.~A., {Page}, M.~J., {Ivison}, R.~J., {et~al.} 2005, \mnras, 360,
  610

\bibitem[{{Strateva} {et~al.}(2005){Strateva}, {Brandt}, {Schneider}, {Vanden
  Berk}, \& {Vignali}}]{Strateva05}
{Strateva}, I.~V., {Brandt}, W.~N., {Schneider}, D.~P., {Vanden Berk}, D.~G.,
  \& {Vignali}, C. 2005, \aj, 130, 387

\bibitem[{{Strickland} {et~al.}(2004){Strickland}, {Heckman}, {Colbert},
  {Hoopes}, \& {Weaver}}]{Strickland04}
{Strickland}, D.~K., {Heckman}, T.~M., {Colbert}, E.~J.~M., {Hoopes}, C.~G., \&
  {Weaver}, K.~A. 2004, \apjs, 151, 193

\bibitem[{{Telesco} {et~al.}(1984){Telesco}, {Becklin}, {Wynn-Williams}, \&
  {Harper}}]{Telesco84}
{Telesco}, C.~M., {Becklin}, E.~E., {Wynn-Williams}, C.~G., \& {Harper}, D.~A.
  1984, \apj, 282, 427

\bibitem[{{Teng} {et~al.}(2005){Teng}, {Wilson}, {Veilleux}, {Young},
  {Sanders}, \& {Nagar}}]{Teng05}
{Teng}, S.~H., {Wilson}, A.~S., {Veilleux}, S., {et~al.} 2005, \apj, 633, 664

\bibitem[{{Ueda} {et~al.}(2003){Ueda}, {Akiyama}, {Ohta}, \& {Miyaji}}]{Ueda03}
{Ueda}, Y., {Akiyama}, M., {Ohta}, K., \& {Miyaji}, T. 2003, \apj, 598, 886

\bibitem[{{Veilleux} {et~al.}(1999){Veilleux}, {Kim}, \&
  {Sanders}}]{Veilleux99}
{Veilleux}, S., {Kim}, D.-C., \& {Sanders}, D.~B. 1999, \apj, 522, 113

\bibitem[{{Veilleux} {et~al.}(2002){Veilleux}, {Kim}, \&
  {Sanders}}]{Veilleux02}
{Veilleux}, S., {Kim}, D.-C., \& {Sanders}, D.~B. 2002, \apjs, 143, 315

\bibitem[{{Veilleux} {et~al.}(1995){Veilleux}, {Kim}, {Sanders}, {Mazzarella},
  \& {Soifer}}]{Veilleux95}
{Veilleux}, S., {Kim}, D.-C., {Sanders}, D.~B., {Mazzarella}, J.~M., \&
  {Soifer}, B.~T. 1995, \apjs, 98, 171

\bibitem[{{Verma} {et~al.}(2002){Verma}, {Rowan-Robinson}, {McMahon}, \&
  {Andreas Efstathiou}}]{Verma02}
{Verma}, A., {Rowan-Robinson}, M., {McMahon}, R., \& {Andreas Efstathiou},
  A.~E. 2002, \mnras, 335, 574

\bibitem[{{Watson} {et~al.}(2009){Watson}, {Schr{\"o}der}, {Fyfe}, {Page},
  {Lamer}, {Mateos}, {Pye}, {Sakano}, {Rosen}, {Ballet}, {Barcons}, {Barret},
  {Boller}, {Brunner}, {Brusa}, {Caccianiga}, {Carrera}, {Ceballos}, {Della
  Ceca}, {Denby}, {Denkinson}, {Dupuy}, {Farrell}, {Fraschetti}, {Freyberg},
  {Guillout}, {Hambaryan}, {Maccacaro}, {Mathiesen}, {McMahon}, {Michel},
  {Motch}, {Osborne}, {Page}, {Pakull}, {Pietsch}, {Saxton}, {Schwope},
  {Severgnini}, {Simpson}, {Sironi}, {Stewart}, {Stewart}, {Stobbart}, {Tedds},
  {Warwick}, {Webb}, {West}, {Worrall}, \& {Yuan}}]{2xmmi}
{Watson}, M.~G., {Schr{\"o}der}, A.~C., {Fyfe}, D., {et~al.} 2009, \aap, 493,
  339

\bibitem[{{Wilman} {et~al.}(2003){Wilman}, {Fabian}, {Crawford}, \&
  {Cutri}}]{Wilman03}
{Wilman}, R.~J., {Fabian}, A.~C., {Crawford}, C.~S., \& {Cutri}, R.~M. 2003,
  \mnras, 338, L19

\bibitem[{{Wilman} {et~al.}(1998){Wilman}, {Fabian}, {Cutri}, {Crawford}, \&
  {Brandt}}]{Wilman98}
{Wilman}, R.~J., {Fabian}, A.~C., {Cutri}, R.~M., {Crawford}, C.~S., \&
  {Brandt}, W.~N. 1998, \mnras, 300, L7

\end{thebibliography}
\end{document}